\documentstyle[10pt,epsf,epsfig,dp_delphititle,rotating,amsmath,lineno]
{dp_delphi}
\makeindex
\def\DpPaperGroup{EP}
\def\DpPaperRef{2002-082}
\def\DpDate{11 February 2003}
\def\DpAuthors{DELPHI Collaboration}
\def\DpSubmit{(Accepted by Eur.Phys.J.C)}
\def\DpTitle{{
A study of the energy evolution of\\ event shape distributions\\ and
their means\\ with the DELPHI detector at LEP
   }}
\def\DpComment{ }
\def\DpEMail{ }

\newcommand{\ecm}{E_{\mathrm{cm}}}

\newcommand{\fig}{Figure~\ref}
\newcommand{\tab}{Table~\ref}
\newcommand{\as}{$\alpha_s$}

\newcommand{\oas}{$\cal O$($\alpha_s^2$)}

\newcommand{\gev}{\mbox{\,Ge\kern-0.2exV}}
\newcommand{\mev}{\mbox{\,Me\kern-0.2exV}}
\newcommand{\bmin}{$B_{\mathrm{min}}$}

\newcommand{\Msum}{M^2_{\mathrm{s}}/E^2_{\mathrm{vis}}}
\newcommand{\Mhigh}{M^2_{\mathrm{h}}/E^2_{\mathrm{vis}}}

\newcommand{\epem}{${\mathrm{e^+e^-}}$}
\newcommand{\qcd}{${\mathrm{e^+e^-}} \rightarrow {\rm Z}/\gamma
\rightarrow {\mathrm{q}}\bar{\mathrm{q}}$\hspace{0.1cm}}

\newcommand{\asb}{${\alpha}_0$}
\newcommand{\beq}{\begin{equation}}
\newcommand{\eeq}{\end{equation}}
\def\jetset{{\sc Jetset}}
\def\pythia{{\sc Pythia}}

\def\ariadne{{\sc Ariadne}}

\def\delsim{{\sc Delsim}}

\def\delphi{{\sc Delphi}}
\def\lep{{\sc Lep}}

\begin{document}
\makeatletter
\newcount\@tempcntc
\def\@citex[#1]#2{\if@filesw\immediate\write\@auxout{\string\citation{#2}}\fi
  \@tempcnta\z@\@tempcntb\m@ne\def\@citea{}\@cite{\@for\@citeb:=#2\do
    {\@ifundefined
       {b@\@citeb}{\@citeo\@tempcntb\m@ne\@citea\def\@citea{,}{\bf ?}\@warning
       {Citation `\@citeb' on page \thepage \space undefined}}%
    {\setbox\z@\hbox{\global\@tempcntc0\csname b@\@citeb\endcsname\relax}%
     \ifnum\@tempcntc=\z@ \@citeo\@tempcntb\m@ne
       \@citea\def\@citea{,}\hbox{\csname b@\@citeb\endcsname}%
     \else
      \advance\@tempcntb\@ne
      \ifnum\@tempcntb=\@tempcntc
      \else\advance\@tempcntb\m@ne\@citeo
      \@tempcnta\@tempcntc\@tempcntb\@tempcntc\fi\fi}}\@citeo}{#1}}
\def\@citeo{\ifnum\@tempcnta>\@tempcntb\else\@citea\def\@citea{,}%
  \ifnum\@tempcnta=\@tempcntb\the\@tempcnta\else
   {\advance\@tempcnta\@ne\ifnum\@tempcnta=\@tempcntb \else \def\@citea{--}\fi
    \advance\@tempcnta\m@ne\the\@tempcnta\@citea\the\@tempcntb}\fi\fi}
 
\makeatother
\begin{titlepage}
\pagenumbering{roman}
\CERNpreprint{\DpPaperGroup}{\DpPaperRef} 
\date{{\small\DpDate}} 
\title{\DpTitle} 
\address{\DpAuthors} 
\begin{shortabs} 
\noindent

\noindent Infrared and collinear safe event shape distributions and their
mean values are determined in \epem\ collisions at  
centre-of-mass energies between 45 and 202\gev.
A phenomenological analysis based on power correction models
including hadron mass effects for both differential distributions
and mean values is presented.
Using power
corrections, $\alpha_s$ is extracted from the mean values and shapes.
In an alternative approach, renormalisation
group invariance (RGI) is used as an explicit constraint, leading
to a consistent description of mean values
without the need for sizeable power corrections. 
The QCD $\beta$-function is precisely measured using this
approach.  From the DELPHI data on Thrust,
including data from low energy experiments, one finds
\begin{center}
$\beta_0 = 7.86 \pm 0.32~~$
\end{center}
for the one loop coefficient of the $\beta$-function
or, assuming QCD,
\begin{center}
$n_{\mathrm{f}} = 4.75 \pm 0.44$
\end{center}
for the number of active flavours. These values agree well with the QCD 
expectation of $\beta_0=7.67$ and $n_{\mathrm{f}}=5$.
A direct measurement of the full logarithmic energy slope excludes light 
gluinos with a mass below 5\gev.

\end{shortabs}
\vfill
\begin{center}
\DpSubmit \ \\ 
\DpComment \ \\
\DpEMail \ \\
\end{center}
\vfill
\clearpage
\headsep 10.0pt
\addtolength{\textheight}{10mm}
\addtolength{\footskip}{-5mm}
\begingroup
%
\newcommand{\DpName}[2]{\hbox{#1$^{\ref{#2}}$},\hfill}
\newcommand{\DpNameTwo}[3]{\hbox{#1$^{\ref{#2},\ref{#3}}$},\hfill}
\newcommand{\DpNameThree}[4]{\hbox{#1$^{\ref{#2},\ref{#3},\ref{#4}}$},\hfill}
\newskip\Bigfill \Bigfill = 0pt plus 1000fill
\newcommand{\DpNameLast}[2]{\hbox{#1$^{\ref{#2}}$}\hspace{\Bigfill}}
\small
\noindent
\DpName{J.Abdallah}{LPNHE}
\DpName{P.Abreu}{LIP}
\DpName{W.Adam}{VIENNA}
\DpName{P.Adzic}{DEMOKRITOS}
\DpName{T.Albrecht}{KARLSRUHE}
\DpName{T.Alderweireld}{AIM}
\DpName{R.Alemany-Fernandez}{CERN}
\DpName{T.Allmendinger}{KARLSRUHE}
\DpName{P.P.Allport}{LIVERPOOL}
\DpName{U.Amaldi}{MILANO2}
% ...
\normalsize
\endgroup
\titlefoot{Department of Physics and Astronomy, Iowa State
     University, Ames IA 50011-3160, USA
    \label{AMES}}
\titlefoot{Physics Department, Universiteit Antwerpen,
     Universiteitsplein 1, B-2610 Antwerpen, Belgium \\
     \indent~~and IIHE, ULB-VUB,
     Pleinlaan 2, B-1050 Brussels, Belgium \\
     \indent~~and Facult\'e des Sciences,
     Univ. de l'Etat Mons, Av. Maistriau 19, B-7000 Mons, Belgium
    \label{AIM}}
\titlefoot{Physics Laboratory, University of Athens, Solonos Str.
     104, GR-10680 Athens, Greece
    \label{ATHENS}}
\titlefoot{Department of Physics, University of Bergen,
     All\'egaten 55, NO-5007 Bergen, Norway
    \label{BERGEN}}
\titlefoot{Dipartimento di Fisica, Universit\`a di Bologna and INFN,
     Via Irnerio 46, IT-40126 Bologna, Italy
    \label{BOLOGNA}}
\titlefoot{Centro Brasileiro de Pesquisas F\'{\i}sicas, rua Xavier Sigaud 150,
     BR-22290 Rio de Janeiro, Brazil \\
     \indent~~and Depto. de F\'{\i}sica, Pont. Univ. Cat\'olica,
     C.P. 38071 BR-22453 Rio de Janeiro, Brazil \\
     \indent~~and Inst. de F\'{\i}sica, Univ. Estadual do Rio de Janeiro,
     rua S\~{a}o Francisco Xavier 524, Rio de Janeiro, Brazil
    \label{BRASIL}}
\titlefoot{Coll\`ege de France, Lab. de Physique Corpusculaire, IN2P3-CNRS,
     FR-75231 Paris Cedex 05, France
    \label{CDF}}
\titlefoot{CERN, CH-1211 Geneva 23, Switzerland
    \label{CERN}}
\titlefoot{Institut de Recherches Subatomiques, IN2P3 - CNRS/ULP - BP20,
     FR-67037 Strasbourg Cedex, France
    \label{CRN}}
\titlefoot{Now at DESY-Zeuthen, Platanenallee 6, D-15735 Zeuthen, Germany
    \label{DESY}}
\titlefoot{Institute of Nuclear Physics, N.C.S.R. Demokritos,
     P.O. Box 60228, GR-15310 Athens, Greece
    \label{DEMOKRITOS}}
\titlefoot{FZU, Inst. of Phys. of the C.A.S. High Energy Physics Division,
     Na Slovance 2, CZ-180 40, Praha 8, Czech Republic
    \label{FZU}}
\titlefoot{Dipartimento di Fisica, Universit\`a di Genova and INFN,
     Via Dodecaneso 33, IT-16146 Genova, Italy
    \label{GENOVA}}
\titlefoot{Institut des Sciences Nucl\'eaires, IN2P3-CNRS, Universit\'e
     de Grenoble 1, FR-38026 Grenoble Cedex, France
    \label{GRENOBLE}}
\titlefoot{Helsinki Institute of Physics, P.O. Box 64,
     FIN-00014 University of Helsinki, Finland
    \label{HELSINKI}}
\titlefoot{Joint Institute for Nuclear Research, Dubna, Head Post
     Office, P.O. Box 79, RU-101 000 Moscow, Russian Federation
    \label{JINR}}
\titlefoot{Institut f\"ur Experimentelle Kernphysik,
     Universit\"at Karlsruhe, Postfach 6980, DE-76128 Karlsruhe,
     Germany
    \label{KARLSRUHE}}
\titlefoot{Institute of Nuclear Physics,Ul. Kawiory 26a,
     PL-30055 Krakow, Poland
    \label{KRAKOW1}}
\titlefoot{Faculty of Physics and Nuclear Techniques, University of Mining
     and Metallurgy, PL-30055 Krakow, Poland
    \label{KRAKOW2}}
\titlefoot{Universit\'e de Paris-Sud, Lab. de l'Acc\'el\'erateur
     Lin\'eaire, IN2P3-CNRS, B\^{a}t. 200, FR-91405 Orsay Cedex, France
    \label{LAL}}
\titlefoot{School of Physics and Chemistry, University of Lancaster,
     Lancaster LA1 4YB, UK
    \label{LANCASTER}}
\titlefoot{LIP, IST, FCUL - Av. Elias Garcia, 14-$1^{o}$,
     PT-1000 Lisboa Codex, Portugal
    \label{LIP}}
\titlefoot{Department of Physics, University of Liverpool, P.O.
     Box 147, Liverpool L69 3BX, UK
    \label{LIVERPOOL}}
\titlefoot{Dept. of Physics and Astronomy, Kelvin Building,
     University of Glasgow, Glasgow G12 8QQ
    \label{GLASGOW}}
\titlefoot{LPNHE, IN2P3-CNRS, Univ.~Paris VI et VII, Tour 33 (RdC),
     4 place Jussieu, FR-75252 Paris Cedex 05, France
    \label{LPNHE}}
\titlefoot{Department of Physics, University of Lund,
     S\"olvegatan 14, SE-223 63 Lund, Sweden
    \label{LUND}}
\titlefoot{Universit\'e Claude Bernard de Lyon, IPNL, IN2P3-CNRS,
     FR-69622 Villeurbanne Cedex, France
    \label{LYON}}
\titlefoot{Dipartimento di Fisica, Universit\`a di Milano and INFN-MILANO,
     Via Celoria 16, IT-20133 Milan, Italy
    \label{MILANO}}
\titlefoot{Dipartimento di Fisica, Univ. di Milano-Bicocca and
     INFN-MILANO, Piazza della Scienza 2, IT-20126 Milan, Italy
    \label{MILANO2}}
\titlefoot{IPNP of MFF, Charles Univ., Areal MFF,
     V Holesovickach 2, CZ-180 00, Praha 8, Czech Republic
    \label{NC}}
\titlefoot{NIKHEF, Postbus 41882, NL-1009 DB
     Amsterdam, The Netherlands
    \label{NIKHEF}}
\titlefoot{National Technical University, Physics Department,
     Zografou Campus, GR-15773 Athens, Greece
    \label{NTU-ATHENS}}
\titlefoot{Physics Department, University of Oslo, Blindern,
     NO-0316 Oslo, Norway
    \label{OSLO}}
\titlefoot{Dpto. Fisica, Univ. Oviedo, Avda. Calvo Sotelo
     s/n, ES-33007 Oviedo, Spain
    \label{OVIEDO}}
\titlefoot{Department of Physics, University of Oxford,
     Keble Road, Oxford OX1 3RH, UK
    \label{OXFORD}}
\titlefoot{Dipartimento di Fisica, Universit\`a di Padova and
     INFN, Via Marzolo 8, IT-35131 Padua, Italy
    \label{PADOVA}}
\titlefoot{Rutherford Appleton Laboratory, Chilton, Didcot
     OX11 OQX, UK
    \label{RAL}}
\titlefoot{Dipartimento di Fisica, Universit\`a di Roma II and
     INFN, Tor Vergata, IT-00173 Rome, Italy
    \label{ROMA2}}
\titlefoot{Dipartimento di Fisica, Universit\`a di Roma III and
     INFN, Via della Vasca Navale 84, IT-00146 Rome, Italy
    \label{ROMA3}}
\titlefoot{DAPNIA/Service de Physique des Particules,
     CEA-Saclay, FR-91191 Gif-sur-Yvette Cedex, France
    \label{SACLAY}}
\titlefoot{Instituto de Fisica de Cantabria (CSIC-UC), Avda.
     los Castros s/n, ES-39006 Santander, Spain
    \label{SANTANDER}}
\titlefoot{Inst. for High Energy Physics, Serpukov
     P.O. Box 35, Protvino, (Moscow Region), Russian Federation
    \label{SERPUKHOV}}
\titlefoot{J. Stefan Institute, Jamova 39, SI-1000 Ljubljana, Slovenia
     and Laboratory for Astroparticle Physics,\\
     \indent~~Nova Gorica Polytechnic, Kostanjeviska 16a, SI-5000 Nova Gorica, Slovenia, \\
     \indent~~and Department of Physics, University of Ljubljana,
     SI-1000 Ljubljana, Slovenia
    \label{SLOVENIJA}}
\titlefoot{Fysikum, Stockholm University,
     Box 6730, SE-113 85 Stockholm, Sweden
    \label{STOCKHOLM}}
\titlefoot{Dipartimento di Fisica Sperimentale, Universit\`a di
     Torino and INFN, Via P. Giuria 1, IT-10125 Turin, Italy
    \label{TORINO}}
\titlefoot{INFN,Sezione di Torino, and Dipartimento di Fisica Teorica,
     Universit\`a di Torino, Via P. Giuria 1,\\
     \indent~~IT-10125 Turin, Italy
    \label{TORINOTH}}
\titlefoot{Dipartimento di Fisica, Universit\`a di Trieste and
     INFN, Via A. Valerio 2, IT-34127 Trieste, Italy \\
     \indent~~and Istituto di Fisica, Universit\`a di Udine,
     IT-33100 Udine, Italy
    \label{TU}}
\titlefoot{Univ. Federal do Rio de Janeiro, C.P. 68528
     Cidade Univ., Ilha do Fund\~ao
     BR-21945-970 Rio de Janeiro, Brazil
    \label{UFRJ}}
\titlefoot{Department of Radiation Sciences, University of
     Uppsala, P.O. Box 535, SE-751 21 Uppsala, Sweden
    \label{UPPSALA}}
\titlefoot{IFIC, Valencia-CSIC, and D.F.A.M.N., U. de Valencia,
     Avda. Dr. Moliner 50, ES-46100 Burjassot (Valencia), Spain
    \label{VALENCIA}}
\titlefoot{Institut f\"ur Hochenergiephysik, \"Osterr. Akad.
     d. Wissensch., Nikolsdorfergasse 18, AT-1050 Vienna, Austria
    \label{VIENNA}}
\titlefoot{Inst. Nuclear Studies and University of Warsaw, Ul.
     Hoza 69, PL-00681 Warsaw, Poland
    \label{WARSZAWA}}
\titlefoot{Fachbereich Physik, University of Wuppertal, Postfach
     100 127, DE-42097 Wuppertal, Germany \\
\noindent
{$^\dagger$~deceased}
    \label{WUPPERTAL}}
\addtolength{\textheight}{-10mm}
\addtolength{\footskip}{5mm}
\clearpage
\headsep 30.0pt
\end{titlepage}
%
\pagenumbering{arabic} 
\setcounter{footnote}{0} %
\large
%
\section{Introduction}
The decrease of the strong coupling parameter, \as,  with increasing 
energy, $E$, or momentum transfer, $Q$,  and the related properties of 
asymptotic freedom and confinement are striking phenomena of  Quantum
Chromodynamics, QCD, the gauge theory of strong interactions.
Besides the measurement of the strong coupling itself, the precise
measurement of its energy dependence is an experimental task of fundamental 
importance. This energy dependence is governed by the $\beta$ function, 
defined as \cite{PDG2000}:

\begin{eqnarray}
\label{beta}
Q\frac{\mathrm{d}\alpha_s}{\mathrm{d}Q} &=& \beta(\alpha_s) \\
\nonumber
&=& -\frac{\beta_0}{2\pi}\alpha_s^2-\frac{\beta_1}{4\pi^2}\alpha_s^3
-\frac{\beta_2}{64\pi^3}\alpha_s^4-\cdots
\end{eqnarray}
The coefficients $\beta_i$ are given in the Equations \ref{betadef}.

In principle, the study of event shape observables (e.g. Thrust)
in \epem\ annihilation as a function of energy permits
these determinations. Event shape observables,
however, are obscured by the effects of non--perturbative hadronisation.
This influence is expected to vanish with increasing energy (going dominantly 
proportional to $1/E$). A similar dependence is  present in the so-called 
infrared renormalons appearing in perturbation theory \cite{beneke99}. 
Both phenomena are often considered to originate from the same physics.

The theoretical analysis of power terms indicates some properties which are
universal to all observables.
A coherent comparison of the power correction models and the proposed 
universality is one topic of this paper.

Power corrections are subject to ambiguities.
They depend on the borderline dividing perturbative 
and non--perturbative physics in the models; this borderline is a matter of
convention.
Usually the perturbative terms are treated in an ${\cal O}(\alpha_s^2)$
approximation, 
the remainder being taken as a power correction.
The inclusion of higher order perturbative
corrections in general will reduce the size of the power terms. 

Power correction models are now considered as established.
However, in view of a precise measurement of the strong coupling, 
power corrections 
and the ambiguity involved are obstructive.
Quantities for which power corrections are minimised should be
emphasised.
Moreover the reason for the success of power corrections and their
magnitudes is ``not yet fully understood''~\cite{sterman}.
Hence a  critical review of other theoretical methods
for many experimental observables is in order.

The second focus of the phenomenological analysis presented in this paper
is on the study of renormalisation group invariant perturbation theory (RGI)
\cite{Dhar-plett,DG,Dhar-pramana}. 
Here the property of complete renormalisation group invariance is 
used, leading to predictions without the freedom arising from
the choice of renormalisation scheme or scale. The theoretical ansatz employed
\cite{DG}, however, only applies to ``fully inclusive'' 
observables depending on a single energy scale, such as
total cross--sections~\cite{Korner:2000xk,Soper:1996ns},  or mean values 
of event shape observables.
A thorough test of this theoretical method is presented here
based on the energy dependence of the means of the distributions of seven event
shape observables.
The convincing success of this test implies that the size
of the power corrections found for the power correction models
in the $\overline{MS}$ renormalisation scheme
can be predicted using RGI perturbation theory.
Furthermore RGI perturbation theory allows a direct measurement 
of the $\beta$-function of QCD avoiding any scheme dependence.

Since the  goal of this analysis is a study of the scale dependence, data 
with a wide range of 
centre--of--mass energies are needed. Therefore data have been used 
from the high energy running of \lep\ up to centre-of-mass energies
of $202\gev$, from the
energies around the Z pole, and also radiative events with a reduced 
centre-of-mass energy of the hadronic system due to
hard photon radiation. Additionally for some part of the analysis low energy 
data from other experiments are included.

The organisation of this paper is as follows: Section~\ref{sec_select}
discusses briefly the apparatus, the data and the analysis of the
\lep 2 data and the radiative events used to extract data below the Z mass
scale.
The observables used throughout the paper are introduced and their 
dependence on the masses of the final state hadrons is discussed.
Finally, procedures to determine systematic uncertainties are specified.
Section~\ref{results} presents the data on shape distributions and their means
and compares them to predictions of prominent fragmentation models.
Section~\ref{res_dw_shape} similarly makes comparisons with
analytic power model predictions \cite{DW,NuclPhysB511_396,hep-ph/9802381},
together with a comparison of the
non--perturbative parameters for the different observables.
For the first time the prediction of power corrections
for the energy-energy correlation (EEC) is compared to experimental data and
evidence for power shifts at the three jet phase space boundary is given.
Section~\ref{sec:powermeans} contains the power correction
analysis for mean values of exponentiating event shape observables based on the prescription of
\cite{PhysLettB352_451,Wicke:1997ja}.
Simple power fits \cite{Abreu:1997mk,hamachermont} are then
presented for all shape observable means and it is shown that 
the size of the power terms
correlates with that of the corresponding second order perturbative
contribution.
Section~\ref{data_DG} briefly recalls the basics of the RGI method
as given in \cite{DG} and confronts it with the data on shape observable means.
It is shown that this method describes the data very well.
The inclusion of power terms in the fit shows no indication of significant 
non--perturbative effects.
Consequently the size of the power terms determined in the
previous chapter is estimated from the RGI method and shown to agree with
experimental data.
By applying the RGI method to different shape observable means as measured 
by \delphi, especially to the data on $\langle {\mathrm{1-Thrust}} 
\rangle$ combined with results of other experiments at low energy, a direct 
precise measurement of the QCD $\beta$-function is obtained.
Finally we summarise and conclude in Section~\ref{summary}.



\section{Detector, data and data analysis\label{sec_select}}
\delphi\ is a hermetic detector with a solenoidal magnetic field of 1.2T.
The tracking detectors, situated in front of the electro-magnetic calorimeters
are a silicon micro-vertex detector (VD),
a combined jet/proportional chamber inner detector (ID),
a time projection chamber (TPC) as the major tracking device, and
the streamer tube detector OD in the barrel region. The forward region
is covered by  silicon mini-strip and
pixel detectors (VFT) and by the drift chamber detectors FCA and FCB.

The electromagnetic calorimeters are the high density projection chamber
HPC in the barrel, and the lead-glass calorimeter FEMC in the forward region.
Detailed information about the design and performance of
\delphi\ can be found in~\cite{NuclInstrMethA303_233,NuclInstrMethA378_57}.

The phenomenological analysis of the event shape data
presented in the following sections relies on the \delphi\ data measured at
the Z-peak, the off-peak energies of 89 and 93\gev,
as well as the published data between the Z peak and 183\gev 
\cite{danielpaper,daniel_promo}. In addition the
data measured at centre-of-mass energies $\sqrt{s}$ between 189 and 202\gev\
and from radiative events at mean hadronic centre-of-mass energies of
45, 66 and 78\gev\ are presented.
The number of events accepted in the analysis at these energies
and the integrated luminosities collected are given in Table \ref{lumies}.

\begin{table}[tb]
\begin{center}
\begin{tabular}{|r@{.}l|r@{.}l|r@{.}l|r|r|r@{.}l|r@{.}l|}\hline
\multicolumn{2}{|c|}{$E_{cm}$} & 
\multicolumn{2}{|c|}{$\sigma$} & 
\multicolumn{2}{|c|}{$\sigma_{eff}$} & 
\multicolumn{1}{|c|}{${\cal L}$} & 
\multicolumn{1}{|c|}{$N_{\mathrm{sel}}$} & 
\multicolumn{2}{|c|}{$\epsilon$} & 
\multicolumn{2}{|c|}{$p$} \\ \hline
\multicolumn{2}{|c|}{$[\gev]$} & 
\multicolumn{2}{|c|}{$[pb]$} & 
\multicolumn{2}{|c|}{$[pb]$} & 
\multicolumn{1}{|c|}{$[pb^{-1}]$} & 
\multicolumn{1}{|c|}{} & 
\multicolumn{2}{|c|}{} & 
\multicolumn{2}{|c|}{} \\ \hline
45&2 & \multicolumn{2}{|c|}{-} & \multicolumn{2}{|c|}{-} & {-} & 650 & 0&255 & 0&842  \\
66&0 & \multicolumn{2}{|c|}{-} & \multicolumn{2}{|c|}{-} & {-} & 1099 & 0&283 & 0&913 \\
76&3 & \multicolumn{2}{|c|}{-} & \multicolumn{2}{|c|}{-} & {-} & 1212 & 0&238
& 0&876 \\ \hline
   89&4 & 9656&& 9509&  & 17.0  & $148\cdot 10^3$ 
                                     & 0&843 & 0&998 \\
   91&3 & 30486& 
               & 30080&
                      & 92.3  & $2.5\cdot 10^6$ 
                                     & 0&845 & 0&999 \\
   93&0 & 14007&
               & 13757& 
                      & 18.6  & $230 \cdot 10^3$ 
                                     & 0&844 & 0&999 \\
  133&2 & 291& & 69&2 & 11.9  & 850  & 0&958 & 0&999 \\
  161&4 & 147& & 32&3 & 10.1  & 282  & 0&711 & 0&992 \\
  172&3 & 121& & 27&5 & 9.9   & 219  & 0&729 & 0&956 \\
  183&1 & 100& & 23&4 & 54.0  & 1035 & 0&733 & 0&930 \\  
  189&2 & 99&8 & 21&1 & 150.7 & 2774 & 0&749 & 0&909  \\
  192&2 & 96&0 & 20&2 & 25.8  & 433  & 0&757 & 0&889 \\
  196&2 & 90&0 & 19&2 & 77.5  & 1288 & 0&754 & 0&876  \\
  200&1 & 85&2 & 18&2 & 80.9  & 1281 & 0&767 & 0&857  \\
  202&1 & 83&3 & 17&7 & 40.0  & 624  & 0&764 & 0&890   \\ \hline
\end{tabular}

\caption{Nominal centre-of-mass energies $E_{cm}$,
cross--sections, without ($\sigma$), and with ISR cut 
($\sigma_{\mathrm{eff}}$), luminosities (${\cal L}$), the number of selected 
events ($N_{\mathrm{sel}}$), the efficiencies ($\epsilon$) and the  purities 
($p$).}
\label{lumies}
\end{center}
\end{table}

\subsection{Selection and analysis of high energy data}
In order to select well measured particles,
the cuts given in the upper part of \tab{cuts} have been applied.
The cuts in the lower part of the table are used to
select \qcd\ events and to
suppress background processes such as two-photon interactions, beam-gas and
beam-wall interactions, leptonic final states and, for the \lep 2 analysis,
initial state radiation (ISR) and four-fermion background.


%
\newcommand{\bmincut}{\leq 0.08}
\newcommand{\sprimecut}{\geq 90 \%\ecm}
\begin{table}[p]
\begin{center}
\renewcommand{\arraystretch}{1.2}
\begin{tabular}{|l|l|}\hline
 neutral particle &   $E \geq 0.5\gev$ \\
 selection        &$ 20^{\circ} \le \theta \le 160^{\circ} $ \\ \hline
 charged particle & $0.4\gev\leq p \leq 100\gev$ \\
 selection   & $ \Delta p / p \leq 1.0$     \\
             & measured track length $\geq 30$ cm \\
             & distance to I.P. in $r\phi$ plane $\leq 4$ cm \\
             & distance to I.P. in $z \leq 10$ cm  \\
             & $20^{\circ} <\theta < 160^{\circ}$  \\
\hline 

 Standard    & $ N_{\mathrm{ch}}\ge 7 $ \\
 event       & $ 30^{\circ} \le \theta_{\mathrm{Thrust}} \le 150^{\circ} $\\
 selection   & $ E_{\mathrm{tot}} \geq  0.50\ecm $  \\
\hline 
 ISR cuts    & $ \sqrt{s'_{\mathrm{rec}}} \sprimecut $ \\ \hline
 WW cuts     & $N_{\mathrm{ch}}>500 \cdot B_{\mathrm{min}} +1.5 $\\
             & $  N_{\mathrm{ch}} \le 42 $ \\
 \hline 
 prompt photon
 selection
             & $E_{\gamma} - 10 \gev < E_W < E_{\gamma} + 5 \gev$ \\
             & $ 11\gev < E_{\gamma} < E_{\mathrm{cm}}/2 $ \\
             & $\alpha_\gamma = 25^{\circ}$ \\
             & $E_{\alpha} < 0.5 \gev$ \\

\hline
\end{tabular}
\caption[gaga]{\label{cuts}
Selection of particles and events. $E$ is the energy, $p$ is the
momentum, $\Delta p$ its error, $r$  the distance to the
beam-axis, $z$  the distance to the beam interaction point (I.P.)
along the beam-axis, $\phi$ and $\theta$ the azimuthal and polar angles
with respect to the beam, $N_{\mathrm{
ch}}$  the number of charged particles, $\theta_{\mathrm{
Thrust}}$ the polar angle of the Thrust axis with respect to the
beam, $E_{\mathrm{tot}}$ the total energy carried by all particles,
$E_{\mathrm{cm}}$ ther nominal \lep\ energy,
$\sqrt{s'_{\mathrm{rec}}}$ the reconstructed hadronic centre-of-mass energy,
\bmin\ is the minimal Jet Broadening (described in Section \ref{obsdefs}),
$E_{\gamma}$ the energy of the detected
photon, $E_W$ the angular energy (see Equation \ref{winkelenergie}),
$a_{\gamma}$ the opening angle of the photon isolation cone
and $E_{\alpha}$ the maximum additional energy deposit within 
this cone.}
\end{center}
\end{table}

 At energies above 91.2\gev,
the high cross--section of the Z resonance peak raises the possibility
of hard ISR allowing the creation of a nearly
on--shell Z boson.
These ``radiative return events'' constitute a large
fraction of all hadronic events.
The initial state photons are
typically  aligned along the beam direction and are rarely
identified inside the detector.
In order to evaluate the effective hadronic centre-of-mass energy
$\sqrt{s^{\prime}_{rec}}$  of an event,
taking ISR into account, the procedure described in \cite{CERN-OPEN98-026}
was applied. It is based on a fit imposing four--momentum conservation to
measured jet four--momenta (including estimates of their uncertainties).
Several assumptions
about the event topology are tested. The decision is taken according to the
$\chi^2$ obtained from the constrained fits with different topologies.

\begin{figure}[tb]
\unitlength1cm
 \unitlength1cm
 \begin{center}
 \begin{minipage}[t]{7.5cm}
           \mbox{\epsfig{file=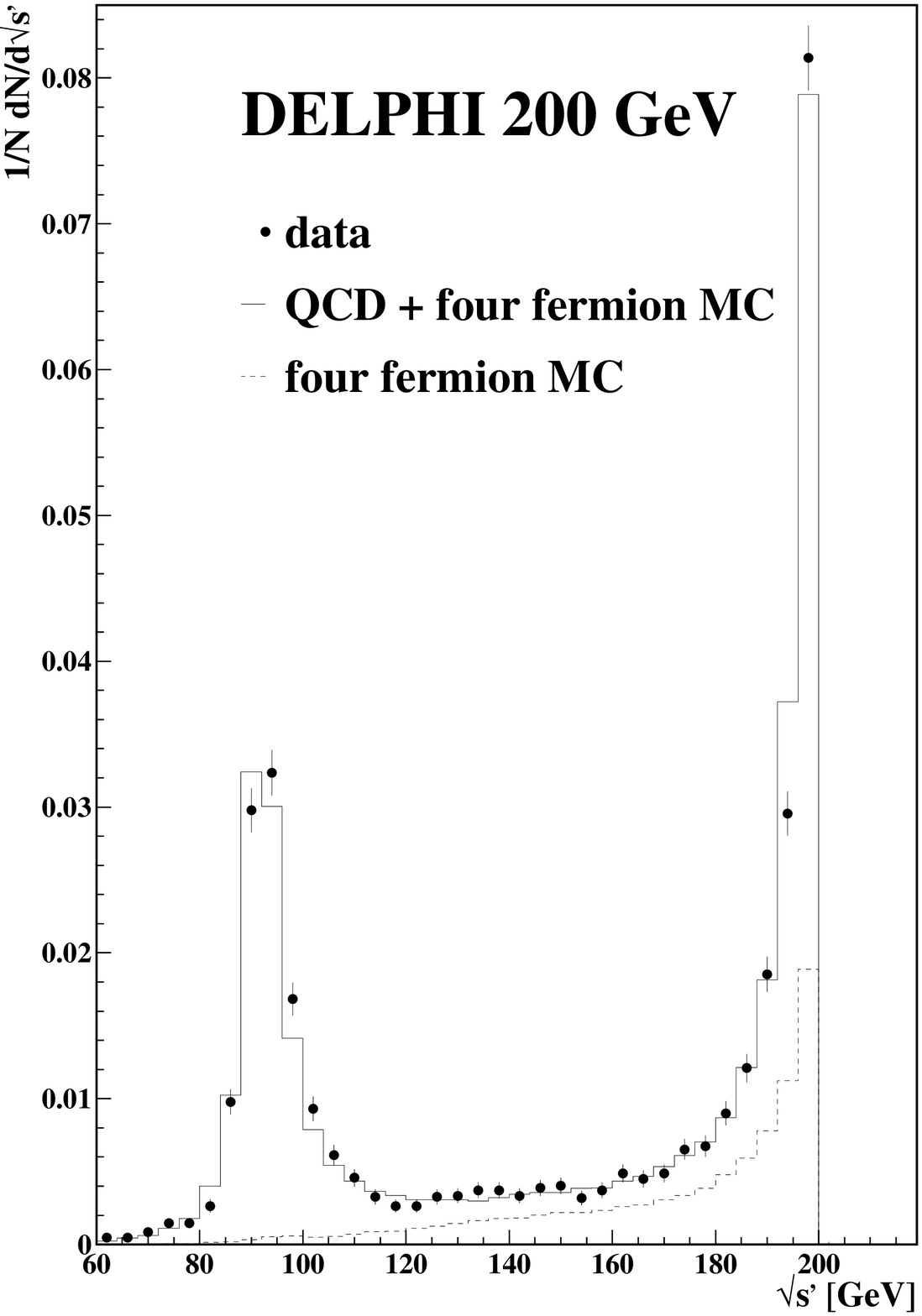,width=7.5cm}}
 \end{minipage}
 \begin{minipage}[t]{7.5cm}
           \mbox{\epsfig{file=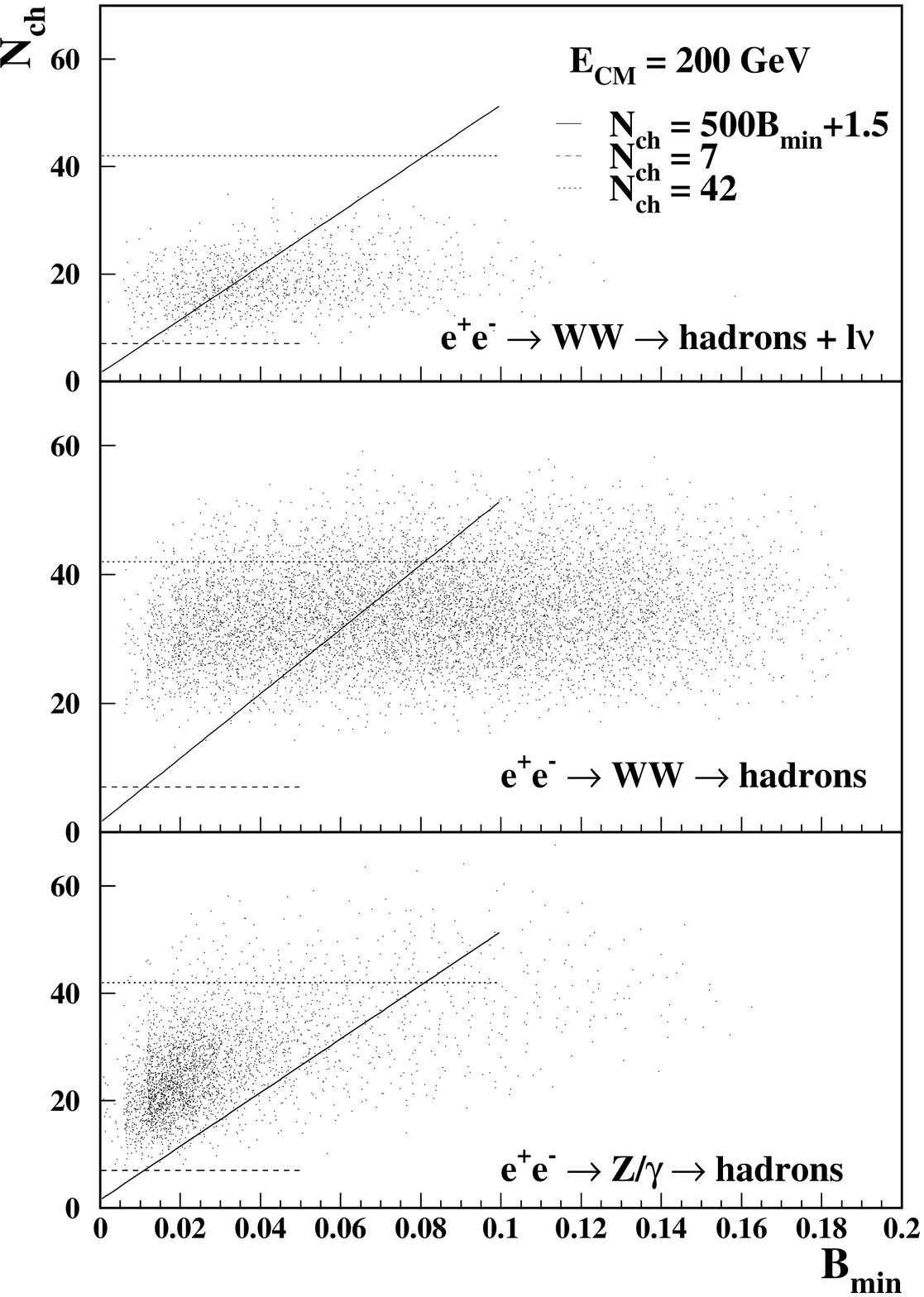,width=7.5cm}}
 \end{minipage}
\end{center}
\caption{\label{plot_isr}
Left: Reconstructed centre-of-mass energy for accepted data before $\sqrt{s^{\prime}}$ cut compared to QCD and four-fermion simulation.
 Right: Simulation of four-fermion background and QCD events in the
$N_{\mathrm{ch}}$--$B_{\mathrm{min}}$ plane. The upper two plots show
the distributions for semi--leptonic and fully hadronic WW events, 
respectively. The lines indicate the cut values chosen.}
\end{figure}

\fig{plot_isr} shows the  spectrum of the calculated
energies for simulated and measured  events passing all but the
$\sqrt{s^{\prime}_{rec}}$ cut for 200\gev\ \epem\ centre-of-mass energy.
The agreement between data and simulation is good for the high
energies relevant to this analysis,
while the peak around $M_{\mathrm{Z}}$ appears to be slightly
shifted in the simulation.
A cut requiring the reconstructed centre-of-mass energy
$\sqrt{s^{\prime}_{rec}}$ to be greater than $0.9 \cdot E_{cm}$
is applied to discard radiative return events (see \tab{cuts}).

Two photon events are strongly suppressed by the cuts. Leptonic
background was found to be negligible in this analysis.

Since the topological signatures of QCD four-jet events and four-fermion
backgrounds such as hadronically decaying ZZ or WW events are similar, no highly
efficient separation of these two classes of events is possible.
Thus any four-fermion rejection implies a severe bias to the shape
distributions of QCD events, which needs to be corrected with
simulation.
By applying a cut on an observable calculated from the narrow event
hemisphere only (like $B_{\mathrm{min}}$, see Section \ref{obsdefs}), 
the bias to event shape observables mainly sensitive
to the wide event hemisphere is reduced.
The two dimensional cut in the $N_{\mathrm{ch}}$--$B_{\mathrm{min}}$ plane
exploits the different correlation between these observables in QCD and
four-fermion 
events (see Figure~\ref{plot_isr}). 
Applying the two dimensional cut almost 90\%
of the four-fermion background can be suppressed.
The remaining four-fermion contribution is estimated by Monte Carlo  generators and
subtracted from the measurement.
The remaining detector and cut effects are unfolded with
simulation.
The influence of detector effects  was studied by passing
generated events (\jetset/\pythia~\cite{Sjostrand:2000wi} using the
\delphi\ tuning described in~\cite{ZPhysC73_11}) through a full
detector simulation (\delsim~\cite{NuclInstrMethA303_233}). These simulated
events are processed with the same reconstruction program and selection
cuts as are the real data.
In order to correct for cuts, detector, and ISR effects a
bin-by-bin acceptance correction $C$, obtained from
\qcd\ simulation, is applied to the data:
\begin{equation}
              C_i = \frac{h(f_i)_{\mathrm{gen, no ISR}}}
                             {h(f_i)_{\mathrm{acc}}} ~~~,
\label{acc}
\end{equation}
where $h(f_i)_{\mathrm{gen, no ISR}}$ represents the contents of
 bin $i$ of the shape distribution $f$
generated with the tuned generator. The calculation includes all stable particles.
The subscript ${\mathrm{ no ISR}}$
indicates that only events without significant ISR ($\sqrt{s}-\sqrt{s'}<0.1\gev$)
enter the distribution.
$h(f_i)_{\mathrm{acc}}$ represents the accepted distribution $f$ as obtained
with the full detector simulation.
%

%
%
\subsection
{Data selection at hadronic        
centre-of-mass energies below ${\boldmath{M_{\rm \bf Z}}}$ 
\label{lowenergy}
}
In order to extend the available energy range below the Z-peak, events
with reduced hadronic centre-of-mass energies due to hard photon radiation
are selected from data taken at centre-of-mass energy of 91\gev\ in 1992
through 95. The method requires an energetic isolated photon to be detected and
is based on the hypothesis that such photons are emitted before or immediately
after the $Z/\gamma$ interaction \cite{hardradiative} and do not interfere 
with the QCD fragmentation
processes. The angular distribution of the initial state radiation 
is aligned along the direction of the beams,
with the result, that most photons go undetected in the very
forward region. In contrast, photons from final state radiation tend to be 
grouped along the direction of the final state partons and can be
detected with better efficiency. As a result the selected events
are dominated by final state radiation.
\begin{figure}[tb]
\unitlength1cm
\begin{center}
\mbox{\epsfig{file=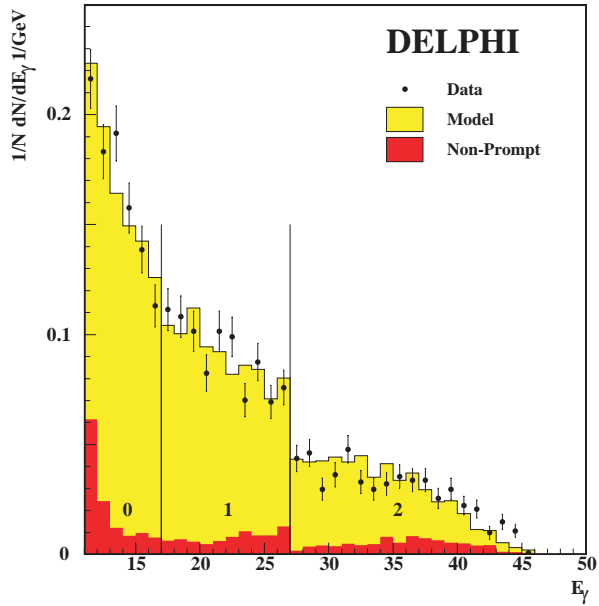,width=8.1cm}}
\end{center}
\caption{\label{photon_plot}
The energy distribution of the photon in the selected isolated photon  events.
The three classes (0,1,2) correspond to a mean centre-of-mass energy of 
76, 66 and 45\gev, respectively. The steps between the classes  result from
different $\pi^0$ rejection cuts.
}
\end{figure}

Many of the photons convert into \epem\ pairs in the material
before the calorimeter. These are reconstructed using the tracking detector
information left by the $\mathrm{e^+}$ and $\mathrm{e^-}$ particles.
Only the reconstructed conversions
before the TPC with electron and positron measurements
in the TPC and conversions behind the TPC with
electron and positron measurements in the OD or the HPC are used.

The largest part of the non-photonic background stems from $\pi^0$'s
decaying into two photons.
Due to the high granularity of the HPC the photon shower can be tested for
a two photon hypothesis. This is done by two methods. The first tries to
divide the cluster into two subclusters and reconstructs the invariant mass
of the decayed particle. The second measures the asymmetry of the energy
distribution in the $\theta\phi$-plane, as
two photons generate a more elliptic  cluster shape. 
The results of these two methods are combined  in a single probabilistic 
variable. Since the angle between the two photons decreases
with the energy, harder cuts had to be made for higher photon energies.

In order to distinguish prompt photons from soft collinear photons
arising in  the later stages of fragmentation and decays, hard cuts on
the photon energy and the isolation with respect to other jets
have to be applied.
Isolation is defined by two criteria. In order to obtain photons at
a large angle with respect to the final state particle a minimum angle
of $20^\circ$ to the jet axes is demanded. The jets are defined by the Durham
cluster algorithm with $y_{\mathrm{cut}}=0.06$. The additional energy deposition
within a $25^\circ$ cone around the jet--axis had to be less than 0.5\gev, 
which reduces background from $\pi^0$ decays.
Electromagnetic punch-through entering the HCAL has been considered.

To test the consistency of the measured photon energy, the following 
cross--check
is performed: the event, exluding the photon, is clustered into two jets and 
the energy of the radiative photon is reconstructed
from the angles between jets $j$,$k$ and photon $i$ through the 
following equation:
\begin{equation} 
E_{\mathrm{W}}=\frac{|\sin \theta_{jk}|}{|\sin \theta_{ji}|+|\sin \theta_{ik}|+|\sin \theta_{jk}|} E_{cm} \quad .
\label{winkelenergie} 
\end{equation}
This reconstructed energy is required to coincide with the photon energy
measured by the calorimeters in the range
\mbox{$E_\gamma-10\gev < E_{\mathrm{W}} < E_\gamma +5\gev$}.

The additional selection criteria for
ISR and final state radiation (FSR) events are summarised in \tab{cuts}.
The energy distribution of the final prompt photon candidates
can be seen in Figure~\ref{photon_plot}.
From selected events the tagged photon is removed, and the event
is boosted into the centre-of-mass frame of the hadronic system.
The boost was determined by the measured photon.
The events are summed up into three intervals in centre-of-mass
energy. The mean value of each sample is taken as the nominal
energy as calculated using the measured radiated photon 
and a correction is applied.

\subsection{Observables and their mass corrections\label{obsdefs}}
The event shape observables used throughout this paper are calculated from the
charged and neutral particles.

The Thrust $T$ \cite{thrust} is defined by
\begin{equation} 
T=\frac{\sum_{k=1}^{N}|\vec{p}_k\cdot\vec{n}_{T}|}
  {\sum_{k=1}^{N}|\vec{p}_k|} \label{thrustdef} ~~~,
\end{equation}
where $\vec{p}_i$ is the momentum vector of particle $i$ and
$\vec{n}_{T}$ is the Thrust axis, which maximizes the above expression.
The observable Major is defined similarly, replacing $\vec{n}_{T}$ with
$\vec{n}_{M}$ which is constrained to be perpendicular to
$\vec{n}_{T}$.
The C-parameter\cite{cparam} is defined by the 
eigenvalues $\lambda$ of the infrared-safe
linear momentum tensor $\Theta^{ij}$:
\begin{xalignat}{3}
\Theta^{ij}&=\frac{1}{\sum_{k=1}^N|\vec{p}_k|}
   \sum_{k=1}^N\frac{p_k^i p_k^j}{|\vec{p}_k|} &
C&=3(\lambda_1\lambda_2+\lambda_2\lambda_3+\lambda_3\lambda_1)  ~~~.
\label{cdef}
\end{xalignat}\noindent
Here $p_k^i$ denotes the $i$-component of $\vec{p}_k$.
Events can be divided into two hemispheres, positive and negative,
by the plane perpendicular
to the Thrust axis $\vec{n}_{T}$. The so--called Jet Masses \cite{jetmass} are then given by
\begin{xalignat}{5}
M_{\pm}^2&= \left( \sum_{ \pm\vec{p}_k \cdot \vec{n}_{T}> 0}p_k \right)^2~~, &
M_{\mathrm{h}}^2&=\max(M^2_+,M^2_-) \label{mhighdef}~~, &
M_{\mathrm{s}}^2&=M^2_++M^2_-~~.
\end{xalignat}\noindent
The Jet Broadenings \cite{broadenings} 
are defined by summing the transverse momenta of the particles:
\begin{xalignat}{5}
B_\pm&=\frac{1}{2\sum_{k=1}^N|\vec{p}_k|}
{\sum_{\pm\vec{p}_k\cdot\vec{n}_{\mathrm{T}}>0}|\vec{p}_k \times 
\vec{n}_{\mathrm{T}}|}~~,&
B_{\mathrm{max}}&=\max(B_-,B_+)~~,&
B_{\mathrm{sum}}&=B_-+B_+~~. \label{bdef}
\end{xalignat}
\noindent
The energy-energy correlation EEC \cite{eecdef} measures the 
correlation of the energy flow in an hadronic event.
It is defined as
\begin{equation}
\frac{d\Sigma(\chi)}{d\cos \chi}
=\frac{1}{N}\frac{\sigma}{\Delta \cos\chi}\sum_{N_{\mathrm{Evt.}}}
\sum_{ij}^{particles} \frac{E_i E_j}{E^2_{\mathrm{vis}}}
\int_{\cos\chi-\frac{\Delta\cos\chi}{2}}^{\cos\chi+\frac{\Delta\cos\chi}{2}}
\delta(\rho-\cos\chi_{ij})d\rho \label{eecdef}~~.
\end{equation}
Here $\chi_{ij}$ denotes the angle between the particles $i$ and $j$.
The jet cone energy fraction JCEF \cite{jcefdef} integrates the energy within a conical shell
of average half-angle $\chi$ around the Thrust axis. It is defined as
\begin{equation}
JCEF(\chi)=\frac{1}{N}\frac{1}{\Delta\chi}
\sum_N\sum_i \frac{E_i}{E_{\mathrm{vis}}}
\int_{\chi-\frac{\Delta\chi}{2}}^{\chi+\frac{\Delta\chi}{2}}
\delta(\chi'-\chi_i)d\chi'~~, \label{jcefdef}
\end{equation}
where $\chi_i$ is the opening angle between a particle and the Thrust axis
pointing
in the direction of the light Jet Mass hemisphere:
\begin{equation}
\chi_i=\cos^{-1}\left( \frac{\vec{p}_i\cdot\vec{n}_T}{|\vec{p}_i|}\right)~~~.
\end{equation}

Evidently the statistical correlation between the event shape observables 
when calculated from the same data is very high. 
But since their QCD predictions are different (e.g. the relative size of 
the second order contributions) studying them  provides an important 
cross--check for QCD.

In the subsequent analysis  QCD effects are
calculated in the massless limit. This is an approximation which
can lead to substantial deviations for some sensitive observables
in certain cases and in the low energy limit in general. In order to
reduce mass effects two approaches are applied:
In \cite{salamwicke} it is proposed that hadron masses
are associated with corrections that are proportional to
$(\ln Q)^A/Q $ with $A\simeq 1.6$, which can be of the same
size as traditional power corrections. The mass
induced power corrections can be separated
into two classes, universal and non-universal, 
where the non-universal can be reduced by a
redefinition of the observables.
The Jet Masses in particular are subject to large mass corrections.
To  suppress  the influence of the hadron masses, new observables
were defined which for massless hadrons are identical to the standard
$\Mhigh$--definition \cite{salamwicke}. They are defined by replacing the 
four--momentum $p$ in the
standard formula:
\begin{eqnarray}
  p = (\vec{p},E) & \longrightarrow &(\vec{p},\left|\vec{p}\right|)\qquad\mbox{($p$-scheme)} \\
  p = (\vec{p},E) & \longrightarrow &(\hat{p}E,E) \qquad\mbox{($E$-scheme)}
\end{eqnarray}
with $\hat{p}$ being the unit vector in direction of $\vec{p}$.

\begin{figure}[p]
\unitlength1cm
\begin{center}
\mbox{\epsfig{file=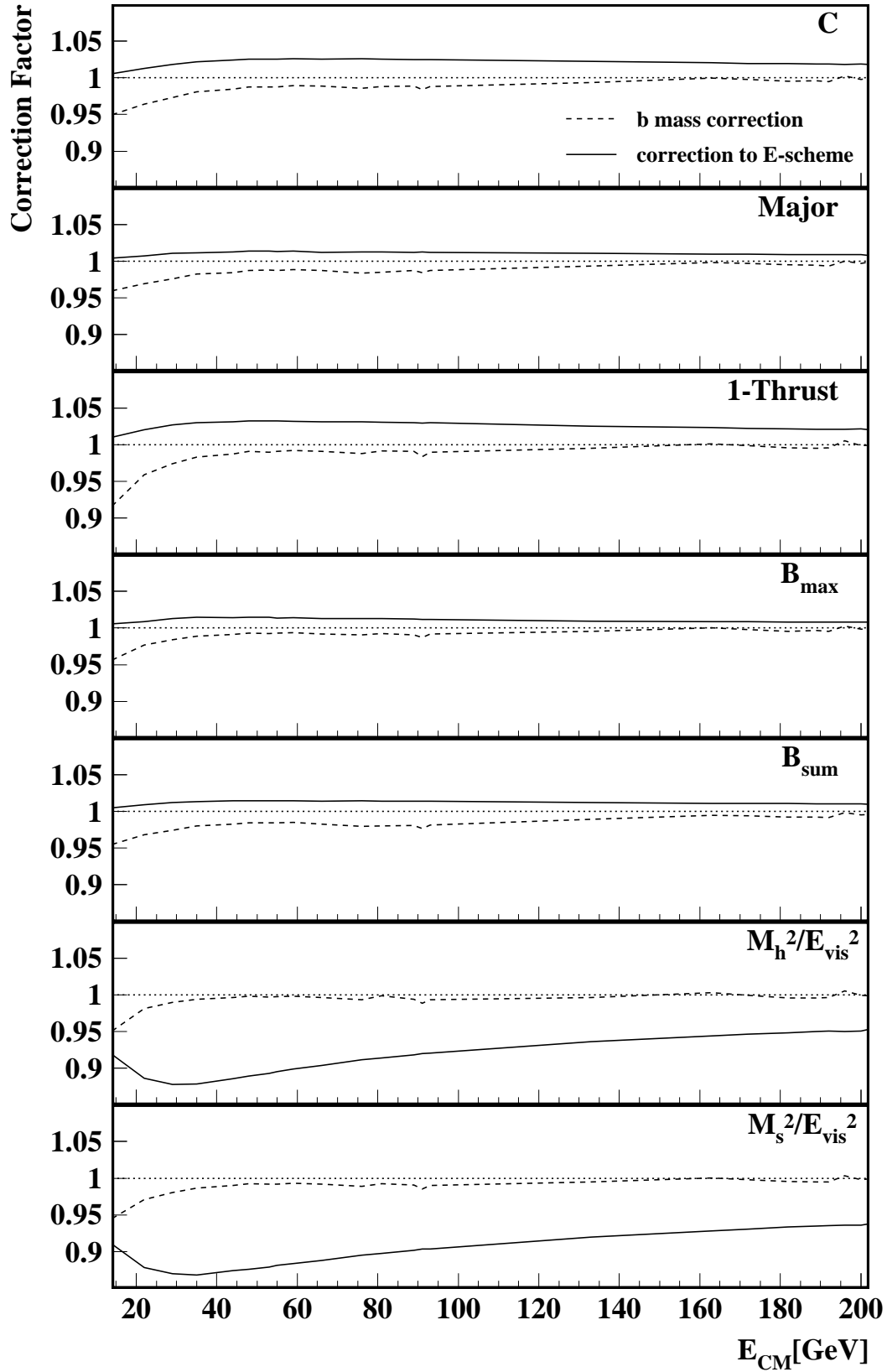,width=14.cm}}
\end{center}
\caption{\label{e_effect}
Corrections due to mass effects:
The full line shows the relative change of the distribution of
the observable due to a change into the E-scheme as a function of the energy.
The dashed line shows the b--mass correction applied to the data.
}
\end{figure}

The resulting observables have the same second order
coefficients and the same power correction coefficient as standard definitions,
because the theoretical calculations were performed for massless particles.
Figure~\ref{e_effect} shows, for shape observable means, the relative size of the
corrections due to a change into the E-scheme as calculated by \pythia\ 6.1.
$\Mhigh$ and $\Msum$ show the biggest changes (Other observables show in 
principle the same effect, but to a negligible extent.).
As a consequence
the E definitions of $\Mhigh$ and $\Msum$ are used wherever possible.

Besides the influence of final state hadron masses the influence of
heavy b--hadron decays has to be taken into account.
In order to correct for the kinematic effects of b--hadron decays,
samples of 100000 b--quark and 100000 light quark events were simulated
using \pythia\ 6.1. for each energy used.
For the calculation of the shape observables all stable particles 
were considered.
A correction was then calculated as the ratio of the event shape 
distribution (or mean) for
light quark to b events considering the energy dependence
of the fraction of b events in \epem annihilation.
The correction is applied to the data throughout the analysis.
It is shown for several event shape means in Figure~\ref{e_effect}.
While the overall effect on the observables
is of the order of a few $\%$ at
the Z--peak, it rises well above $5\%$ at low energy and
influences the evolution of the observables.

\subsection{Systematic uncertainties\label{systematics} and definition of 
average values \label{average}}
In order to estimate systematic uncertainties of the corrected data
distributions and of the quantities derived from them, the event selection and
the correction procedure were varied.
For each variation the analysis was repeated. The individual deviations from
the central result were added in quadrature and considered as the systematic
uncertainty.

The following variations were made in the event selection: the cut in
the charged multiplicity was modified by $\pm$1 unit, the cut in the polar
angle of the event Thrust axis was modified from $25^{\circ}$ to $35^{\circ}$,
and the cut on the observed visible energy was varied between 0.45 and 0.55.
For the high energy data the $\sqrt{s'_{\mathrm{rec}}}$ cut was lowered to 0.8.
For data at centre-of-mass energies above the WW threshold the weight of
$B_{\mathrm{min}}$ in the cut relation was lowered to 480 from 500
and the WW cross--section was
increased conservatively by 5\%.

When hadronisation corrections are included in the analysis
the predictions of \ariadne\ were used instead of the standard choice \pythia.
Model parameters are as given in~\cite{ZPhysC73_11}.
Additionally $\pm 10\%$ was taken of the hadronisation correction as
the  systematic uncertainty. For the kinematically dominated b--hadron mass
$\pm 20\%$ of the correction was taken as the systematic uncertainty.

For all renormalisation scale dependences the scale $f$ was varied between
half and twice the central value.

The fit results for the different observables are summarized by quoting two
kinds of average values: the unweighted mean value with the R.M.S. as first 
error and the\\ (error--)weighted mean value with the simple 
average of the individual statistical errors as the
first error. The second error in both cases is the systematic uncertainty 
which  has been  propagated from the individual results. Quoting the R.M.S. 
indicates  the size of  theoretical uncertainties, while the average 
statistical error
of the weighted mean value indicates the statistical significance of the 
results. Following from  our definition, the statistical error of the weighted 
mean value is bigger than some individual statistical errors.
A treatment of statistical correlation has not been performed, since 
our errors are dominantly systematic. Furthermore the 
statistical correlation between e.g. the event shape means is high
($\ge$ 0.8), thus the gain in reducing the statistical error would be 
negligible. Using these highly correlated observables is useful for other 
reasons: It provides a cross--check for QCD and indicates theoretical 
uncertainties, since their perturbative expansions show different properties 
(e.g. a different size of the second order contribution).

\section{Comparison of experimental results to fragmentation models
\label{results}}
\begin{figure}[tbp]
 \begin{center}
  \vspace{-2.5cm}
  \unitlength1cm
  \begin{minipage}[t]{7.0cm}
    \mbox{\epsfig{file=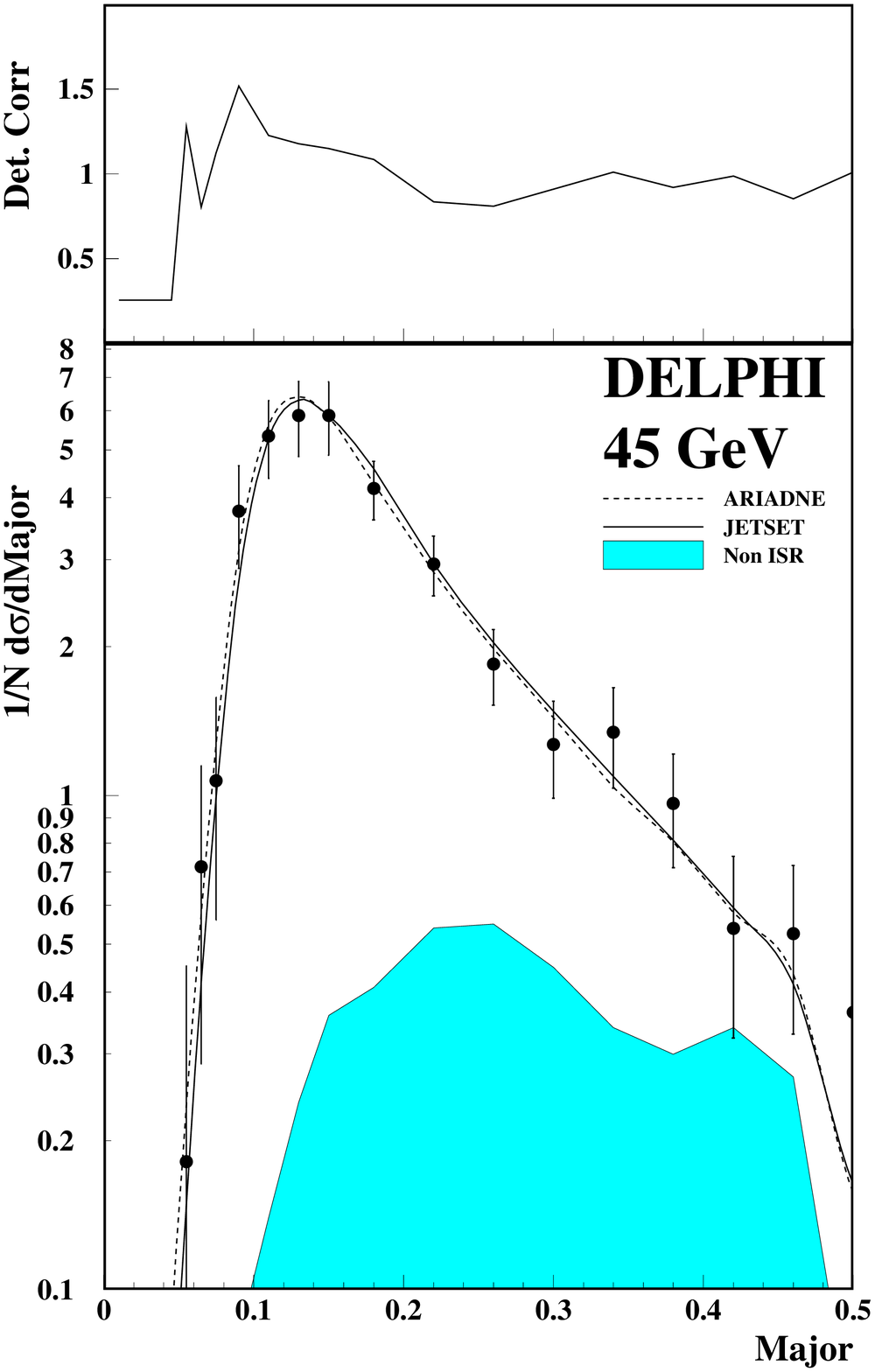,width=7.5cm}}
  \end{minipage}
  \begin{minipage}[t]{7.0cm}
    \mbox{\epsfig{file=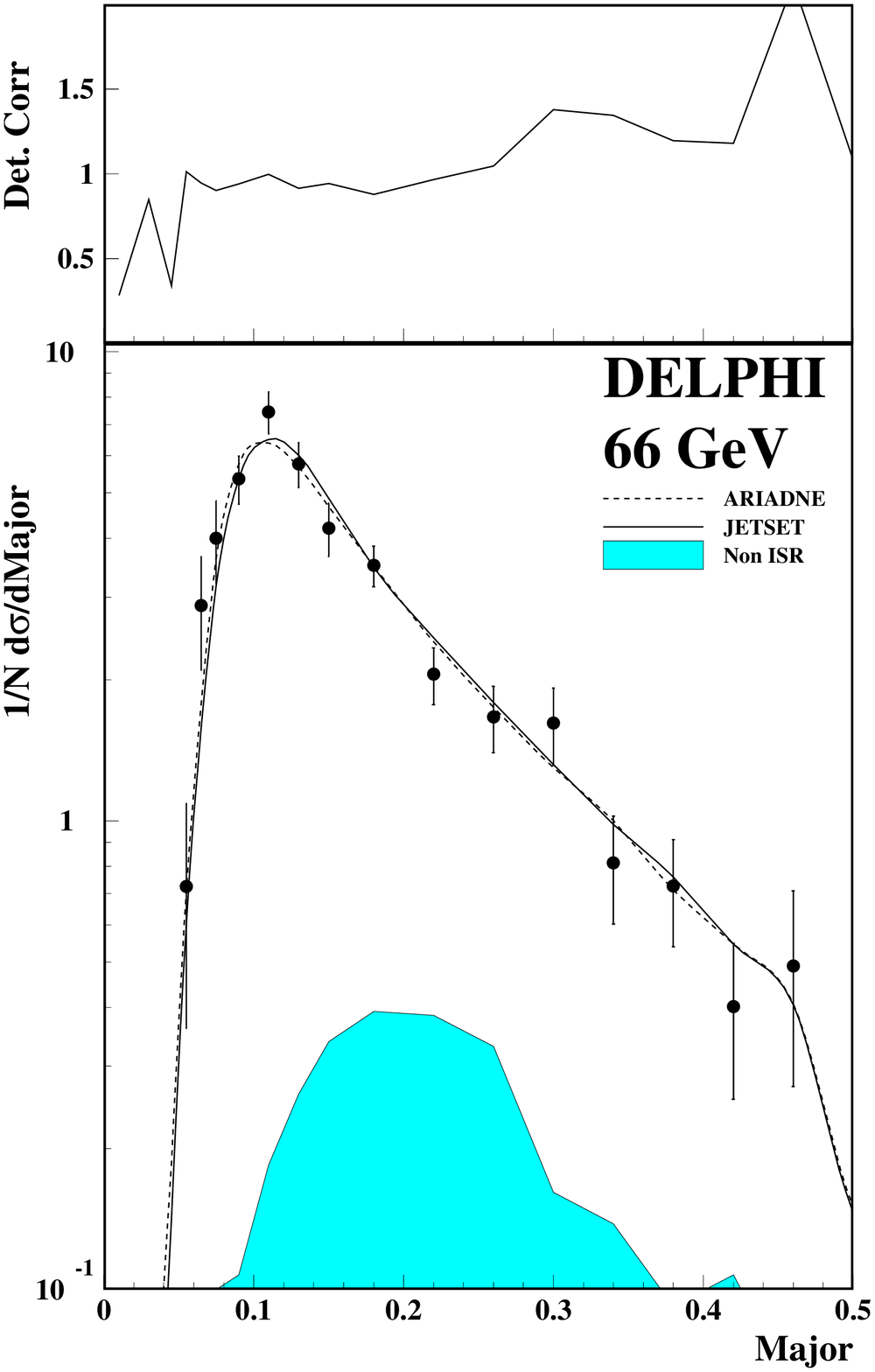,width=7.5cm}}
  \end{minipage}
 \end{center}

\vspace*{-1.5cm}
 \begin{center}
  \unitlength1cm
  \begin{minipage}[t]{7.0cm}
    \mbox{\epsfig{file=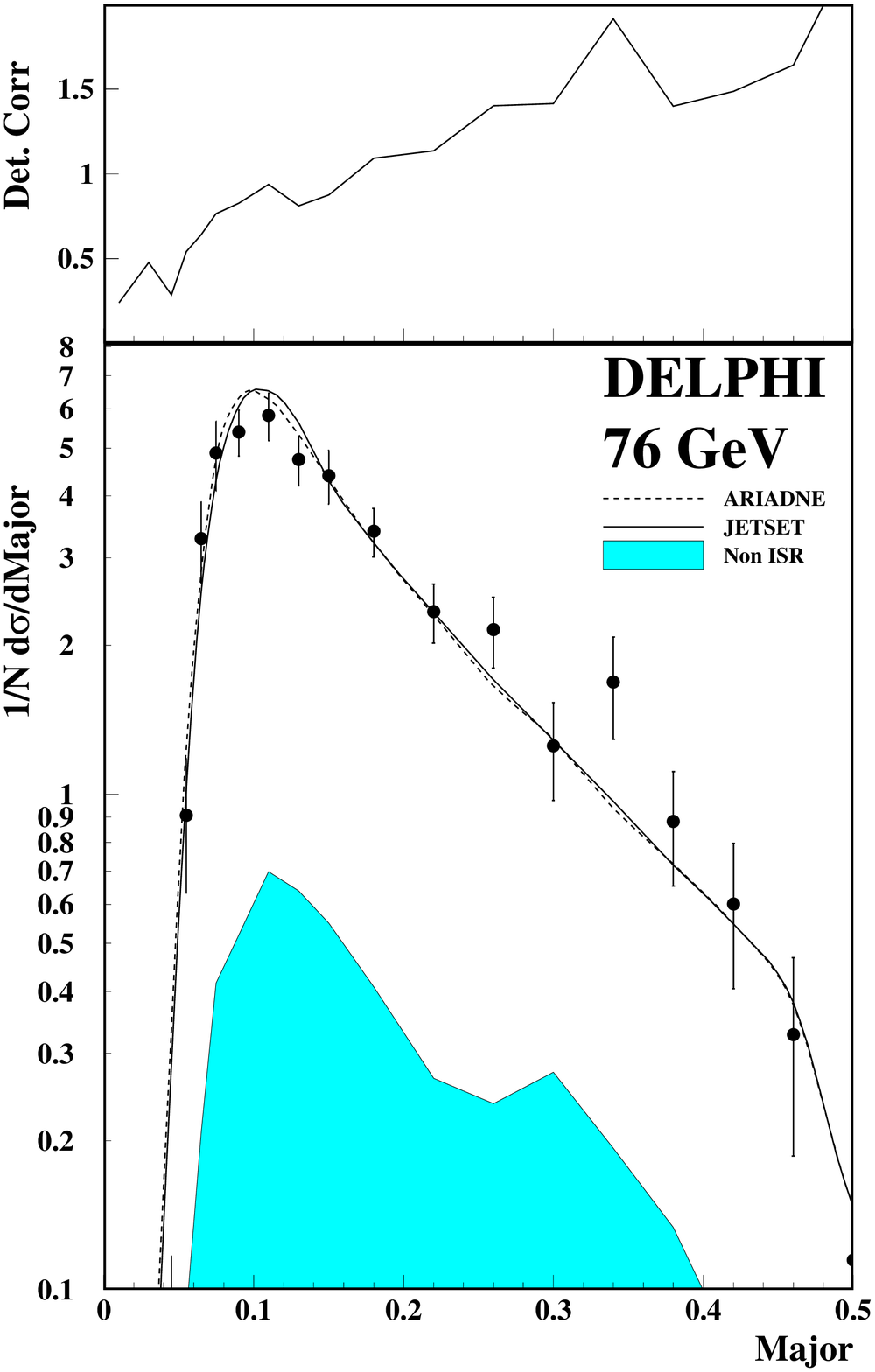,width=7.5cm}}
  \end{minipage}
  \begin{minipage}[t]{7.0cm}
    \mbox{\epsfig{file=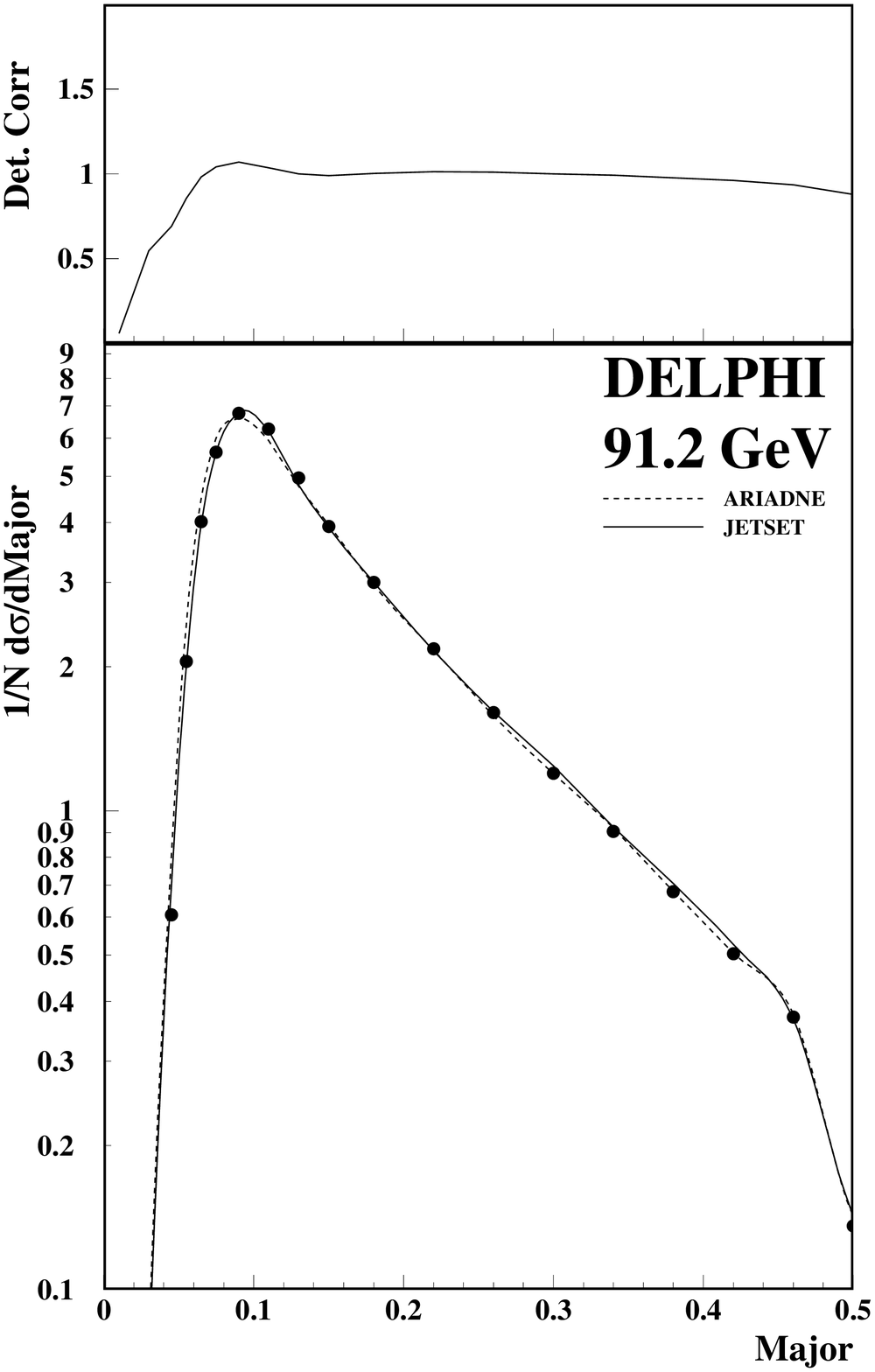,width=7.5cm}}
  \end{minipage}
 \end{center}

 \caption{
Distributions of the observable Major for centre-of-mass energies of
45, 66, 76 and 91.2\gev\ compared to predictions of \jetset\ and \ariadne.
In each plot the upper chart shows the size of the detector correction, defined
as $\frac{MC_{\mathrm{gen}}}{MC_{\mathrm{acc}}}$. 
The grey area indicates the distribution of non-radiative background events
which has been subtracted.
 }\label{majorshapes_1}
\end{figure}

\begin{figure}[tbp]
 \begin{center}
  \vspace{-2.5cm}
  \unitlength1cm
  \begin{minipage}[t]{7.0cm}
    \mbox{\epsfig{file=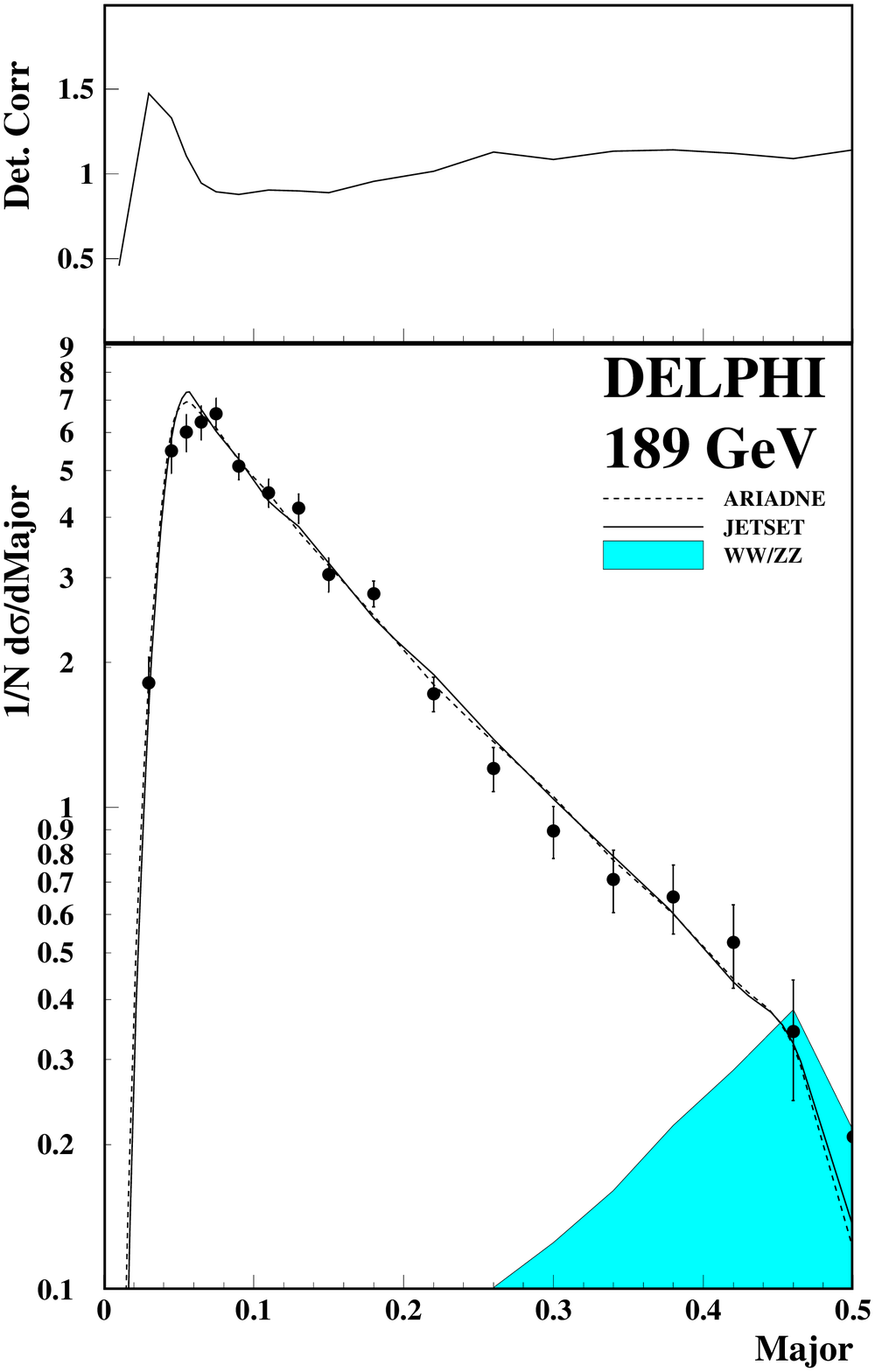,width=7.5cm}}
  \end{minipage}
  \begin{minipage}[t]{7.0cm}
    \mbox{\epsfig{file=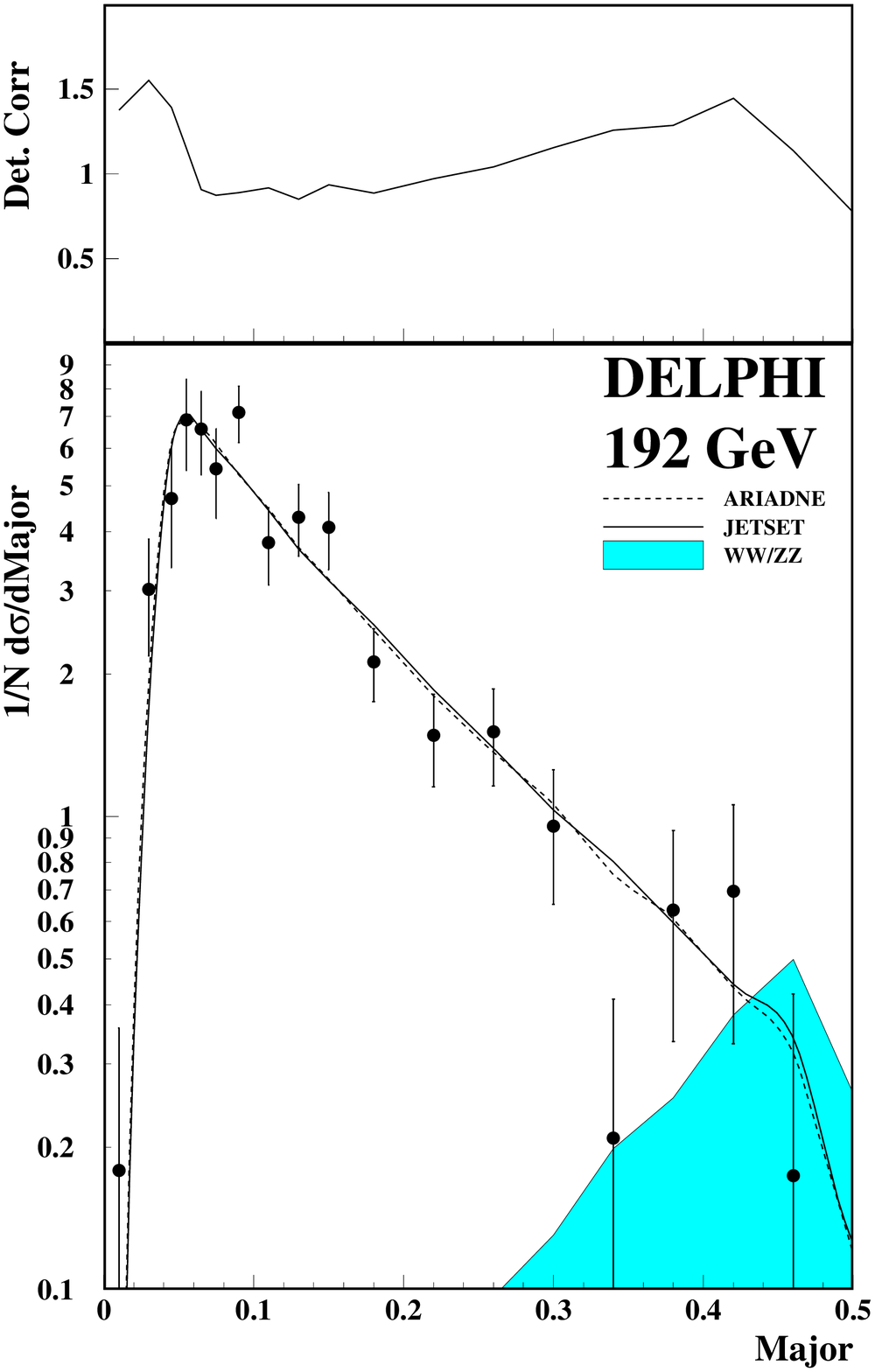,width=7.5cm}}
  \end{minipage}
 \end{center}

\vspace*{-1.5cm}
 \begin{center}
  \unitlength1cm
  \begin{minipage}[t]{7.0cm}
    \mbox{\epsfig{file=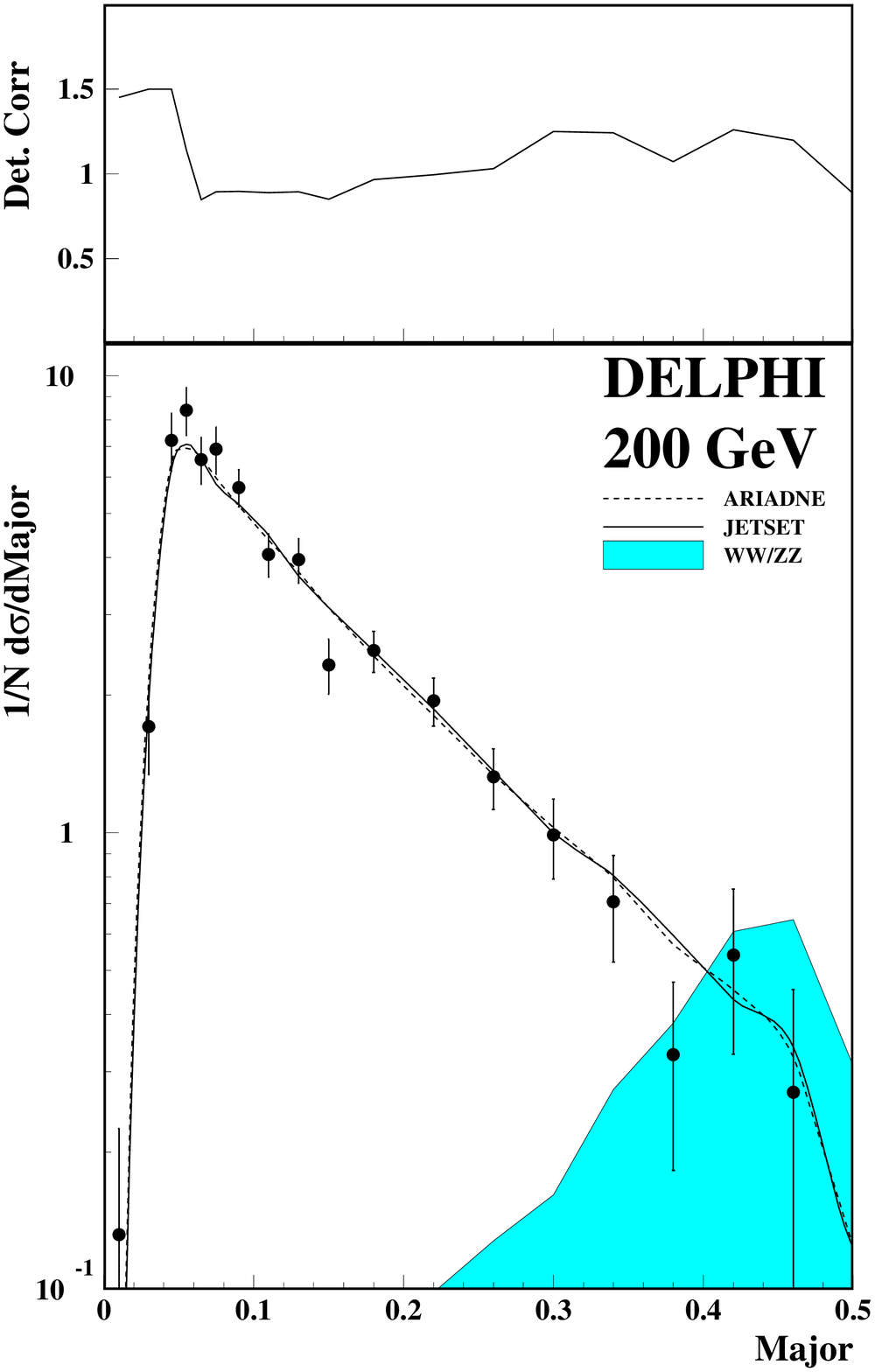,width=7.5cm}}
  \end{minipage}
  \begin{minipage}[t]{7.0cm}
    \mbox{\epsfig{file=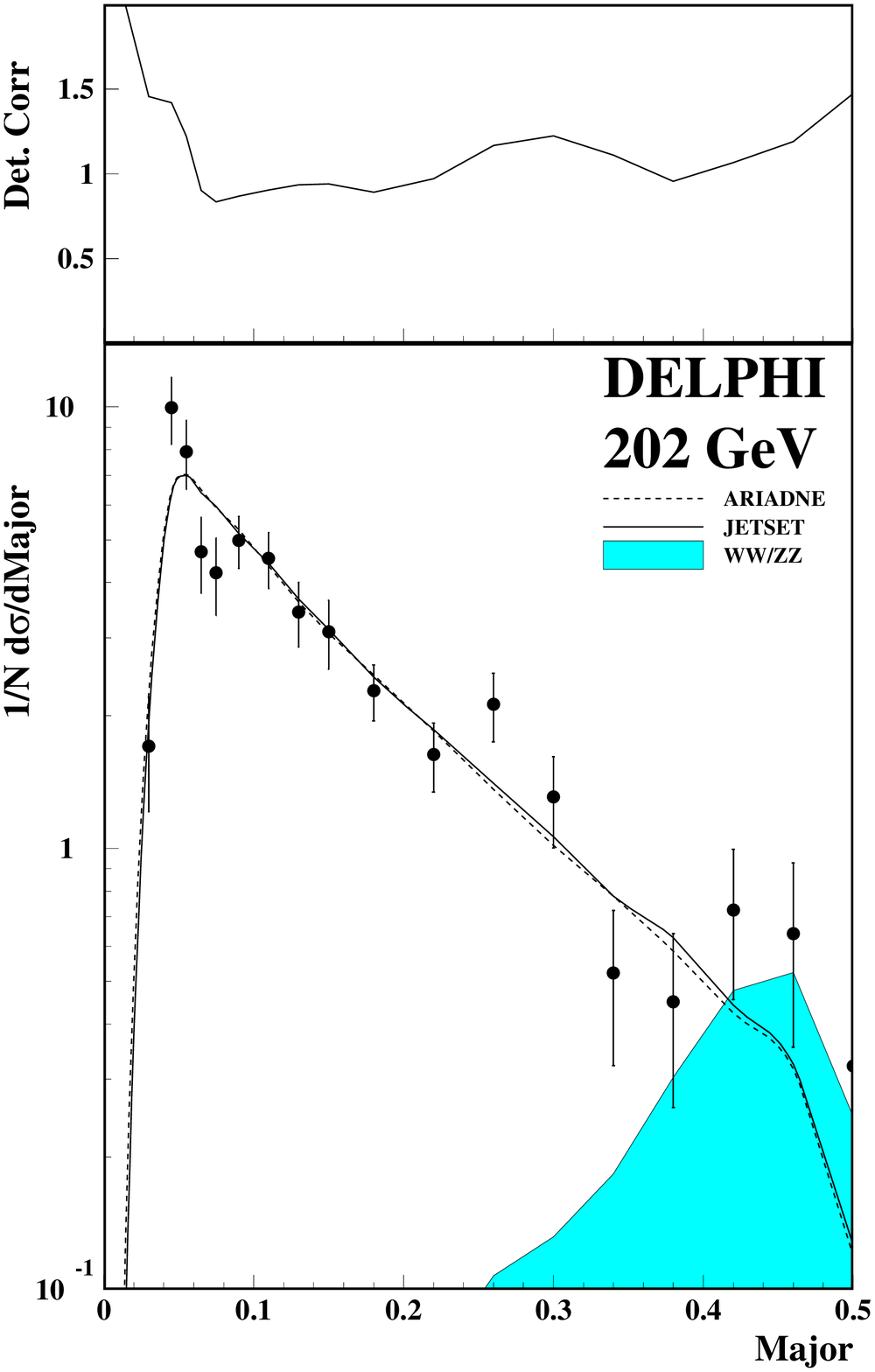,width=7.5cm}}
  \end{minipage}
 \end{center}

 \caption{\label{majorshapes_2}
Distributions of the observable Major for centre-of-mass energies of
189, 192, 200 and 202\gev\
compared to predictions of \jetset\ and \ariadne.
In each plot the upper chart shows the size of the detector correction, defined
as $\frac{MC_{\mathrm{gen}}}{MC_{\mathrm{acc}}}$. 
The grey area indicates the distribution of WW and ZZ background events which
has been subtracted.}
\end{figure}
Although among the oldest event shape measures, few data are
available for the observable Major.
Therefore in the Figures~\ref{majorshapes_1} and \ref{majorshapes_2}
the Major distributions are
shown for several energies between 45\gev\ and 202\gev\
compared to predictions of the \jetset\ and \ariadne\ Monte Carlo models.
Except the lowest energy data at 45\gev\ both simulations are almost
indistinguishable. Data and simulation agree very well.
The agreement between data and models is similarly good for other
observables \cite{dr:reinhardt}.

\begin{figure}[hp]
 \begin{center}
  \vspace{-2.5cm}
  \unitlength1cm
  \begin{minipage}[t]{7.0cm}
    \mbox{\epsfig{file=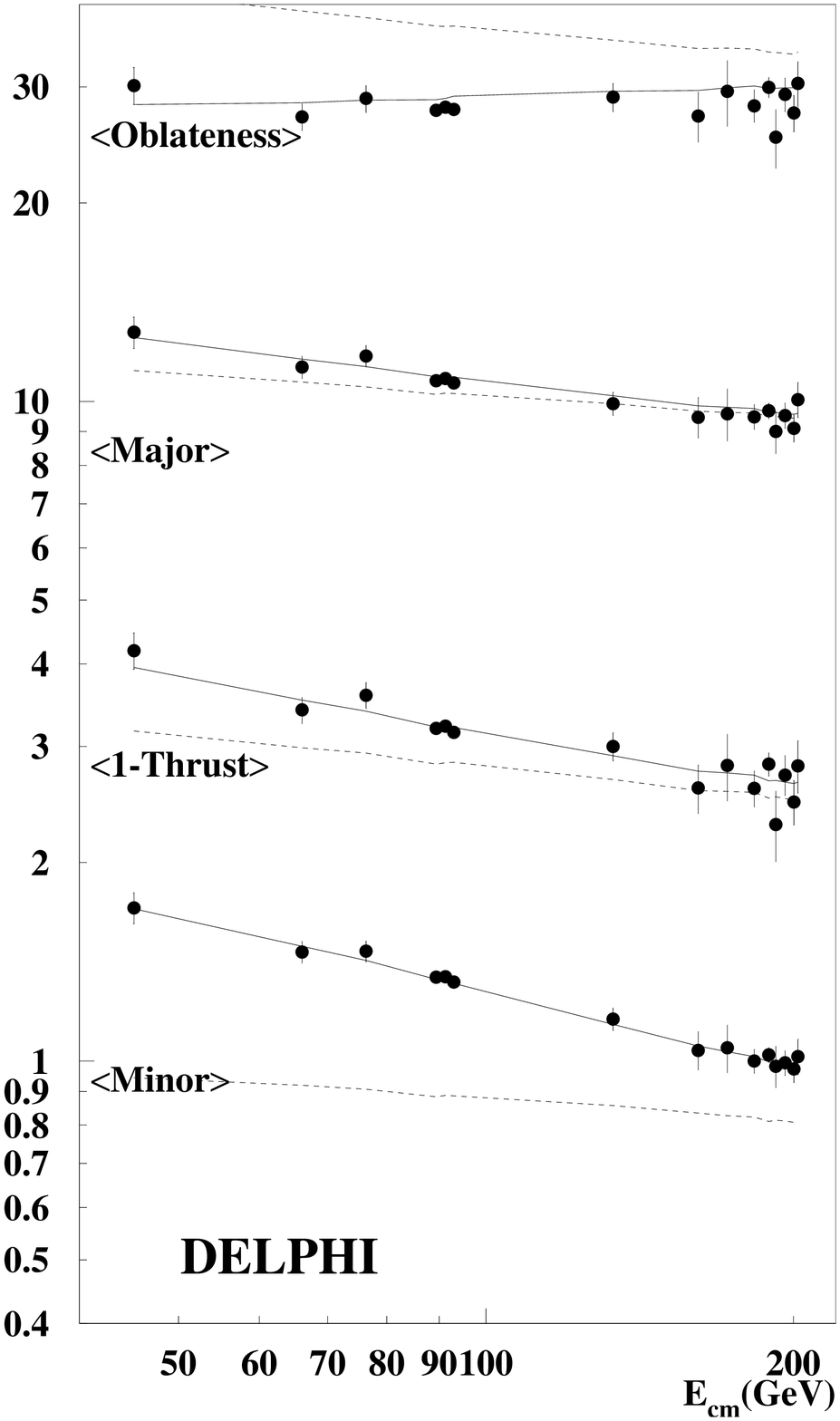,width=7.5cm}}
  \end{minipage}
  \begin{minipage}[t]{7.0cm}
    \mbox{\epsfig{file=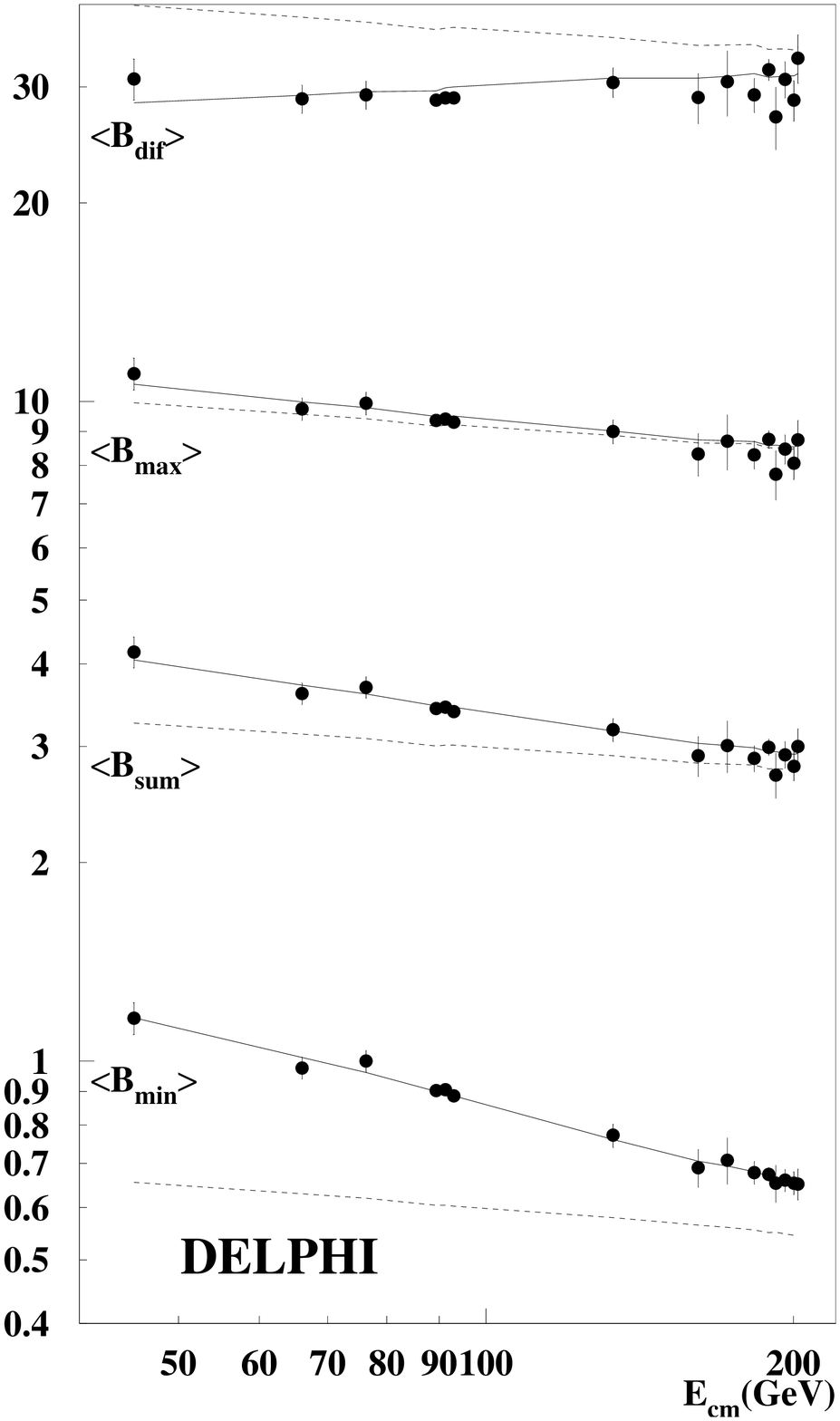,width=7.5cm}}
  \end{minipage}
 \end{center}

\vspace*{-1.cm}
 \begin{center}
  \unitlength1cm
  \begin{minipage}[t]{7.0cm}
    \mbox{\epsfig{file=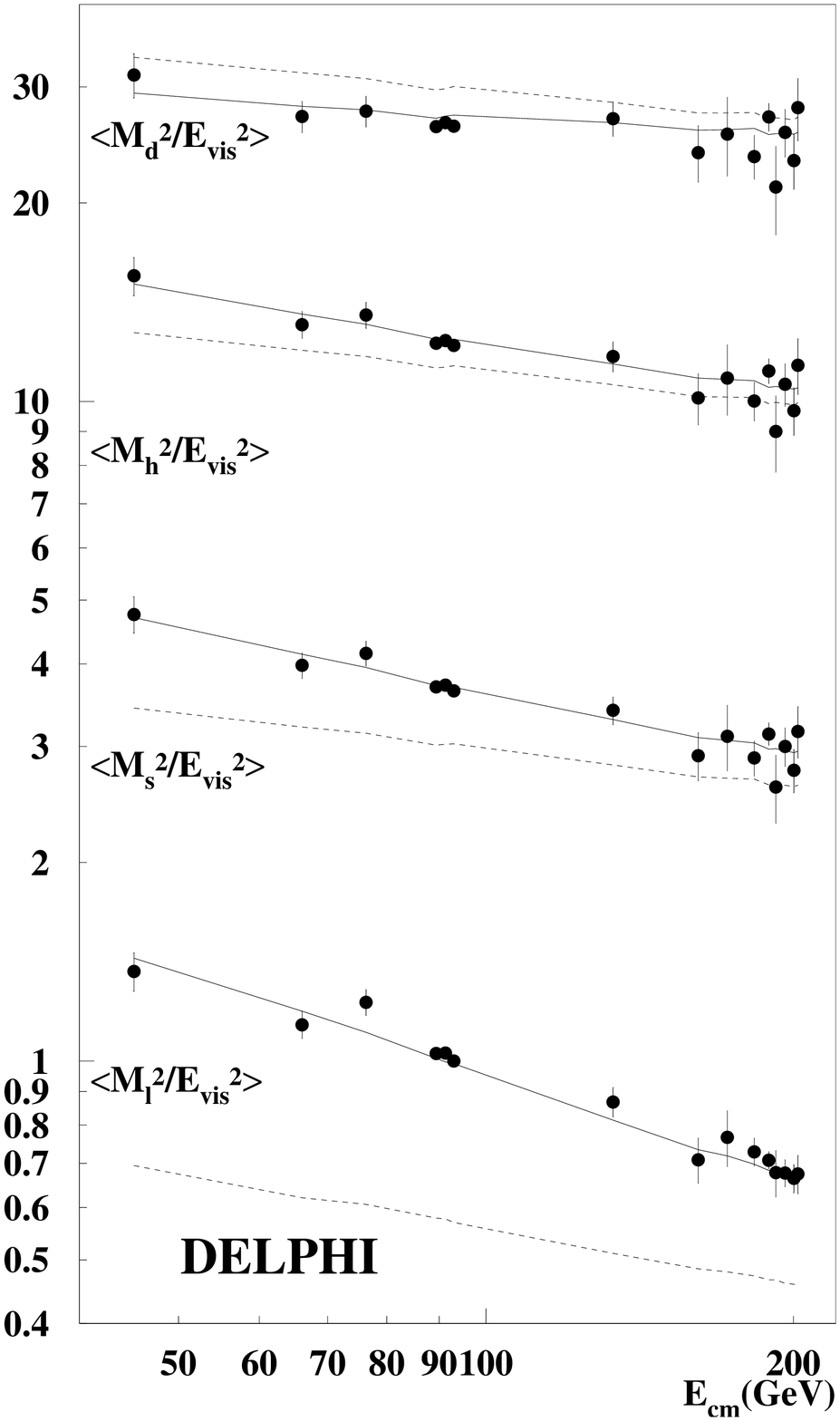,width=7.5cm}}
  \end{minipage}
  \begin{minipage}[t]{7.0cm}
    \mbox{\epsfig{file=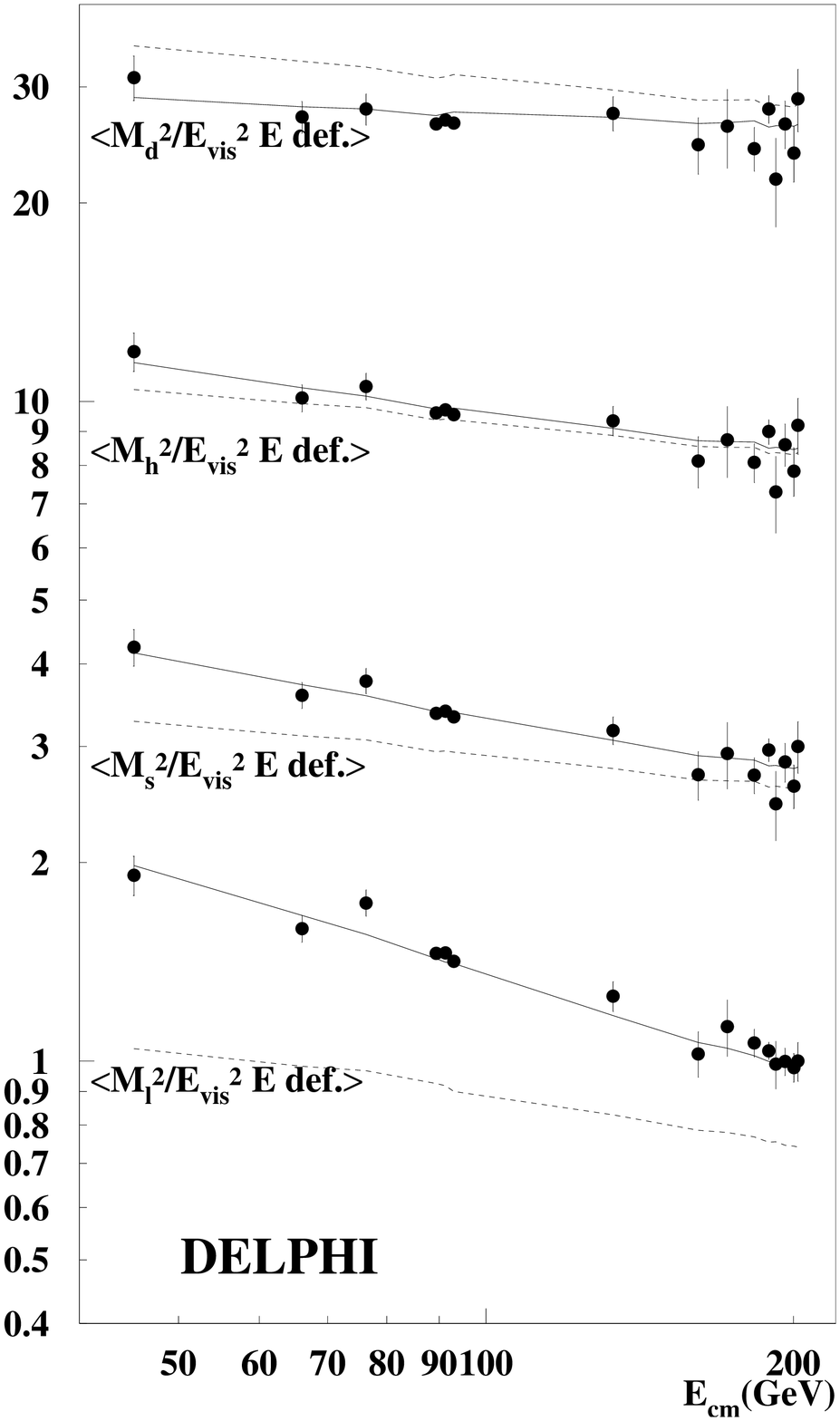,width=7.5cm}}
  \end{minipage}
 \end{center}
 \caption{\label{means}
Event shape means for different observables in comparison to
\pythia\ 6.1 predictions. The full line shows the hadron level,
the dashed line the parton level.
 }
\end{figure}

Figure~\ref{means} shows mean values in the energy range between 45
and 202\gev\ for several observables, including the standard
and the E definition of the Jet Masses.
For comparison, results
from \jetset\ simulations are shown. Again good agreement
between data and model is observed. The dotted line in Figure~\ref{means}
represents the shape observable mean at the parton level
\footnote{Parton and hadron level refer to the simulation of
the hadronic final state. 
Parton level is before, hadron level after hadronisation has taken place.}. 
It is seen that the hadronisation
correction, that is the difference between hadron and parton level curves is
smallest for the observables $\langle{\mathrm{Major}}\rangle$,
$\langle M_{\mathrm{h}}^2/E_{\mathrm{vis}}^2\rangle$, and $\langle B_{\mathrm{max}}\rangle$.
Also the slope of the parton and hadron level agrees best in these cases.
On the other hand for the ``subtracted'' observables like
$ \langle {\mathrm{Oblateness}}\rangle$, which have originally been
constructed to compensate for hadronisation effects, show increased
hadronisation corrections. In these cases the correction can have
opposite sign to the other observables and sometimes even the sign of the
slope of the energy evolution is opposite for parton and hadron level.

The behaviour of the hadronisation correction indicates a clear
preference for observables such as $\langle {\mathrm{Major}} \rangle$,
$ \langle M_{\mathrm{h}}^2/E_{\mathrm{vis}}^2\rangle$, and $\langle
B_{\mathrm{max}}\rangle$ which are mainly sensitive to the
hard gluon radiation in the events.
It should, however, be noted that in these cases some technical problems
may exist in the calculation of resummed theoretical predictions as
soft gluon radiation may lead to a badly controlled exchange of the
wide and narrow event hemispheres~\cite{salamwicke}.

\section{Power corrections to differential distributions \label{res_dw_shape}}
When comparing event shape data to perturbative calculations 
in general corrections
for the effects of non-perturbative hadronisation are applied.
One 
approach to hadronisation corrections 
is the renormalon induced power correction model proposed
in~\cite{DW}. In this model the
origin of non--perturbative effects is determined by Borel transforming
the observables and fixing the singularities found on the real axis.
For several differential distributions and in the simplest
approximation this results in shifting the distribution
${\cal D}_f(f)$ of the observable $f$:
\begin{equation}
\begin{split}
{\cal D}_f(f)&={\cal D}_f^{pert}(f-{\mathcal P} \cdot c_f)~~,
\\
{\cal P}&=\frac{4C_F}{\pi^2} {\cal M}
\frac{\mu_I}{E_{cm}}
  \left[{\alpha}_0(\mu_I) - \alpha_s(\mu)
        - \beta_0\frac{\alpha_s^2(\mu)}{2\pi}
        \left(\ln{\frac{E_{cm}}{\mu_I}} +
\frac{K}{\beta_0} + 1 \right) \right]~~, \\
K&=(67/18-\pi^2/6)C_A-5n_{\mathrm{f}}/9~~. \label{eq:Pshift}
\end{split}
\end{equation}
$C_F$ and $C_A$ are the QCD colour factors.
\asb\ is a non--perturbative parameter accounting for the
contributions to the event shape below an infrared matching scale
$\mu_{\mathrm{I}}$.
The Milan factor ${\cal M}$ for three active flavours in the
non--perturbative region is 1.49 \cite{milan2000}.

Since the derivation of the coefficient $c_f$ is based on resummation,
there are only predictions for exponentiating observables available.
$c_f$ is an observable--dependent constant, which is identical
to the coefficient in the predictions for event shape observable means.
These are:
\begin{center}
\begin{tabular}{|c|c|c|c|c|} \hline
  Observable & ${\mathrm{1-T}}$ & $\mathrm{C}$ & 
$M^2_{\mathrm{h}}/E_{\mathrm{vis}}^2$ & $M^2_{\mathrm{s}}/E_{\mathrm{vis}}^2$ \\ \hline
  $c_f$ & 2 & $3\pi$ & 1 & 2 \\ \hline
\end{tabular}
\end{center}
In order to show all formulae in a coherent fashion, we use
the definition of \cite{PDG2000} for the coefficients of the
$\beta$-function:
\begin{xalignat}{5}
\beta_0&=\frac{33-2n_{\mathrm{f}}}{3}~~, &
\beta_1&=\frac{153-19n_{\mathrm{f}}}{3}~~, &
\beta_2&=2857-\frac{5033}{9}n_{\mathrm{f}}+\frac{325}{27}n_{\mathrm{f}}^2~~.
\label{betadef}
\end{xalignat}

\subsection{Power corrections for the Jet Broadenings 
${\boldmath{B_{\mathrm max}}}$ and ${\boldmath{B_{\mathrm sum}}}$}

Unlike the former observables,
the Jet Broadenings cannot be  sufficiently described by simple shifts,
as the shift becomes a function
of the Jet Broadening.
Early predictions neglected the recoil of the quark due to the gluon emission, 
which proved to be an important effect.
Improved  calculations \cite{hep-ph/9812487} take this  mismatch  into account.
For the wide Jet Broadening $B_{\mathrm{max}}$ the correction coefficient has
the form

\begin{equation}
c_{B_{\mathrm{max}}}(B_{\mathrm{max}})=\frac{1}{2}\left(1/\ln B_{\mathrm{max}}
+\eta_0-2-\rho
({\mathcal R}')+\chi({\mathcal R}')+\psi(1+{\mathcal R}')+\psi(1)\right)~~,
\end{equation}
where
\begin{xalignat}{3}
{\mathcal R}'&=2C_F
\frac{\alpha_s(B_{\mathrm{max}}Q)}{\pi}(\ln B_{\mathrm{max}}^{-1}-\frac{3}{4})~~, &
       \eta_0&=-0.6137~~, \nonumber \\
\psi(z)&=\frac{d}{dz}\Gamma(z)~~, &
\rho(a) &=\int_0^1 dz \left( \frac{1+z}{2z\lambda(a)}\right)^{-a}\ln z(1+z)~~,
       \nonumber \\
\chi(a)&=\frac{2}{a}\left([\lambda(a)]^a-1 \right)~~, &
[\lambda(a)]^{-a} &\equiv \int_0^1 dz \left( \frac{1+z}{2z} \right)^{-a}~~.
\nonumber
\end{xalignat}
For the total Jet Broadening $B_{\mathrm{sum}}$ the correction factor is
\begin{equation}
c_{B_{\mathrm{sum}}}(B_{\mathrm{sum}})=\frac{1}{2} 
\left(2c_{B_{\mathrm{max}}}(B_{\mathrm{sum}})+
2[\psi(1+2{\mathcal R}')-\psi(1+{\mathcal R}')] +H({\mathcal B}^{-1})\right)
\end{equation}
with
\begin{xalignat}{3}
H(x)&=\int_x^{z_0}\frac{dz}{z}\exp\{{{\mathcal R}(x)-{\mathcal R}(z)}\}
  \frac{\Gamma(1+2{\mathcal R}')}{\Gamma(1+2{\mathcal R}'+{\mathcal R}'(z))}~~,&
{\mathcal B}&=\frac{2B_{\mathrm sum}}{e^{\gamma_E}\lambda({\mathcal R}')}~~,
 \nonumber \\
\intertext{with $\gamma_E=0.5772$ being the Euler constant 
and $z_0$ given by the position of the Landau Pole of the two 
loop radiator ${\mathcal R}$:} 
{\mathcal R}(x)&=-\frac{4C_F}{\beta_0}
             \left[ \left({\mathcal L}-\frac{3}{4}\right)
         \ln \left( 1-\frac{\ln x}{\mathcal L}\right)+\ln x \right]~~,
& {\mathcal L}&=\frac{2\pi}{\beta_0\alpha_s
\left(1+K \frac{\alpha_s}{2 \pi} \right) }~~. \nonumber
\end{xalignat}

\subsection{Power corrections for the Energy Energy Correlation EEC}

The power corrections for the EEC have been calculated in
\cite{dw_eec}.
Unlike for the other observables, there is no simple factorisation of
the perturbative and non--perturbative components possible.
Instead the non--perturbative coefficient is a part of the radiator function.
The dominating non--perturbative part is based on the quark-gluon radiation and in
the limit of large angles has a $Q^{-\gamma}$ like behaviour,
with $\gamma\simeq 0.32$ for $n_{\mathrm{f}}=5$. As a result one gets
\begin{equation}
\frac{d\Sigma}{d \cos \chi}=C(\alpha_s)\frac{(1+\tan^2\frac{\chi}{2})^3}{4}
\frac{Q^2}{2}\int_0^\infty b db ~J_0(bQ\tan\frac{\chi}{2})
e^{-{\mathcal R}^{(PT)}(b)-\frac{1}{2}b^2\sigma}\gamma_E
(1-2b\lambda)~~,
\end{equation}
where the perturbative radiator is defined by
\begin{multline}
{\mathcal R}^{(PT)}(b)=-\frac{16\pi C_F}{\beta_0^2} \cdot
\left[ \frac{1}{\alpha_s}(\ln(1-\ell)+\ell) \right .\\
\left . -\frac{3\beta_0}{8\pi}
\ln(1-\ell)
+\frac{\beta_1}{2\pi\beta_0}
\left(\frac{1}{2}\ln^2(1-\ell)+\frac{\ln(1-\ell)}{1-\ell}
+\frac{\ell}{1-\ell} \right) \right]~~,\nonumber
\end{multline}
\begin{equation}
\ell=\beta_0\frac{\alpha_s}{2\pi}\ln\frac{bQe^{\gamma_E}}{2}~~.
 \nonumber
\end{equation}
The linear non--perturbative correction $-2b\lambda$ stems from the correlation
between quarks and soft gluons, where $\lambda$ characterises
the non--perturbative interaction at small momentum scales:
\begin{equation}
\lambda=\frac{4C_F}{\pi^2}{\mathcal M}
\mu_I [{\alpha}_0(\mu_I)-{\alpha}_{0,0}^{PT}(\mu_I)] ~~~.
\end{equation}
The coefficient $C(\alpha_s)$ has the form
\begin{equation}
C(\alpha_s)=1-C_F \left( \frac{11}{2}+\frac{\pi^2}{3}
\right) \frac{\alpha_s}{2\pi} \nonumber~~~.
\end{equation}
The radiator has its own non--perturbative component $\sigma$:
\begin{equation} 
\sigma=\frac{C_F}{2\pi}\mu_I^2 \left\{ \left(
\ln \frac{Q^2}{\mu_I^2}-\frac{1}{2} \right) [ {\alpha}_{1}(\mu_I)
-{\alpha}_{1,0}^{PT}(\mu_I)]+{\alpha}_{1,1}(\mu_I)-
{\alpha}_{1,1}^{PT}(\mu_I) \right\} ~~~.
\end{equation}
The treatment of this non--perturbative component is unclear, as it is much weaker than
the linear part and it is missing the quadratic part from the quark-gluon
correlation.
The ${\alpha}_{p,q}^{PT}(\mu_I)$ are normalisation factors:
\begin{equation} 
{\alpha}_{p,q}^{PT}(\mu_I)=\alpha_s+\frac{\beta_0}{2\pi}
\left( \ln\frac{Q}{\mu_I}+\frac{K}{\beta_0}+\frac{q+1}{p+1}
\right) \alpha_s^2 ~~~.
\end{equation}
As a result one has to deal with three non--perturbative
parameters: ${\alpha}_{0},{\alpha}_{1}$
and ${\alpha}_{1,1}$, where ${\alpha}_{0}$ is equivalent to the
non--perturbative parameter of the other observables.
A phenomenological prediction of ${\alpha}_{1}$ is given in \cite{dw_eec}
from DIS experiments and the theoretical predictions are
${\alpha}_{1}=0.45$ and ${\alpha}_{1,1}=0.55$.
Since the formula for the EEC is a large angle approximation it is important
to choose a proper fit region.
Below $120^\circ$ the prediction becomes invalid,
while above $170^\circ$ the influence of higher order logarithmic terms can 
no longer
be neglected.

\subsection{Comparison with data}

\begin{figure}[htb]
\begin{minipage}[t]{8cm}
\mbox{\epsfig{file=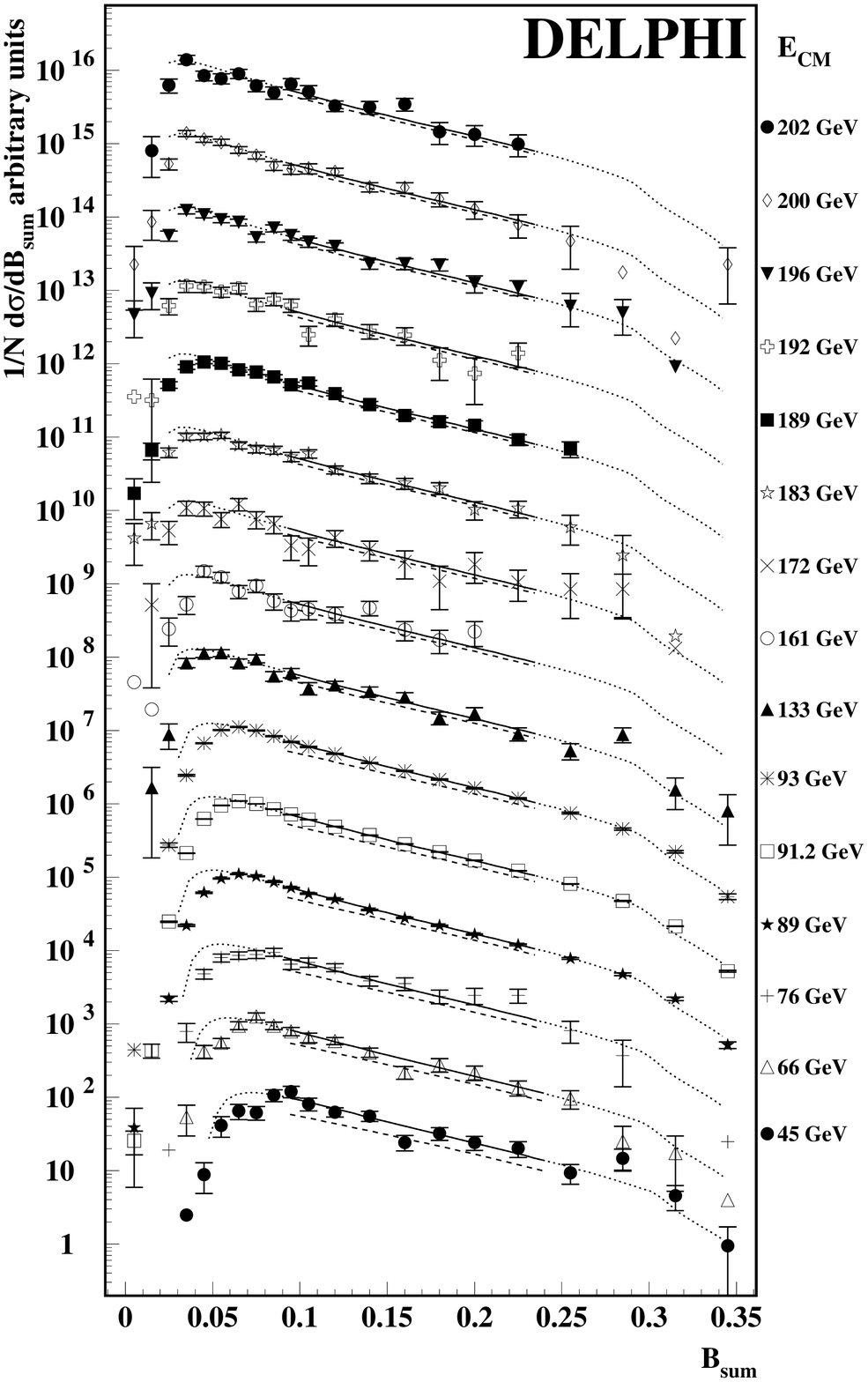,width=8cm}}
\end{minipage}
\begin{minipage}[t]{8cm}
\mbox{\epsfig{file=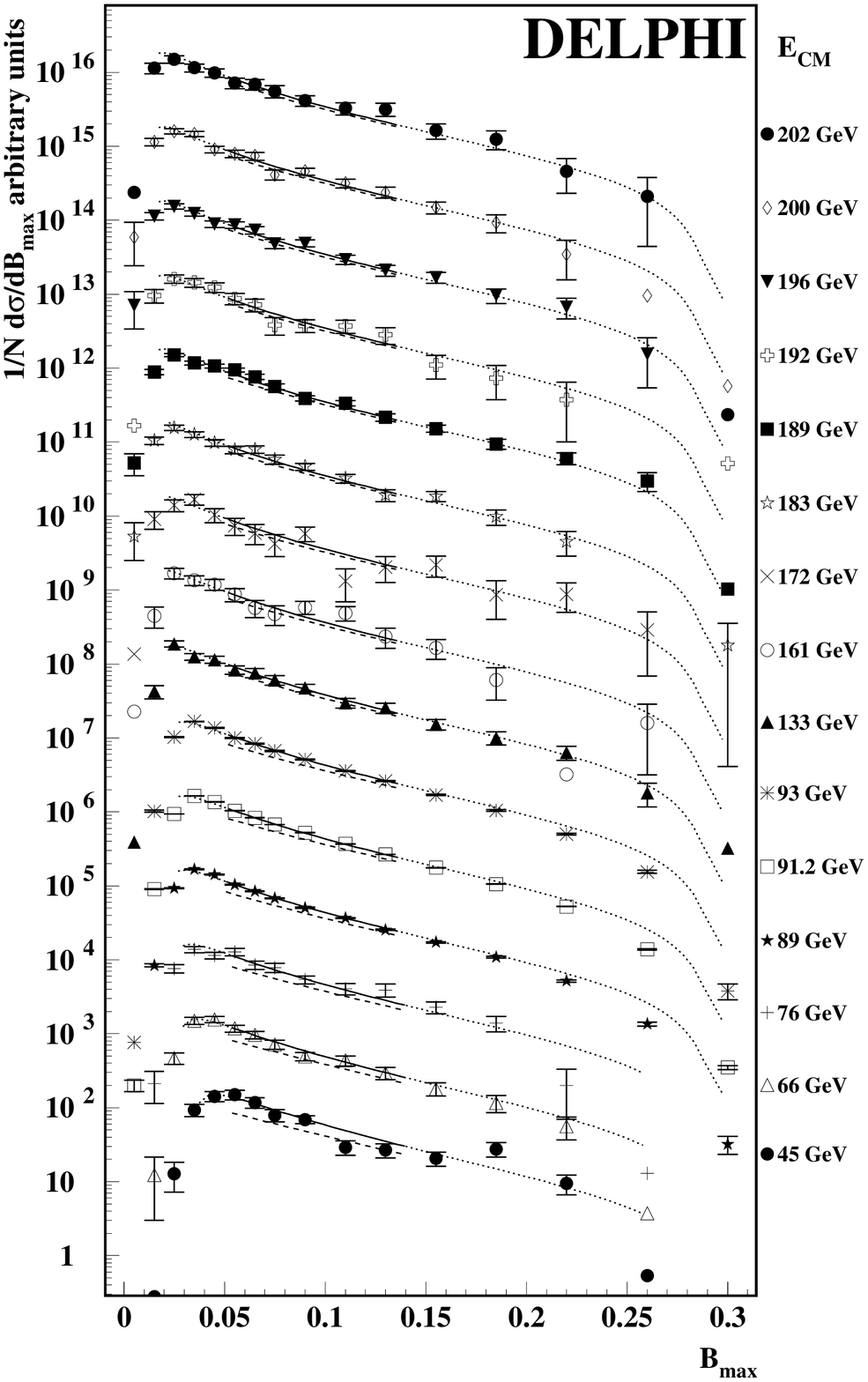,width=8cm}}
\end{minipage}
\caption{
Jet Broadening distributions as measured by DELPHI
for centre-of-mass energies between 45 and 202\gev. 
The full line indicates the power model fit in the
fit range, while the dotted line shows the extrapolation beyond
the fit range.
The dashed line shows the result after subtraction of the
power correction.}\label{12c_pic} 
\end{figure}

\begin{figure}[htb]
\begin{minipage}[t]{8cm}
\mbox{\epsfig{file=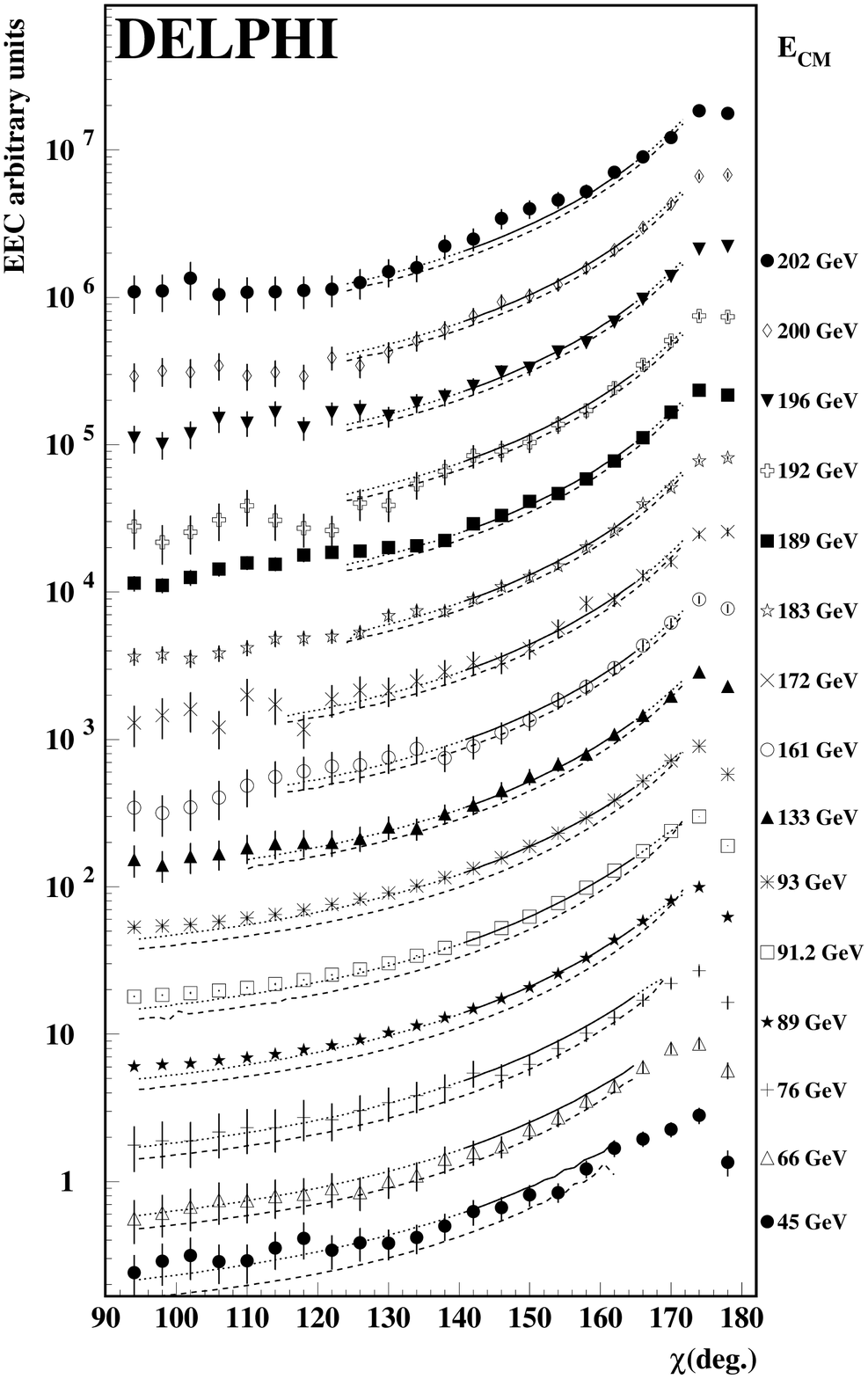,width=8cm}}
\end{minipage}
\begin{minipage}[t]{8cm}
\mbox{\epsfig{file=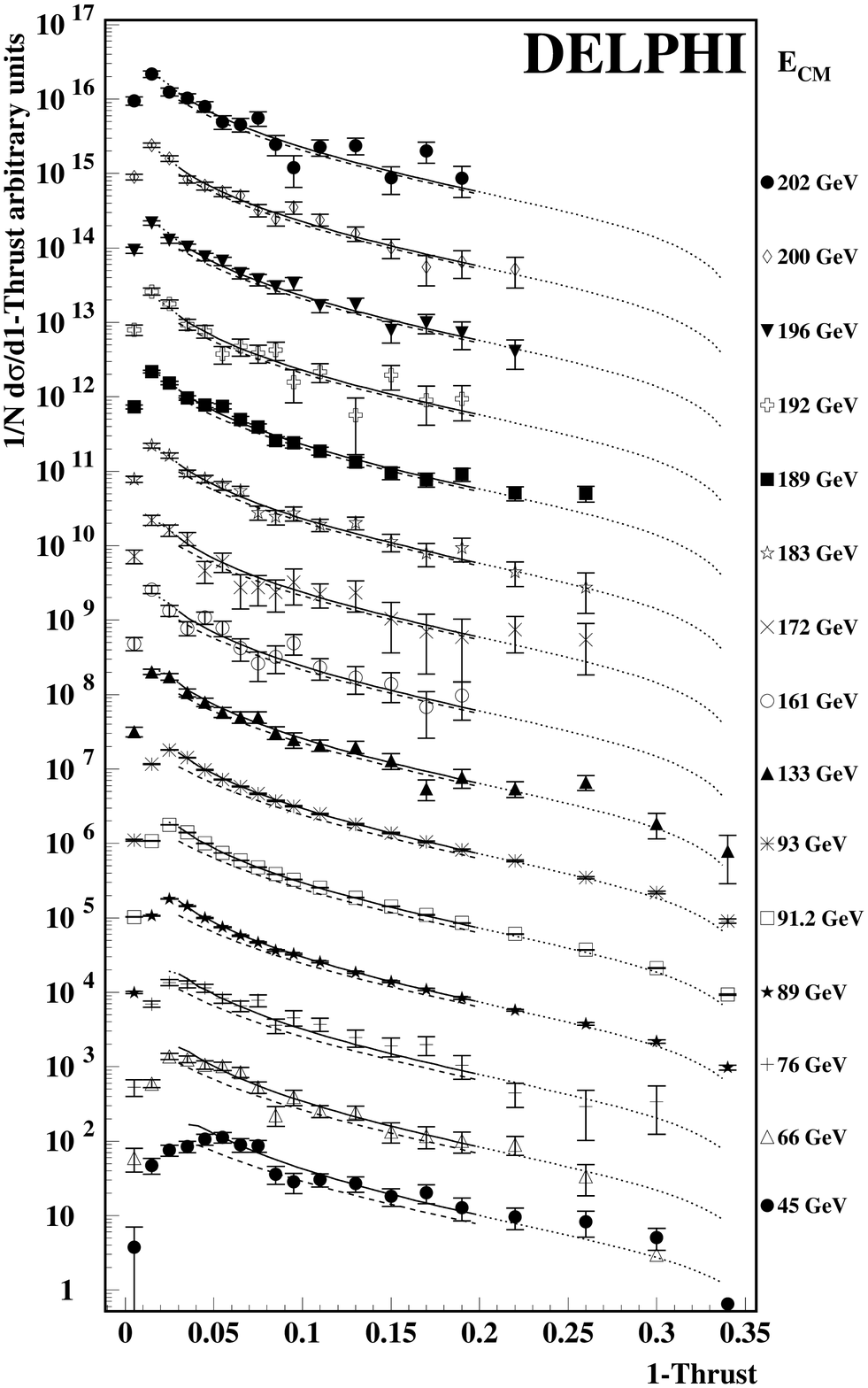,width=8cm}}
\end{minipage}
\caption{
${\mathrm{EEC}}$ and ${\mathrm{1-Thrust}}$ distributions as measured by DELPHI
for centre-of-mass energies between 45 and 202\gev. 
The full line indicates the power model fit in
fit range, while the dotted line shows the extrapolation beyond
the fit range.
The dashed line shows the result after subtraction of the
power correction.}
\label{eecfit}
\end{figure}

\begin{table}[tbh]
\begin{center}
\renewcommand{\arraystretch}{1.2}
\begin{tabular}{|c|r@{.}l@{~(}r@{.}l@{)~}|r@{.}l@{~(}r@{.}l@{)~}|}\hline
Observable &\multicolumn{4}{c|}{lower} &\multicolumn{4}{c|}{upper} \\ \hline
$1-T$& 0&03 & 0&02 &0&2 &0&24  \\
$B_{\mathrm{max}}$& 0&05 & 0&04 &0&14 &0&16  \\
$B_{\mathrm{sum}}$& 0&09 & 0&08 &0&17 &0&19 \\
$M_{\mathrm{h}}^2/E_{\mathrm{vis}}^2$& 0&03 & 0&02 &0&12 &0&14  \\
$M_{\mathrm{hE}}^2/E_{\mathrm{vis}}^2$& 0&03 & 0&02 &0&12 &0&14  \\
$M_{\mathrm{hp}}^2/E_{\mathrm{vis}}^2$& 0&03 & 0&02 &0&12 &0&14  \\
$M_{\mathrm{s}}^2/E_{\mathrm{vis}}^2$& 0&03 & 0&02 &0&12 &0&14  \\
$M_{\mathrm{sE}}^2/E_{\mathrm{vis}}^2$& 0&03 & 0&02 &0&12 &0&14  \\
$M_{\mathrm{sp}}^2/E_{\mathrm{vis}}^2$& 0&03 & 0&02 &0&12 &0&14  \\
$\mbox{C-parameter}$& 0&2 & 0&16 &0&68 &0&72  \\
\hline
\end{tabular}
\end{center}
\caption{\label{fit_interval}Fit intervals used for the fit of power corrections
to event shape distributions. The variations of the fit interval used for systematic studies
are shown in brackets.}
\end{table}

\begin{table}[tb]
\begin{center}
\renewcommand{\arraystretch}{1.2}
\begin{tabular}{|c|r@{.}l@{$\pm$}r@{.}l@{$\pm$}r@{.}l@{
}r@{.}l|r@{.}l@{$\pm$}r@{.}l@{$\pm$}r@{.}l@{ }r@{.}l|r@{/}l|}\hline
Observable &\multicolumn{8}{c|}{$\alpha_s(M_Z)$}
&\multicolumn{8}{c|}{$\alpha_0(\mu_I=2\gev )$}
&\multicolumn{2}{c|}{$\chi^2/ndf$} \\ \hline
$1-T$& 0&1154 & 0&0002 &0&0017 &+0&0004 & 0&543 & 0&002 &0&014 &+0&013 &
291 & 180 \\ 
$B_{\mathrm{max}}$& 0&1009 & 0&0003 &0&0016 &$-$0&0018 & 0&571 & 0&005
&0&031 & +0&021 & 106 &  90 \\ 
$B_{\mathrm{sum}}$& 0&1139 & 0&0006 &0&0015 &$-$0&0035 & 0&465 & 0&005
&0&013 &+0&008 &  88 &  75 \\ 
$M_{\mathrm{h}}^2/E_{\mathrm{vis}}^2$& 0&1076 & 0&0001 &0&0013 & +0&0003 &
0&872 & 0&000 &0&026 &+0&005 & 158 &  90 \\ 
$M_{\mathrm{hE}}^2/E_{\mathrm{vis}}^2$& 0&1056 & 0&0003 &0&0006 &+0&0001 &
0&692 & 0&007 &0&010 &+0&010 & 120 &  90 \\ 
$M_{\mathrm{hp}}^2/E_{\mathrm{vis}}^2$& 0&1055 & 0&0004 &0&0010 &+0&0001 &
0&615 & 0&009 &0&022 &+0&010 & 130 &  90 \\ 
$M_{\mathrm{s}}^2/E_{\mathrm{vis}}^2$& 0&1190 & 0&0004 &0&0030 & +0&0001 &
0&734 & 0&004 &0&034 &+0&009 & 66  &  45 \\ 
$M_{\mathrm{sE}}^2/E_{\mathrm{vis}}^2$& 0&1166 & 0&0004 &0&0028 & +0&0002 &
0&583 & 0&004 &0&027 &+0&007 &  60 &  45 \\ 
$M_{\mathrm{sp}}^2/E_{\mathrm{vis}}^2$& 0&1156 & 0&0005 &0&0010 & +0&0001 &
0&536 & 0&005 &0&010 &+0&008 &  54 &  45 \\ 
$\mbox{C-Parameter}$& 0&1097 & 0&0004 &0&0032 &$-$0&0008 & 0&502 & 0&005
&0&047 &+0&021 & 191 & 180 \\ \hline
weighted mean& 0&1078 & 0&0005 &0&0013 &$-$0&0012 & 0&546 & 0&005 &0&022
&+0&013 & \multicolumn{2}{c|}{} \\ 
unweighted mean& 0&1110 & 0&0055 &0&0007 &$-$0&0008 & 0&559 & 0&073 &0&009
&+0&013 & \multicolumn{2}{c|}{} \\ 
\hline
EEC & 0&1171 & 0&0018 & 0&0004 &\multicolumn{2}{c|}{}& 0&483 & 0&040 & 0&011 & \multicolumn{2}{c|}{}& 53 &90 \\
\hline
\end{tabular}
\end{center}
\caption{\label{shape_fit}Determination of \as\ and \asb\ from a
fit to  event shape distributions.
Only \delphi\ measurements are included in the fit.
The first error is the statistical
uncertainty from the fit, the second one is the systematic uncertainty, 
the third the difference with respect to the R matching scheme. 
Only E--definition Jet Masses have been taken for the means. For the 
definition of the mean values see section \ref{average}.}
\end{table}

The derivation of the power correction predictions in general relies on the
resummation of logarithmically divergent terms.
The validity of these predictions is thus 
limited to a kinematical region close
to the two jet regime.
This has been taken into account when choosing the fit intervals indicated
in Table~\ref{fit_interval}.
Additionally it was required that the corrections applied to the data 
as well as the fit results obtained were stable.
The experimental systematics have been determined as discussed in
Section~\ref{systematics}. In addition, changes of the fit ranges were applied
as tabulated in Table~\ref{fit_interval}.
All systematic studies enter into the specified systematic uncertainty.
Also the so called R matching was applied for the perturbative prediction
instead of the standard $\log$R matching \cite{logr}. It is notable that 
the change of the
matching scheme has a significant influence on the size of the power corrections.

\begin{figure}[htb]
\begin{center}
\mbox{\epsfig{file=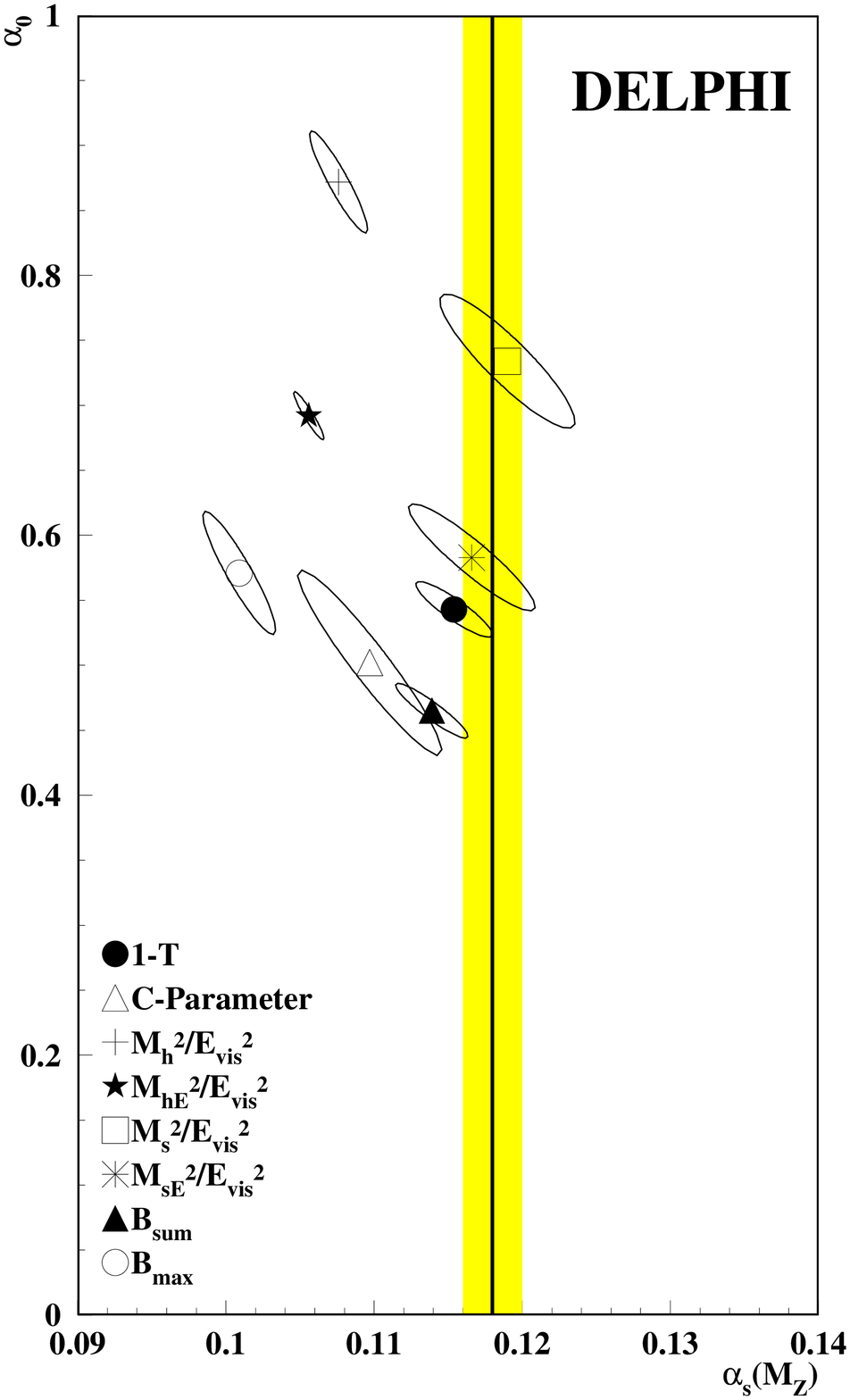,width=7.4cm}}
\end{center}
\caption{
Results of the fits to shape distributions in the $\alpha_s$--$\alpha_0$ plane.
The band indicates the world average of \as.
} \label{shapecorrelation}
\end{figure}

Examples of fits compared to the data on the Jet Broadenings, 
${\mathrm{1-Thrust}}$ and the ${\mathrm{EEC}}$,
are shown in Figures \ref{12c_pic} and \ref{eecfit}.
Results of the fitted parameters for all observables determined 
from \delphi\ data are given in Table~\ref{shape_fit}.
The expected correlation of the fit parameters \as\ and $\alpha_0$ 
is displayed in Figure \ref{shapecorrelation}.
The $\chi^2/ndf$ of the fit is acceptable when systematic uncertainties
are included. The \as\ values tend to be rather low compared to the world
average value of $\alpha_s(M_{\mathrm Z})=0.1181\pm0.002$ \cite{PDG2000} 
for most observables.
The non--perturbative parameter ${\alpha}_0$ is higher for the classical
$M_{\mathrm{h}}^2/E_{\mathrm{vis}}^2$
as expected due to the influence of hadron mass effects.
For the other observables the results for ${\alpha}_0$ agree within
a relative uncertainty of about 20\%.

The result of the fit of the EEC is shown in Figure~\ref{eecfit}~(left).
The influence of the non--perturbative parameters
${\alpha}_1$ and ${\alpha}_{1,1}$
is found to be much smaller than the influence of ${\alpha}_0$.
The precision of the experimental data is insufficient to  determine
${\alpha}_{1,1}$.
A three parameter fit neglecting ${\alpha}_{1,1}$ yielded:
$\alpha_s=0.1173\pm 0.0021 \pm 0.0008 $,
${\alpha}_0=0.478\pm 0.046 \pm 0.017$ and
${\alpha}_1=0.005\pm 0.026 \pm 0.025$
with a $\chi^2/ndf=52.7/90$.
Since this fit indicates that the non--perturbative part of the radiator 
${\alpha}_1$
can be neglected, an additional
two parameter fit was performed, resulting in:
$\alpha_s=0.1171\pm 0.0018 \pm 0.0004 $,
${\alpha}_0=0.483\pm 0.040 \pm 0.011$,
with  $\chi^2/ndf=53/91$.
The ${\alpha}_0$ value for the EEC is consistent with the values
determined from the other observables (see Table~\ref{shape_fit}).

\subsection{Measurement of the non--perturbative shift from the Sudakov
Shoulder}

\begin{figure}[htb]
\begin{center}
\mbox{\epsfig{file=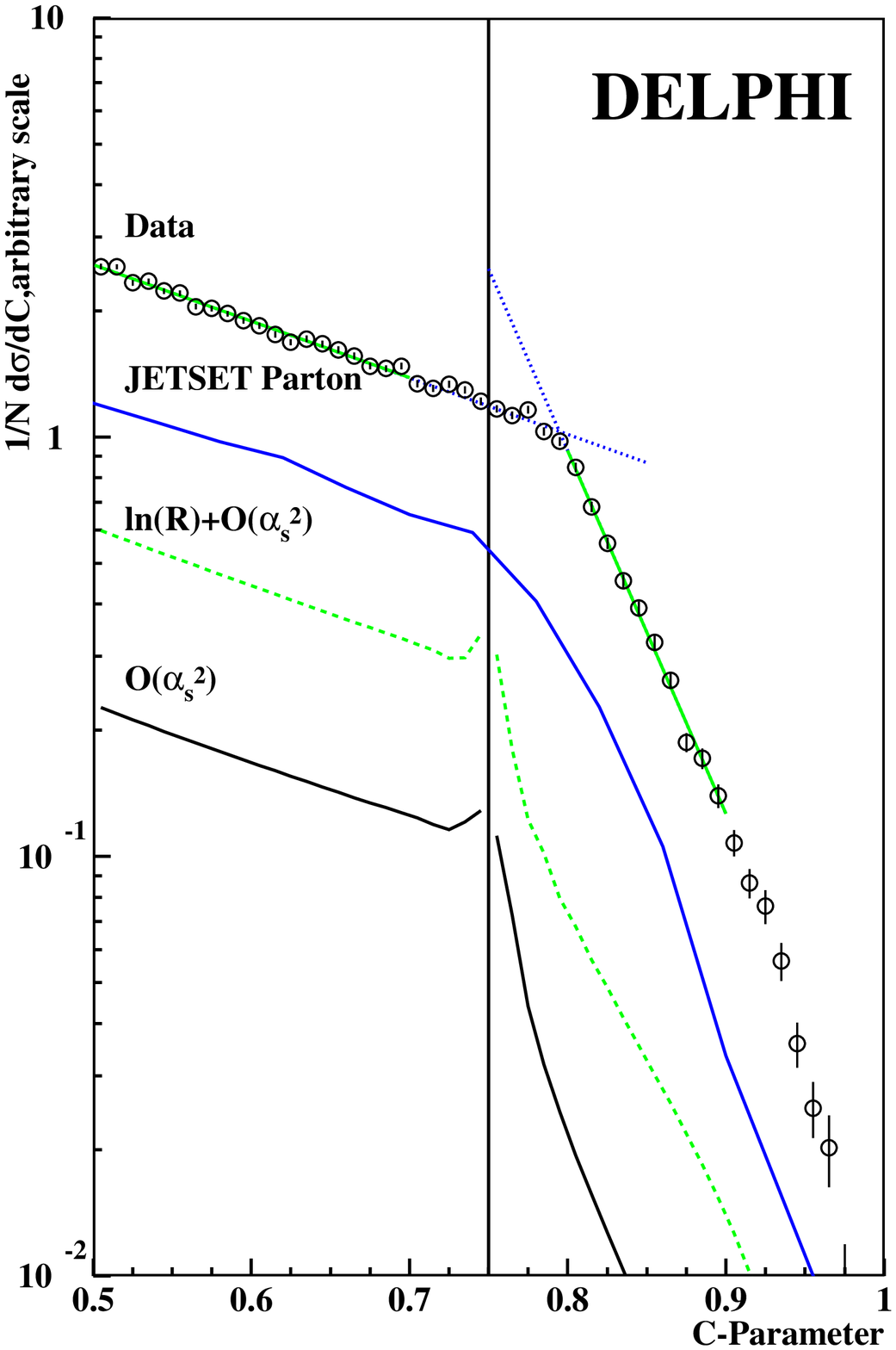,width=9.cm}}
\end{center}
\caption{Determination of the shift of the Sudakov shoulder
in comparison to predictions. The vertical positions are arbitrarily scaled.
The fit range (straight line) and the extrapolation to the intersection
 point (dotted line) are shown. The vertical line denotes the three jet
 limit of~0.75~.} \label{Sudakov}
\end{figure}

So far all predictions apply near to the two jet region. For some
observables, however, the predictions seem to hold even in the far
three jet region.
In order to provide evidence for this observation 
the three jet limit of several observables was studied.

Despite the fact that QCD event shape observables are constructed 
so as to be infrared- and
collinear safe, there can still be infinities at accessible points
in phase space. The finiteness is only restored after the resummation
of divergent terms to all orders. The resulting structure is called a
Sudakov shoulder \cite{sudshoulder}. The most common case is
the phase space boundary of the three jet region, which introduces
a visible edge into the distributions. The position can be calculated
and is, for example,  $2/3$ for 1-Thrust and $3/4$ for the C-parameter.
A simple inspection of the distributions and the corresponding model curves
shows that the shoulder is
typically shifted to higher values, and that power corrections describe
the shift rather well.

The shift for the C parameter can be measured by
fitting the slope of the logarithmic distribution on both sides of 
the shoulder.
The intersection of these fits is a good approximation to the
shoulder position.
The result of the fit can be seen in Figure~\ref{Sudakov}, the fitted
position of the shoulder is at C$=0.794\pm 0.016_{stat} \pm 0.001_{sys}$ corresponding to
a shift of  $+0.044 \pm 0.016_{stat} \pm 0.001_{sys}$ with respect to the
nominal position.
Using a value of $c_{\mathrm{C}}=3\pi$ for the C parameter, a value of
$\alpha_0=0.476 \pm 0.097_{stat} \pm 0.0015_{sys}$
is obtained from Equation~\ref{eq:Pshift}.
This result is well consistent with the result obtained from
the fit of the overall distribution
$\alpha_0=0.502 \pm 0.005_{stat} \pm 0.047_{sys}$
suggesting a constant
shift over the whole three jet region in the case of the C parameter.

\section{Power corrections in event shape means\label{sec:powermeans}}
The mean values of event shape variables are defined as:
\begin{equation}
\left< f \right> ~~=~~
\frac{1}{\sigma_{\mathrm{tot}}}\int f\frac{df}{d\sigma}d\sigma ~~=~~
\frac{1}{N_{\mathrm{evt.}}} \sum_{i=1}^{N_{\mathrm{evt.}}} f_i~~~.
\nonumber
\end{equation}
We have calculated them from the detector corrected and binned distributions.
Hence they are fully inclusive quantities depending on a single energy
scale only, and are well suited for low statistics analyses
as the statistical uncertainty is minimised by using all events.
Though the characteristics of the event shape observables may differ
in specific regions of the value of the observable, global properties 
can be assessed
from the energy dependence of the mean value.

\subsection{The Dokshitzer and Webber ansatz \label{dw_mean}}

The analytical  power ansatz \cite{PhysLettB352_451,hep-ph/9510283} including the Milan
factor \cite{NuclPhysB511_396,hep-ph/9802381}
is used to determine \as\ from mean event shapes.
This ansatz provides an additive non--perturbative term 
$\left< f_{\mathrm{pow}} \right>$
to the perturbative \oas\ QCD prediction $\left< f_{\mathrm{pert}} \right>$,
\begin{equation}
\left< f \right> =
\left< f_{\mathrm{pert}} \right> + \left< f_{\mathrm{pow}} \right>~~,
\label{eq_f}
\end{equation}
where the 2nd order perturbative prediction can be written as
\begin{equation}
\left< f_{\mathrm{pert}} \right> = A   \frac{\alpha_s(\mu)}{2\pi}+
         \left(A\cdot \beta_0 \ln\frac{\mu}{E_{cm}} + B\right)
                \left(\frac{\alpha_s(\mu)}{2\pi}\right)^2{\rm ,}
\label{eq_fpert_o2}
\end{equation}
A and B are known
coefficients~\cite{NuclPhysB178_412,CERN89-08vol1,event2} and
$\mu$ is the renormalisation scale.
The power correction is given by
\begin{equation}
\left < f_{\mathrm{pow}}\right >  =  c_{f} {\cal P}~~,
\label{eq_fpow_dw}
\end{equation}
where ${\cal P}$ is as defined in Equation~\ref{eq:Pshift}.

The observable-dependent coefficient $c_f$ is identical 
for shapes and means. In the case of the jet
broadenings $c_{f}$ cannot be described as a constant.
Here the non--perturbative contribution is proportional to
$1/(Q\sqrt{\alpha_s(Q)})$~\cite{hep-ph/9812487}:
\begin{equation}
c_{f}= c_{\mathrm{B}}
\left( \frac{\pi\sqrt{c_{\mathrm{B}}}}{2\sqrt{C_{F} \alpha_{\overline{MS}}
\left( 1+K\frac{\alpha_{\overline{MS}}}{2\pi}\right)
}}+\frac{3}{4}  -\frac{\beta_0 c_{\mathrm{B}}}{6C_{F}}+\eta_0  \right)~~~,
\label{eq_fpow_dw_bmax}
\end{equation}
where $c_{\mathrm{B}}$ is 1/2 in the case of $\left<B_{\mathrm{max}}\right>$
and 1 for $\left<B_{\mathrm{sum}}\right>$, $\eta_0=-0.6137056$.

In the following analysis the infrared matching scale $\mu_{\mathrm{I}}$
was set to 2\gev, as suggested in \cite{PhysLettB352_451},
and the renormalisation scale $\mu$ was set to ${E_{cm}}$.
%
\begin{table}[tb]
\begin{center}
\renewcommand{\arraystretch}{1.2}
\begin{tabular}{|c|r@{.}l@{$\pm$}r@{.}l@{$\pm$}r@{.}l|r@{.}l@{$\pm$}r@{.}l@{$\pm$}r@{.}l|r@{.}l@{/}l|}\hline
Observable &\multicolumn{6}{c|}{$\alpha_0(\mu_I=2 \gev)$} &\multicolumn{6}{c|}{$\alpha_s(M_Z)$} &\multicolumn{3}{c|}{$\chi^2/ndf$} \\ \hline
$\langle 1-T \rangle$  & 0&491 & 0&016 &0&009& 0&1241 & 0&0015 &0&0031& 26&5 &  41 \\
$\langle \mbox{C-Parameter} \rangle$  & 0&444 & 0&020 &0&008& 0&1222 & 0&0020 &0&0030& 11&6 &  23 \\
$\langle M_{\mathrm{h}}^2/E_{\mathrm{vis}}^2 \rangle$  & 0&601 & 0&058 &0&012& 0&1177 & 0&0030 &0&0018& 14&1 &  27 \\
$\langle M_{\mathrm{hp}}^2/E_{\mathrm{vis}}^2\rangle $  & 0&300 & 0&222 &0&127& 0&1185 & 0&0104 &0&0057& 10&1 &  15 \\
$\langle M_{\mathrm{hE}}^2/E_{\mathrm{vis}}^2\rangle $  & 0&339 & 0&229 &0&129& 0&1197 & 0&0107 &0&0058& 9&5 &  15 \\
$\langle M_{\mathrm{s}}^2/E_{\mathrm{vis}}^2 \rangle$  & 0&544 & 0&160 &0&093& 0&1335 & 0&0118 &0&0074& 7&2 &  15 \\
$\langle M_{\mathrm{sp}}^2/E_{\mathrm{vis}}^2 \rangle$  & 0&378 & 0&138 &0&084& 0&1288 & 0&0104 &0&0067& 8&7 &  15 \\
$\langle M_{\mathrm{sE}}^2/E_{\mathrm{vis}}^2 \rangle$  & 0&409 & 0&143 &0&086& 0&1304 & 0&0107 &0&0069& 8&2 &  15 \\
$\langle B_{\mathrm{max}} \rangle$  & 0&438 & 0&041 &0&027& 0&1167 & 0&0018 &0&0007& 10&1 &  23 \\
$\langle B_{\mathrm{sum}} \rangle$  & 0&463 & 0&032 &0&009& 0&1174 & 0&0021 &0&0020& 8&8 &  23 \\ \hline
weighted mean & 0&468 & 0&080 &0&008& 0&1207 & 0&0048 &0&0026& \multicolumn{2}{c}{} &  \\
unweighted mean & 0&431 & 0&048 &0&039& 0&1217 & 0&0046 &0&0030& \multicolumn{2}{c}{} &  \\

\hline
\end{tabular}
\end{center}
\caption{\label{tab_mess_fit}
Determination of \asb\ and \as\ from a
fit to a large set of event shape mean values measured from different
experiments~\cite{collection_eventshapes}.
For $\ecm\ge M_{\mathrm{Z}}$
only \delphi\ measurements are included in the fit.
The first error is the statistical
uncertainty from the fit, the second one is the systematic uncertainty.
For the mean values only the E--definition Jet Masses have been used. For the 
definition of the mean values see section \ref{average}.}
\end{table}

\begin{figure}[t]
\unitlength1cm
 \unitlength1cm
 \begin{center}
 \begin{minipage}[t]{7.5cm}
    \mbox{\epsfig{file=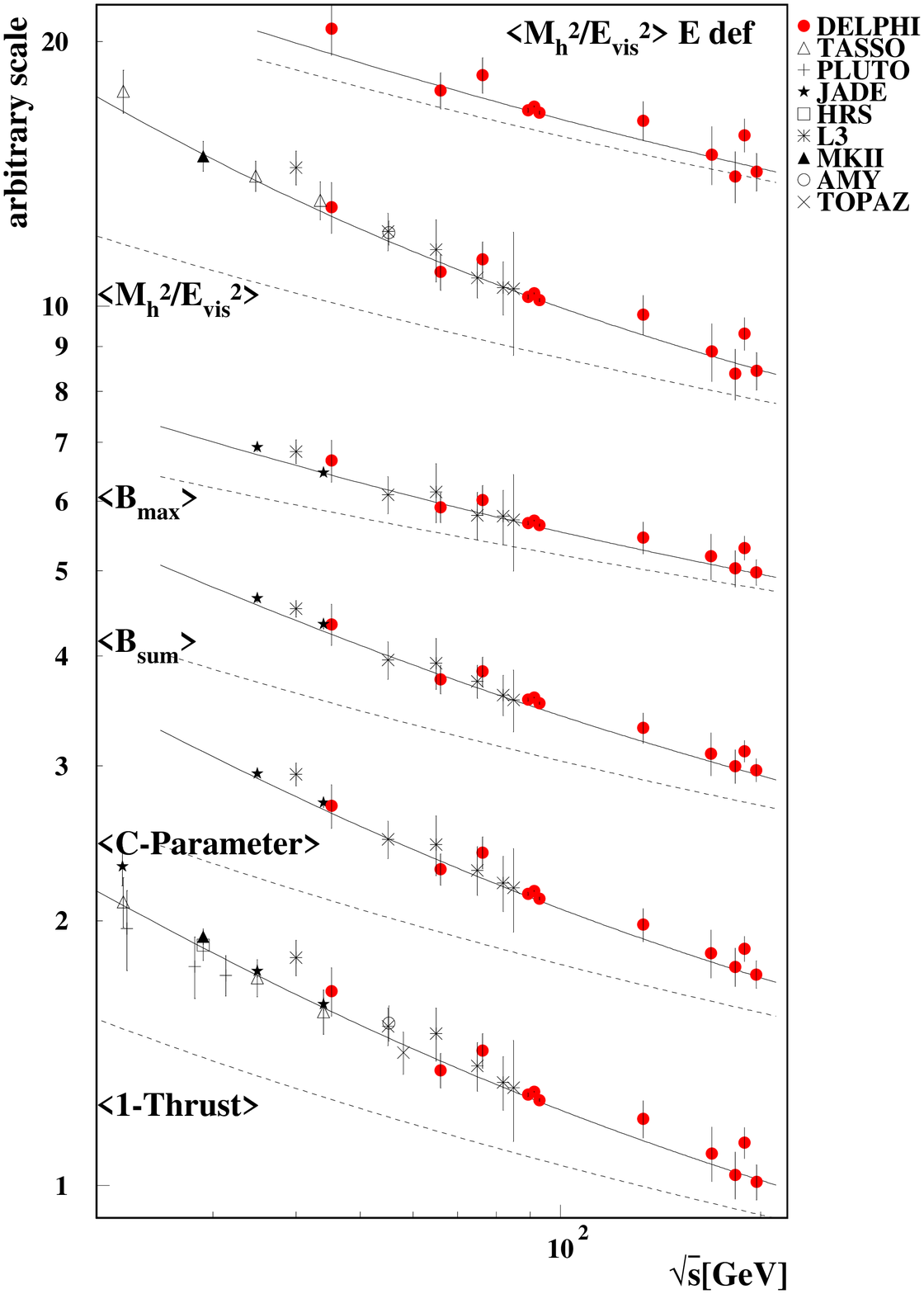,width=8.1cm}}
 \end{minipage}
 \begin{minipage}[t]{7.5cm}
    \mbox{\epsfig{file=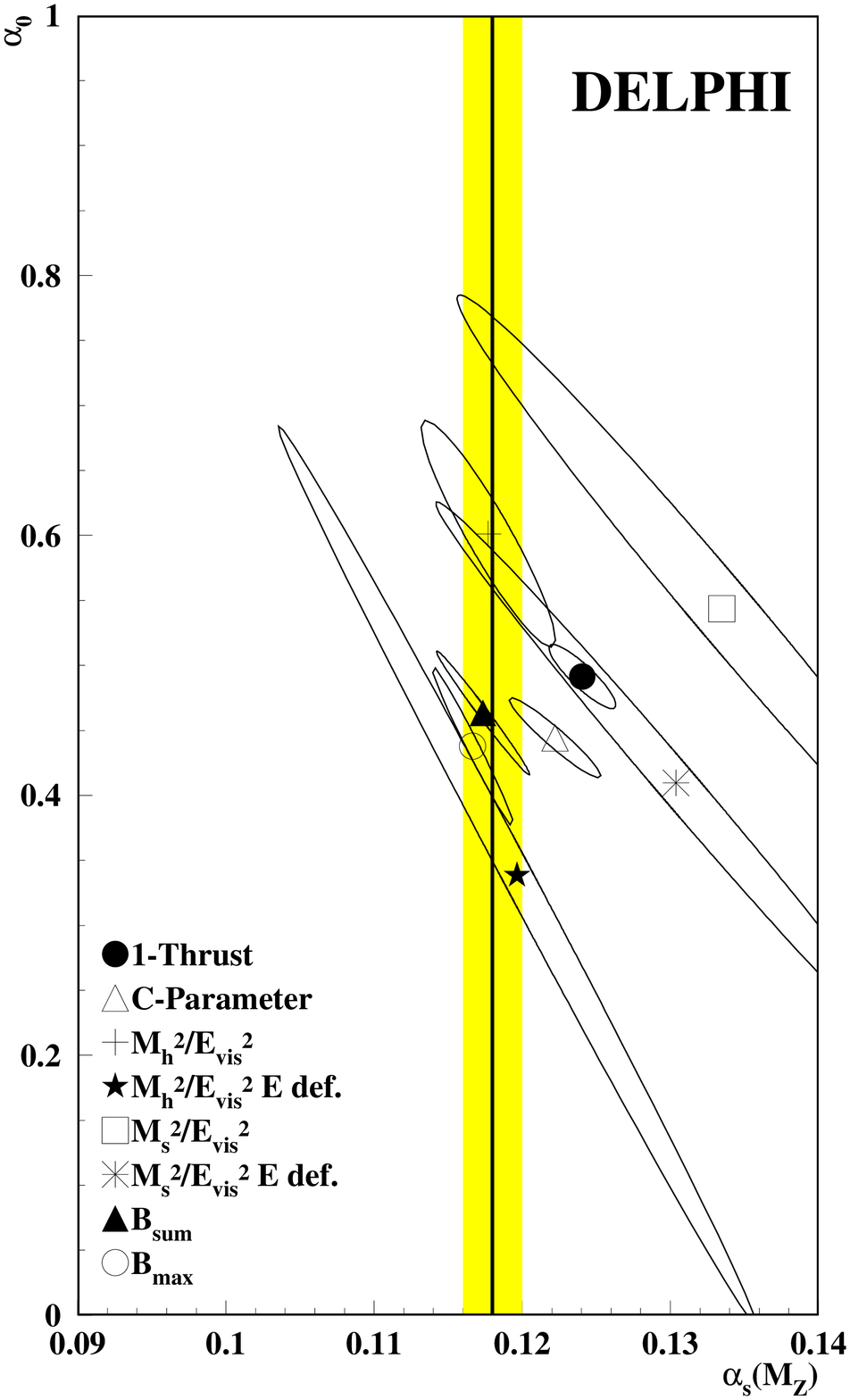,width=8.1cm}}
 \end{minipage}
\end{center}
\caption{\label{fig_mess_fit}
Left: Measured mean values of
$\left<1-T\right>$,
$\left<M_{\mathrm{h}}^2/E_{\mathrm{vis}}^2\right>$,
$\left<C\right>$,
$\left<B_{\mathrm{sum}}\right>$ and
$\left<B_{\mathrm{max}}\right>$ as a function of the centre-of-mass
energy. For clarity some of the high energy data have been merged.
The solid lines present the results of the fits with
Equations~(\protect\ref{eq_f}--\protect\ref{eq_fpow_dw}), 
the dotted lines show the
perturbative part only.
Right: Results of the Dokshitzer-Webber
fits in the $\alpha_s$--$\alpha_0$ plane. The band
indicates the world average of $\alpha_s$.}
\end{figure}

A combined fit of \as\ and $\alpha_0$ to a large set of
measurements\footnote{We concentrated on up--to--date results.}
at different energies~\cite{collection_eventshapes}
has been performed. In the $\chi^2$ calculation, statistical and systematic
uncertainties were considered.
For $\ecm\ge M_{\mathrm{Z}}$, only \delphi\ measurements
were included in the fit.
Figure~\ref{fig_mess_fit} (left) shows the measured mean values of
$\left<1-T\right>$, $\left< M_{\mathrm{hE}}^2/E_{\mathrm{vis}}^2\right>$
(standard-- and E definition),
$\left< C \right>$,
$\left< B_{\mathrm{max}} \right>$ and
$\left< B_{\mathrm{sum}} \right>$ as a function of
the centre-of-mass energy together with the results of the fit.
The fit values of \as\ and $\alpha_0$ are summarised in
Table~\ref{tab_mess_fit} and
displayed in the \as--$\alpha_0$ plane in Figure~\ref{fig_mess_fit} (right).
The systematic uncertainty was obtained as described
in Section~\ref{systematics}. 
In addition $\mu_{\mathrm{I}}$ was varied from $1 \gev$ to $3\gev$.
Both uncertainties were added in quadrature.

The $\alpha_s$ values obtained from these fits are consistent with each other
and in good agreement with the world average
$\alpha_s(M_{\mathrm Z})=0.118\pm0.002$ \cite{PDG2000}.
The extracted $\alpha_0$ values are around 0.5 as expected
in~\cite{hep-ph/9802381,hep-ph/9510283}.
However, the predicted universality (e.g. the observable independence) is 
satisfied on a 25\% level only.
This problem remains even though only the  Jet Mass in the E--definition,
$\left< M_{\mathrm{hE}}^2/E_{\mathrm{vis}}^2\right>$,
is considered which avoids a strong additional energy 
dependence due to the influence of hadron masses.
The \as\ values are higher and the \asb\ values are lower than the
corresponding results from event shape distributions (compare Figures
\ref{shapecorrelation} and \ref{fig_mess_fit}).

\subsection{Simple power corrections \label{all_pow}}
\begin{table}[tbp]
\begin{center}
\renewcommand{\arraystretch}{1.2}
\begin{tabular}{|c|r@{.}l@{$\pm$}r@{.}l@{$\pm$}r@{.}l|r@{.}l@{$\pm$}r@{.}l@{$\pm$}r@{.}l|r@{.}l@{/}l|}\hline
Observable &\multicolumn{6}{c|}{$C_1$}  &\multicolumn{6}{c|}{$\alpha_s(M_Z)$} &\multicolumn{3}{c|}{$\chi^2/ndf$} \\ \hline
$\langle 1-T\rangle$  & 0&483 & 0&085 &0&048& 0&1312 & 0&0019 &0&0032& 26&3 &  41 \\
$\langle \mbox{Major}\rangle$  & 0&405 & 0&979 &0&595& 0&1166 & 0&0071 &0&0043& 9&8 &  15 \\
\hline
$\langle \mbox{C-Parameter}\rangle$  & 2&242 & 0&469 &0&249& 0&1278 & 0&0024 &0&0032& 11&0 &  23 \\
\hline
$\langle M_{\mathrm{h}}^2/E_{\mathrm{vis}}^2\rangle$  & 0&502 & 0&119 &0&027& 0&1210 & 0&0033 &0&0018& 15&1 &  27 \\
$\langle M_{\mathrm{s}}^2/E_{\mathrm{vis}}^2\rangle$  & 0&736 & 0&689 &0&408& 0&1375 & 0&0132 &0&0081& 7&1 &  15 \\
$\langle M_{\mathrm{hE}}^2/E_{\mathrm{vis}}^2\rangle$  & 0&130 & 0&434 &0&246& 0&1215 & 0&0113 &0&0061& 9&3 &  15 \\
$\langle M_{\mathrm{sE}}^2/E_{\mathrm{vis}}^2\rangle$  & 0&341 & 0&617 &0&373& 0&1342 & 0&0119 &0&0075& 8&1 &  15 \\
\hline
$\langle B_{\mathrm{max}}\rangle$  & 0&241 & 0&075 &0&018& 0&1203 & 0&0016 &0&0009& 9&2 &  23 \\
$\langle B_{\mathrm{sum}}\rangle$  & 0&593 & 0&159 &0&050& 0&1236 & 0&0018 &0&0022& 7&8 &  23 \\
\hline
$\langle \mbox{EEC}_{70^\circ-110^\circ}\rangle$  & 0&285 & 0&637 &0&541& 0&1307 & 0&0102 &0&0088& 18&8 &  15 \\
$\langle \mbox{EEC}_{30^\circ-150^\circ}\rangle$  & 0&022 & 0&470 &0&691& 0&1395 & 0&0051 &0&0082& 41&2 &  15 \\
\hline
$\langle \mbox{JCEF}_{110^\circ-160^\circ}\rangle$  & 0&011 & 0&676 &0&954& 0&1201 & 0&0049 &0&0070& 31&4 &  15 \\
\hline
weighted mean &\multicolumn{2}{c}{} &\multicolumn{2}{c}{} &\multicolumn{2}{c|}{}& 0&1250 & 0&0054 &0&0024& \multicolumn{2}{c}{} &  \\
unweighted mean & \multicolumn{2}{c}{} &\multicolumn{2}{c}{} &\multicolumn{2}{c|}{}& 0&1250 & 0&0058 &0&0032& \multicolumn{2}{c}{} &  \\
\hline
\end{tabular}
\end{center}
\caption{\label{c1_result}
Determination of $C_1$ and \as\ from a fit to a large set of 
measurements of different
experiments~\cite{collection_eventshapes}.
For $\ecm\ge M_{\mathrm{Z}}$
only \delphi\ measurements are included in the fit.
The first quoted error is the 
uncertainty from the fit, the second one is the systematic uncertainty.
In calculating the mean values, the Jet masses using standard definitions, EEC and JCEF have been omitted. For the 
definition of the mean values see section \ref{average}.}
\end{table}
\begin{table}[tbp]
\begin{center}
\renewcommand{\arraystretch}{1.2}
\begin{tabular}{|c|r@{.}l@{$\pm$}r@{.}l@{$\pm$}r@{.}l|r@{.}l@{/}l|}\hline
Observable &\multicolumn{6}{c|}{$C_1$}  \\ \hline
$\langle 1-T\rangle$  & 1&093 & 0&016 &0&289 \\
$\langle \mbox{Major}\rangle$  & 0&390 & 0&020 &0&170\\
\hline
$\langle \mbox{C-Parameter}\rangle$  & 4&330 & 0&063 &0&713 \\
\hline
$\langle M_{\mathrm{h}}^2/E_{\mathrm{vis}}^2\rangle$  & 0&649 & 0&023 &0&076 \\
$\langle M_{\mathrm{s}}^2/E_{\mathrm{vis}}^2\rangle$  & 1&839 & 0&051 &0&136\\
$\langle M_{\mathrm{hE}}^2/E_{\mathrm{vis}}^2\rangle$  & 0&327 & 0&018 &0&072 \\
$\langle M_{\mathrm{sE}}^2/E_{\mathrm{vis}}^2\rangle$  & 1&265 & 0&033 &0&136\\
\hline
$\langle B_{\mathrm{max}}\rangle$  & 0&406 & 0&008 &0&046 \\
$\langle B_{\mathrm{sum}}\rangle$  & 1&199 & 0&012 &0&190 \\
\hline
$\langle \mbox{EEC}_{70^\circ-110^\circ}\rangle$  & 1&177 & 0&021 &0&163 \\
$\langle \mbox{EEC}_{30^\circ-150^\circ}\rangle$  & 2&066 & 0&015 &0&227\\
\hline
$\langle \mbox{JCEF}_{110^\circ-160^\circ}\rangle$  & 0&484 & 0&026 &0&223\\
\hline
\end{tabular}
\end{center}
\caption{\label{c1_fixed_result}
Determination of $C_1$ with a fixed $\Lambda_{\overline{MS}}=0.250$~\gev\
from a fit to a set of measurements of different
experiments~\cite{collection_eventshapes}.
For $\ecm\ge M_{\mathrm{Z}}$
only \delphi\ measurements are included in the fit.
The first uncertainty is the 
uncertainty from the fit, the second one is the systematic uncertainty. 
For the  definition of the mean values see section \ref{average}.}
\end{table}

Power corrections in the Dokshitzer--Webber framework can only be calculated 
for the  set of exponentiating observables. The experimental evidence for 
corrections which show $1/Q$ behaviour is however not restricted 
to this type of observables.
The tube model indicates the existence of power corrections on simple
phase space assumptions.

In order to determine approximate power corrections for all observables
measured the ``simple power correction'' ansatz is used.
Here an additional power term $\left<f_{pow}\right>=C_1/Q$ is
added to the ${\cal O}(\alpha_s^2)$ perturbative expansion of the observable,
with $C_1$ being an observable dependent, unknown constant.
The disadvantage of this simple ansatz is that a
double counting in the infrared region of the observables is not corrected 
for as in the Dokshitzer--Webber approach.

Two different types of fits were performed to the DELPHI data
using this simple model.
Firstly, in order to investigate whether this simple model yields 
sensible values for \as\ at all, both parameters,  
$\Lambda_{\overline{MS}}$ and $C_1$, were left
free in the fit.
The results obtained 
are given in Table~\ref{c1_result}.
The average value of these \as\ results and the corresponding
R.M.S. obtained only from the fully inclusive observables are
\as$=0.1250\pm 0.0058$ for the unweighted mean and \as$=0.1250\pm 0.0054$
for the weighted mean.
The reasonable values for \as\ as well as the acceptable
$\chi^2/ndf$ of $~6/7$ of the averaging support the approximate
validity of this simple power correction model.
Secondly, in order to get comparable estimates of the size of the
power correction for the different observables,
$\Lambda_{\overline{MS}}=0.250$\gev\ was chosen,
leaving only $C_1$ as a free parameter.
The fitted values of $C_1$ are contained in Table~\ref{c1_fixed_result}, where the
total experimental uncertainty for $C_1$ is given.

The ratio of the power model parameter normalised to the first order
perturbative coefficient, $C_1/A$, is plotted against the ratio of
second to first order perturbative coefficient $B/A$
in Figure~\ref{abparameter2}.
The normalisation to $A$ is made in order to make the observables
directly comparable.
Experimentally a clear correlation
between the genuine non--perturbative parameter $C_1$ and the
purely perturbative parameter $B$ is observed.
This strong correlation indicates that the term $C_1/Q$ should not be interpreted as
purely non--perturbative.

\section{Interpretation of event shape means using
RGI perturbation theory \label{data_DG}}
Today the $\overline{MS}$ scheme  is commonly used for the
representation of perturbative calculations of physical observables.
In consequence predictions for power corrections have also been given
in this scheme.
Nonetheless the $\overline{MS}$ scheme is only one of an infinite set
of equally well suited schemes.

A previous \delphi\ analysis using experimentally optimised scales for the
determination of \as\ from event shape distributions  \cite{siggi} has shown
that the experimentally optimised scales of different observables
are correlated with the corresponding effective charge, ECH \cite{grunberg},
or principle of minimal sensitivity, PMS \cite{stephenson}, scales.
The renormalisation group invariant (RGI) approach \cite{DG,Korner:2000xk}
uses the same central equations as the method of 
effective charge \cite{grunberg},
however the motivation and philosophy differ.
The derivation of the RGI method makes no reference to any
renormalisation scheme whatsoever.
In the following sections the RGI predictions \cite{DG,beneke,maxwell}
for fully inclusive shape observable means are explored.

\subsection{Theoretical background of RGI}
Instead of expanding an observable $R$ into a perturbative series in
\as$(Q)$, the starting point of the RGI method is
a $\beta$-function like expansion of $\frac{dR}{dQ}$ in $R$:
\begin{equation}
Q\frac{dR}{dQ}=-bR^2(1+\rho_1R+\rho_2R^2+\ldots)=b\rho(R)~~.
\label{RGE}
\end{equation}
Note that here $R=2\langle f \rangle/A$ is normalised 
such that the perturbative expansion in $\alpha_s/\pi$ begins with
unit coefficient.
It can be shown \cite{DG,grunberg}, that the coefficients
$\rho_i$ are scheme invariant and that the scale dependence cancels
out completely. 
The  $\rho_i$ can be calculated
from the coefficients $r_1,r_2$ of the perturbative expansion:
\begin{xalignat}{3}
\rho_1&=c_1~~,  & \rho_2=c_2+r_2-r_1c_1-r_1^2~~,
\end{xalignat}
\begin{xalignat}{9}
b&=\frac{\beta_0}{2}~~, &
c_1&=\frac{\beta_1}{2\beta_0}~~, &
c_2&=\frac{\beta_2}{32\beta_0}~~, &
r_1&=\frac{B}{2A}~~, &
r_2&=\frac{C}{4A}~~.
\nonumber
\end{xalignat}
The coefficients $A$ and $B$ are defined in Equation \ref{eq_fpert_o2},
$C$ is the corresponding third order coefficient.

The solution of Equation~\ref{RGE} is
equivalent to the well known implicit equation for \as:
\begin{equation}
b\ln\frac{Q}{\Lambda_R}=
\frac{1}{R}-\rho_1\ln\left(1+\frac{1}{\rho_1R}\right)+\int_0^R dx
\left(\frac{1}{\rho(x)}+\frac{1}{x^2(1+\rho_1x)}\right)~~~.
\label{imprunning}
\end{equation}
Here $\Lambda_R$ is an R-specific scale parameter.
In next-to-leading (NLO) order \as\ the integrand vanishes and the solution
of this 
equation is identical to a scale which sets the NLO contribution of the perturbative series
to zero.
Using the so called Celmaster Gonzalves equation~\cite{celmaster},
$\Lambda_R$ can be converted into $\Lambda_{\overline{MS}}$ without any loss
of precision:
\begin{equation}
\frac{\Lambda_R}{\Lambda_{\overline{MS}}}=
e^{r_1/b}\left(\frac{2c_1}{b}\right)^{-c_1/b}~~~.
\label{celmaster}
\end{equation}

In a study of event shape observables a test of the validity of 
RGI perturbation theory
is currently limited to NLO. It should therefore be verified that higher order
corrections are small 
and that the NLO $\beta$-function (e.g. the inclusion up to the 
$\rho_1$ term in Equation \ref{RGE}) is a sufficient approximation.
A check on higher order contributions is implied by a consistency check
of the \as\ values measured from different observables.

It is important to note that the above derivation only holds for
observables that depend  on one single energy scale, such as 
fully inclusive observables.
Selections or cuts in the observable introduce additional
scales that have to be included into equation \ref{RGE}.
Thus it is not to be expected that the aforementioned simple form of RGI is 
valid, say,
for $y_{cut}$ dependent jet rates or bins of event shape distributions.
The application of the RGI method to ranges of the EEC or JCEF as performed
in the following sections is thus not fully justified, except
by the success of the comparison to data.
The RGI method may also to some extent apply  here as the intervals chosen
are rather wide and represent an important fraction of the events.
Moreover it should be noted that the total integrals over the EEC or JCEF are
normalised to 2 or 1, respectively.
RGI can also not be unambiguously calculated for observables like
$M_d^2=M_h^2-M_l^2$, as $M_l^2=\mathrm{min}(M_+^2,M_-^2)$ is only known to 
leading order.

It is possible to include power corrections into RGI. In
\cite{beneke} it is shown that the existence of non--perturbative corrections
leads to a predictable asymptotic behaviour of the renormalisation
group equation. This can be included into the equation as a
modification of the $\rho$ function~\cite{Campbell:1998qw}:
\begin{equation}
\rho(x) \to \rho(x)-\frac{K_0}{b}x^{-c_1/b}e^{-1/bx}~~~,
\label{beneq}
\end{equation}
where $K_0$ is a free, unknown parameter that determines the size of
the non--perturbative correction.
The correction is approximately equal to a simple $C_1/Q$ power correction
in the $\overline{MS}$ scheme with~\cite{Campbell:1998qw}:
\begin{equation}
C_1=-{K_0}\cdot e^{r_1/b}
\cdot \left( \frac{b}{2}\right)^{c_1/b} \cdot \Lambda_{\overline{MS}}~~~.
\label{c1-k0}
\end{equation}

As RGI pertubation theory and the ECH method are based on the same basic 
equation, the choice of the ECH renormalisation scheme is implicit in RGI 
perturbation theory. Therefore it may be controversial whether measurements of 
$\alpha_s$ performed using RGI are renormalisation scale or scheme independent.
However, with respect to the $\beta$ function (e.g. its leading coefficients, 
$\beta_0$ and $\beta_1$) the situation is different. Their 
measurement based on Equation 
\ref{RGE} is free of any scheme ambiguity since  this relation holds in 
{\em any} 
renormalisation scheme. Moreover $\beta_0$ and $\beta_1$  are  renormalisation 
scheme invariant quantities. Dhar and Gupta summarize their discussion with 
the  following words:``(...) we have shown that in a renormalizable 
massless field theory with a single dimensionless coupling constant, 
{\em only} the derivative $\rho (R)$ of a physical quantity R with respect to 
an external scale is well defined and unambiguously calculable''\cite{DG}. 
         
\subsection{Comparing RGI with power corrections to data}
\begin{figure}[bt]
\mbox{\epsfig{file=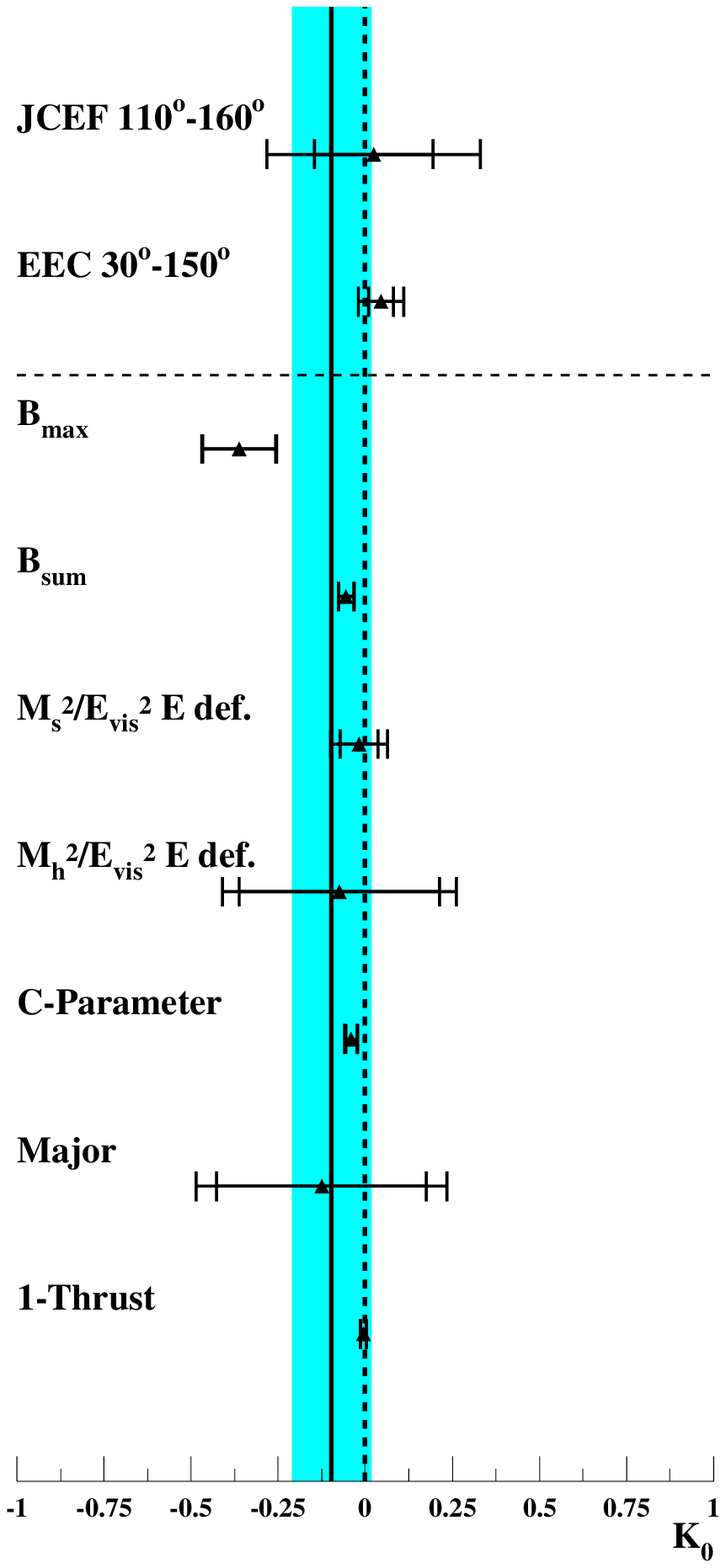,width=7cm}}
\mbox{\epsfig{file=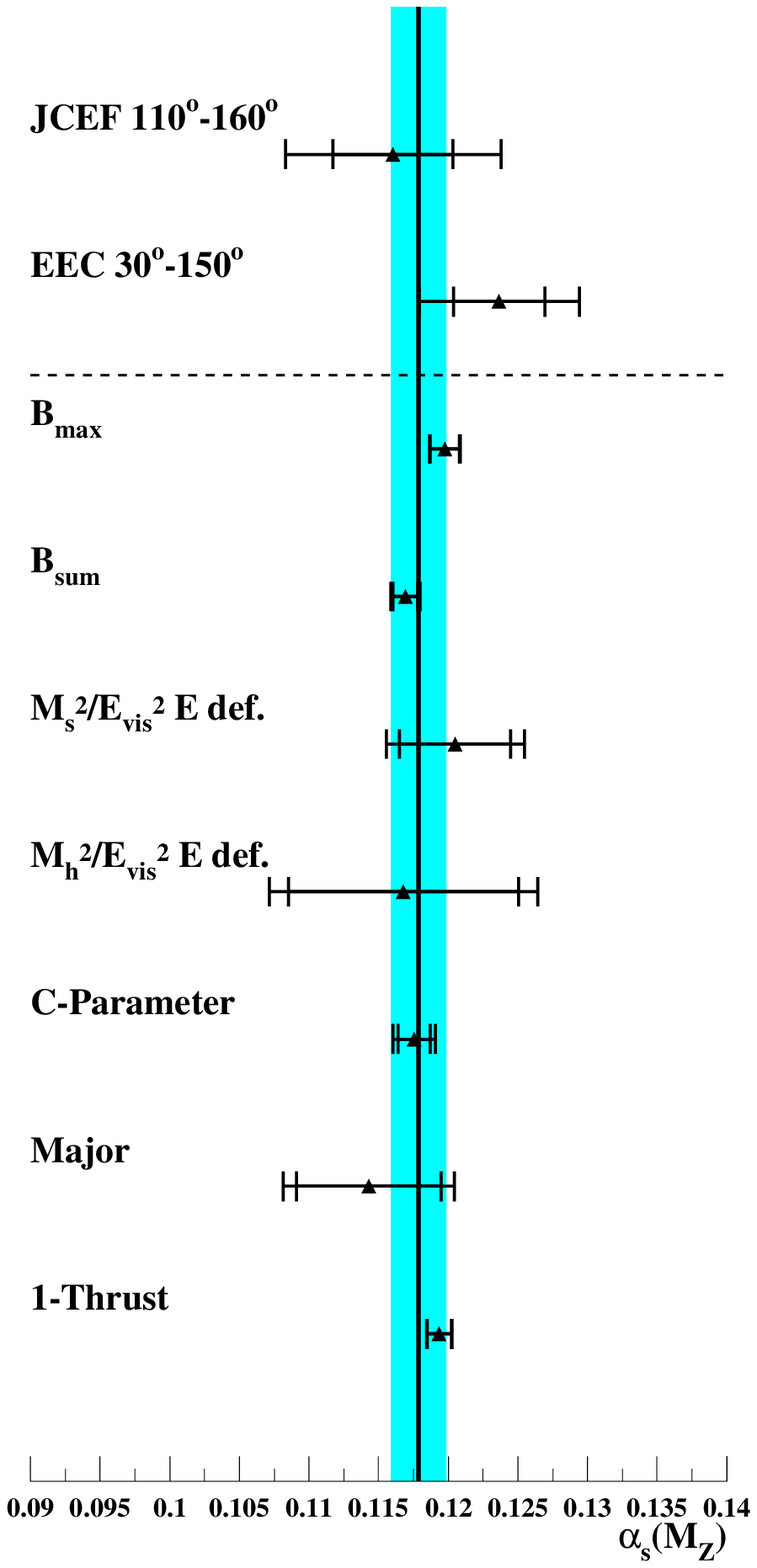,width=7cm}}
\caption{
Results for a combined fit of $\Lambda_R$ and
the power correction parameter $K_0$ using Equation~\ref{beneq}.
Left: The results for the non--perturbative parameter $K_0$. Right: The results
for $\alpha_{\mathrm{s}}(M_{\mathrm{Z}})$ deduced from $\Lambda_{\mathrm{R}}$.
The straight line shows the unweighted mean, the shaded band the
R.M.S. obtained from the fully inclusive observables only.}
\label{bgraf}
\end{figure}

\begin{table}[tb]
\begin{center}
\begin{tabular}{|c|r@{.}l@{$\pm$}r@{.}l@{$\pm$}r@{.}l|r@{.}l@{$\pm$}r@{.}l@{$\pm$}r@{.}l|r@{.}l@{/}l|}\hline 
Observable &\multicolumn{6}{c|}{$K_0$} &\multicolumn{6}{c|}{$\alpha_s(M_{\mathrm{Z}})$} &\multicolumn{3}{c|}{$\chi^2/ndf$} \\ \hline 
$\langle 1-T\rangle$  & $-$0&005 & 0&008 &0&003& 0&1194 & 0&0009 &0&0003& 31&0 &  39 \\ 
$\langle \mbox{Major}\rangle$  & $-$0&126 & 0&302 &0&197& 0&1143 & 0&0052 &0&0033& 9&8 &  13 \\ 
\hline 
$\langle \mbox{C-Parameter}\rangle$  & $-$0&040 & 0&016 &0&009& 0&1175 & 0&0012 &0&0010& 11&0 &  21 \\ 
\hline 
$\langle M_h^2/E_{vis}^2\rangle$  & $-$0&112 & 0&028 &0&006& 0&1219 & 0&0014 &0&0005& 19&6 &  25 \\ 
$\langle M_s^2/E_{vis}^2\rangle$  & $-$0&034 & 0&072 &0&038& 0&1249 & 0&0065 &0&0035& 7&1 &  13 \\ 
$\langle M_{hE}^2/E_{vis}^2\rangle$  & $-$0&074 & 0&288 &0&172& 0&1168 & 0&0083 &0&0050& 9&3 &  13 \\ 
$\langle M_{sE}^2/E_{vis}^2\rangle$  & $-$0&018 & 0&054 &0&061& 0&1205 & 0&0040 &0&0029& 11&0 &  13 \\ 
\hline 
$\langle B_{max}\rangle$  & $-$0&362 & 0&105 &0&021& 0&1198 & 0&0010 &0&0005& 8&8 &  21 \\ 
$\langle B_{sum}\rangle$  & $-$0&055 & 0&022 &0&004& 0&1169 & 0&0009 &0&0005& 7&7 &  21 \\ 
\hline 
$\langle \mbox{EEC}_{70^\circ-110^\circ}\rangle$  & 0&008 & 0&095 &0&075& 0&1174 & 0&0065 &0&0051& 18&8 &  13 \\ 
$\langle \mbox{EEC}_{30^\circ-150^\circ}\rangle$  & 0&045 & 0&036 &0&054& 0&1236 & 0&0033 &0&0047& 40&8 &  13 \\ 
\hline 
$\langle \mbox{JCEF}_{110^\circ-160^\circ}\rangle$  & 0&024 & 0&171 &0&254& 0&1160 & 0&0043 &0&0064& 31&4 &  13 \\ 
\hline 
weighted mean & $-$0&018 & 0&114 &0&068& 0&1184 & 0&0031 &0&0035& \multicolumn{3}{c|}{~} \\ 
unweighted mean & $-$0&097 & 0&114 &0&057& 0&1179 & 0&0020 &0&0013& \multicolumn{3}{c|}{~}  \\ 
\hline 
\end{tabular} 
\end{center}
\caption{
Results for a  RGI plus non--perturbative parameter  
fit to a large set of measurements of different
experiments~\cite{collection_eventshapes}.
For $\ecm\ge M_{\mathrm{Z}}$
only \delphi\ measurements are included in the fit.
The first uncertainty is the statistical
uncertainty from the fit, the second one is the systematic uncertainty.
For the mean values only the E--definition Jet Masses have been used
and both EEC and JCEF have been omitted. For the 
definition of the mean values see section \ref{average}.\label{tab8}
} \end{table}

Using the same data as in the case of simple power corrections,
a combined fit of $\Lambda_{\mathrm{R}}$ and $K_0$ is performed
to the RGI with power correction theory using
Equation~\ref{beneq}.
A correction is applied for the influence of the b--mass but
not for further hadronisation effects.
RGI plus power correction describes the behaviour of the data well
for all observables considered including EEC and JCEF and
leads to a consistent result for \as.
The fit results are given in Table \ref{tab8} and  Figure~\ref{bgraf}.
The uncertainties for the jet masses in E--definition, Major
as well as for JCEF and EEC are large as no data from low energy
experiments are available in these cases.
The result for $\langle {\mathrm{1-T}}  \rangle$ agrees reasonably with the
comparable analysis presented in \cite{Campbell:1998qw}.
Note that in \cite{Campbell:1998qw} no b--mass correction was applied.
Small differences are understood as being due to the b--mass correction and 
the  differing choice of input data.
Combining the \as\ results for the fully inclusive observables
yields \as$=0.1179\pm 0.0020 \pm 0.0013$ for the unweighted mean and
\as$=0.1184\pm 0.0031 \pm 0.0035$
for the weighted mean with a $\chi^2/ndf$ of $7.0/5$.
The spread of the \as\ results from the fit of RGI with power corrections 
is only half of the size as for the simple power correction case.

The fact that the results for $K_0$ are
small (most are compatible with zero) is surprising.
The smallness of the $K_0$ results in combination with the
consistent \as\ values from all 9 different observables casts doubt
on the interpretation of the measured $\left <f_{pow}\right>$
contribution
(see Equation~\ref{eq_f}) as a genuine non--perturbative term.
In view of the observed consistency
one is led to presume a better approximation of the data by RGI 
perturbation theory
and the $\overline{MS}$ scheme appears as an unfortunate choice.
This conjecture implies that the power terms measured for shape observable 
means  in the previous section  mainly parameterise terms
which can in principle be calculated perturbatively.

To elucidate this conjecture further, Thrust is taken as example.
Here for a fixed \as=0.118,~~$C_1/Q=(1.108\pm 0.017){\rm \gev}/Q$ is found as
the simple power correction in the $\overline{MS}$ scheme.
Using Equation~\ref{c1-k0} this translates into $K_0=-0.32\pm 0.006$.
In the RGI plus power term approach $K_0=-0.005\pm 0.008$ is observed, however.
Contrary to the simple power model or the ansatz of Equation~\ref{eq_fpow_dw},
here the power contribution is insignificant.
From the influence of hadron masses, a small kinematical dependence 
similar to an inverse power law is to be expected.
However, the absolute size of this mass correction cannot yet be safely 
calculated~\cite{salamwicke}.


\subsection{Comparing pure RGI to data\label{rgi_compare}}
In light of the negligible power term ($\propto K_0$) observed in the
previous section, a comparison of the data and the pure RGI perturbation
theory appears interesting. No
hadronisation correction is included and $\Lambda_{\mathrm{R}}$ is the 
only free parameter of the theory.
This implies that $\Lambda_R$ can be precisely determined from each individual
shape observable mean measurement allowing for stringent tests of RGI by comparing the
energy dependence of individual observables or by comparing the
$\Lambda_{\overline{MS}}$ or $\alpha_s({M_Z})$ values obtained from different
observables.
Note that these parameters can be deduced from  $\Lambda_{\mathrm{R}}$ without loss of
precision (see Equation \ref{celmaster}).

Observed deviations between different energy points of an observable should
mainly be due to hadronisation or mass corrections. From the above section
these corrections are already known to be small ($\simeq \rm few~\%$).
Differences in \as\ values between different observables represent
different power corrections as well as missing higher order terms of RGI
perturbation theory.
Clearly the \as\ value for each observable will be dominated by the precise
measurements at the Z pole.



\begin{figure}[tb]
\unitlength1cm
 \unitlength1cm
 \begin{center}
 \begin{minipage}[t]{7.5cm}
    \mbox{\epsfig{file=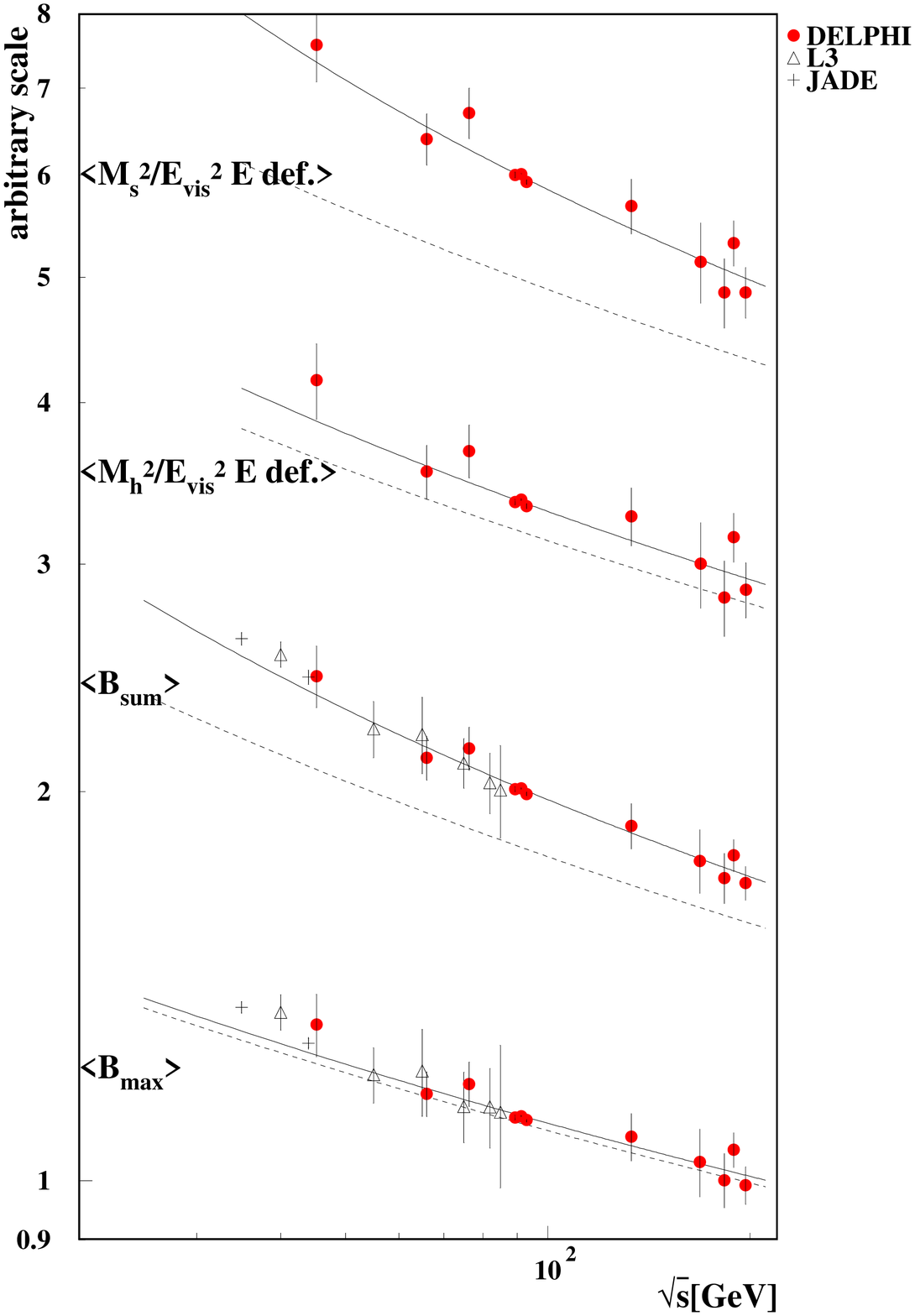,width=8.1cm}}
 \end{minipage}
 \begin{minipage}[t]{7.5cm}
    \mbox{\epsfig{file=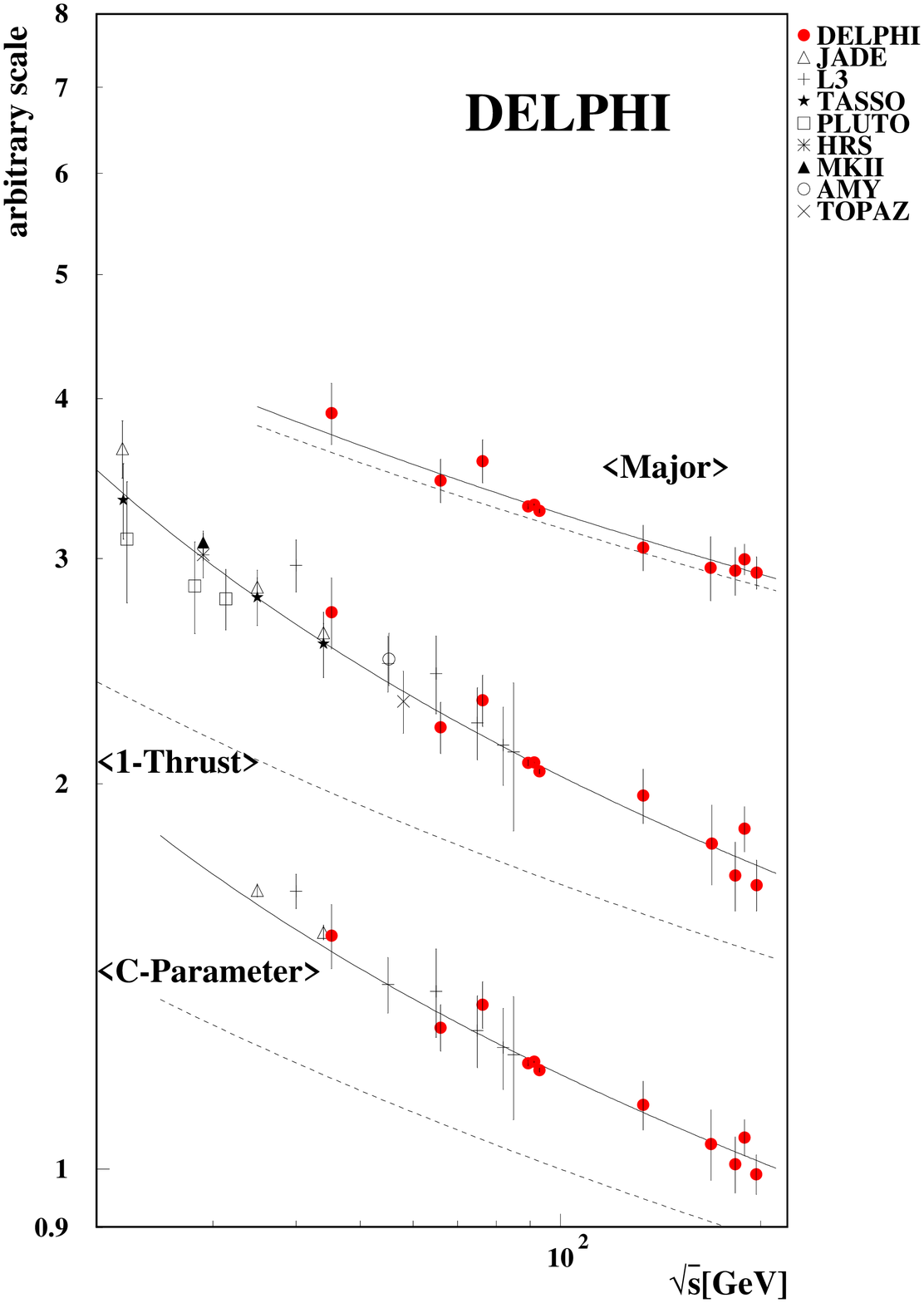,width=8.1cm}}
 \end{minipage}
\end{center}
\caption{Comparison of the data on event shape means with the
prediction of pure RGI perturbation theory (full line). For clarity some of the
high energy data have been merged. The dashed line represents the 
$\overline{MS}$ expectation with the same $\alpha_s(M_Z)$.}
\label{purRGIplot}
\end{figure}

Figure~\ref{purRGIplot} compares the energy dependence of the fully inclusive
event shape means considered in this analysis to the RGI prediction.
The data shown are the same as explained in the simple power correction
section.
The energy dependence of all observables is well represented by RGI theory.
The $\chi^2/ndf$ of the fits is acceptable for all observables (see Table~\ref{simpleRGIresults})
 and similar to the RGI plus power correction case.
The observed values of \as\ of the different observables agree reasonably well,
with the exception of the standard definition of $M_{\mathrm{h}}^2/E_{\mathrm{vis}}^2$ or
$M_{\mathrm{s}}^2/E_{\mathrm{vis}}^2$ (not shown here).
For these observables mass terms are known to be important and
reflect into higher \as\ values compared to the results using the E-scheme.

\begin{figure}[tb]
\mbox{\epsfig{file=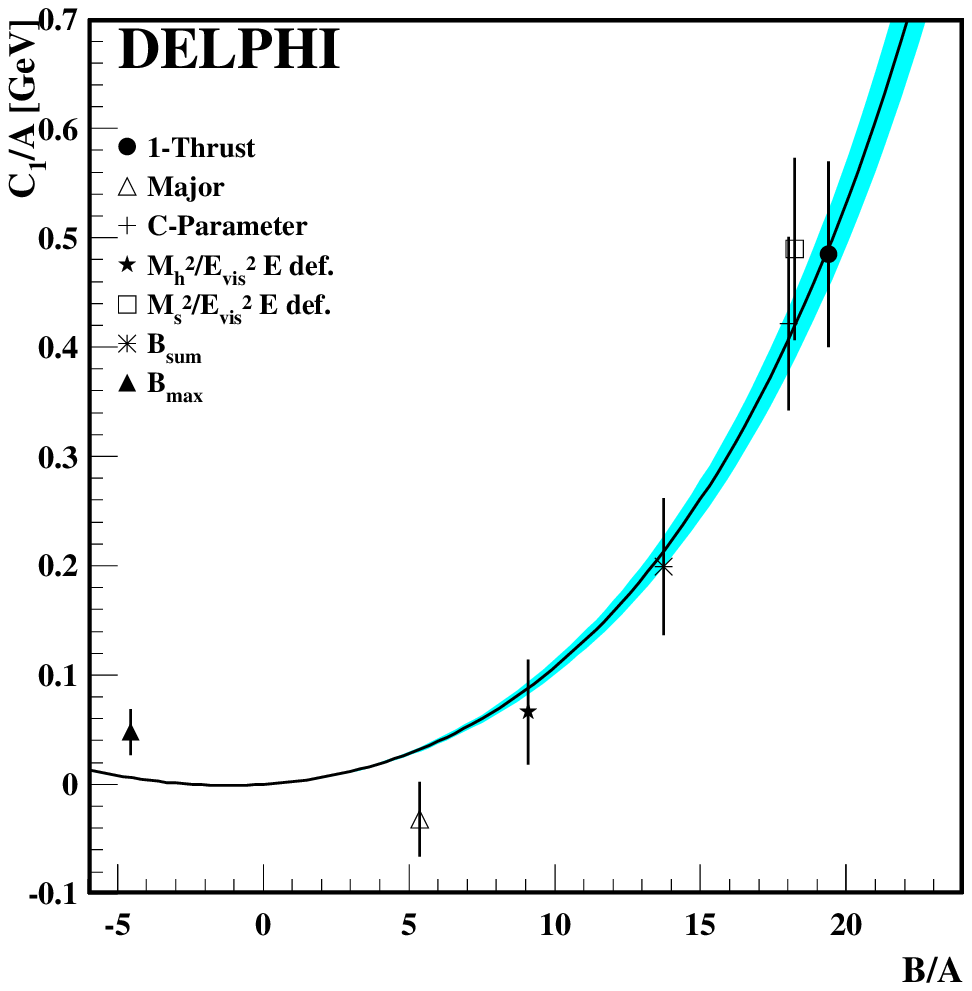,width=7.5cm}}
\mbox{\epsfig{file=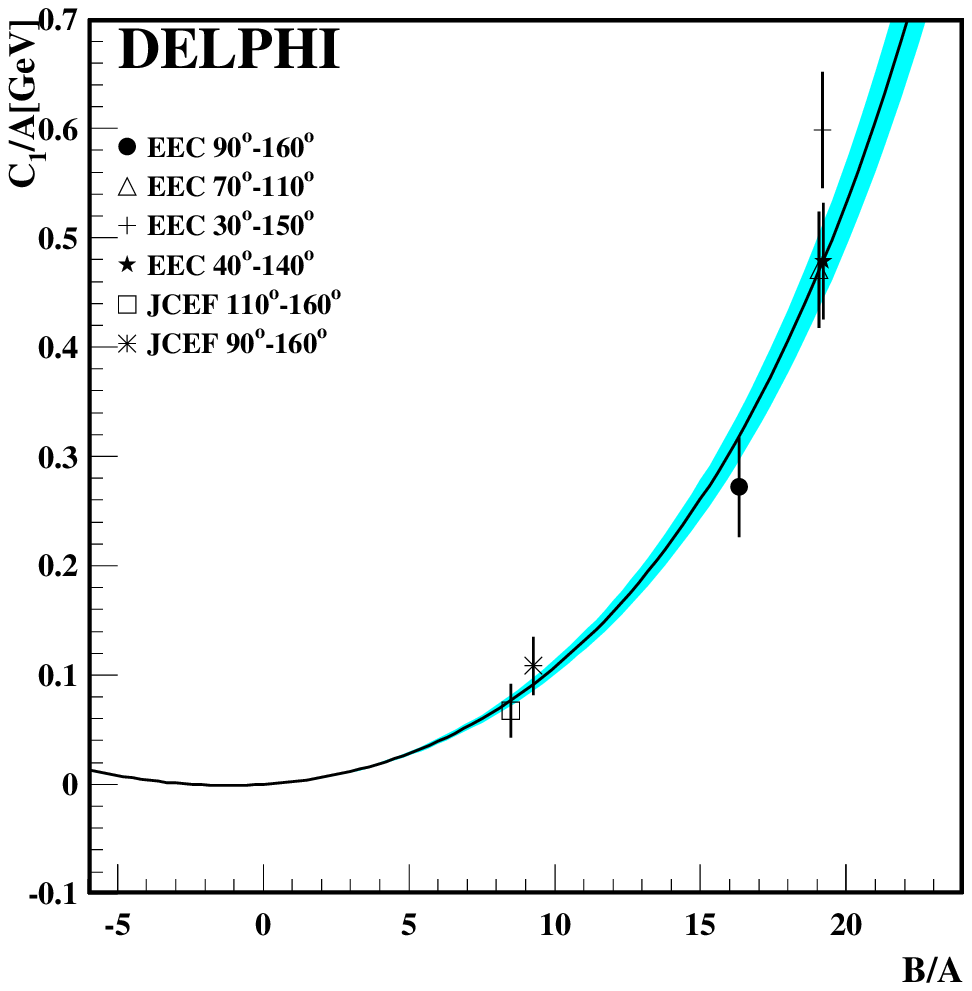,width=7.5cm}}
\caption{Comparison of the results for $C_1/A$ obtained from a
fit with fixed $\Lambda_{\overline{MS}}=0.250\gev$ and the prediction
from pure RGI.
The shaded band shows the variation of $\Lambda_{\overline{MS}}$
between 0.220\gev and 0.280\gev.
Left: fully inclusive observables. Right: different integration
intervals of EEC and JCEF. Note that the data points are correlated.}
\label{abparameter2}
\end{figure}

\begin{figure}[p]
{
\unitlength1cm
\begin{picture}(15,20)
\put(15,0){
    \begin{rotate}{90}
    \begin{minipage}[t]{20cm}
    \mbox{\epsfig{file=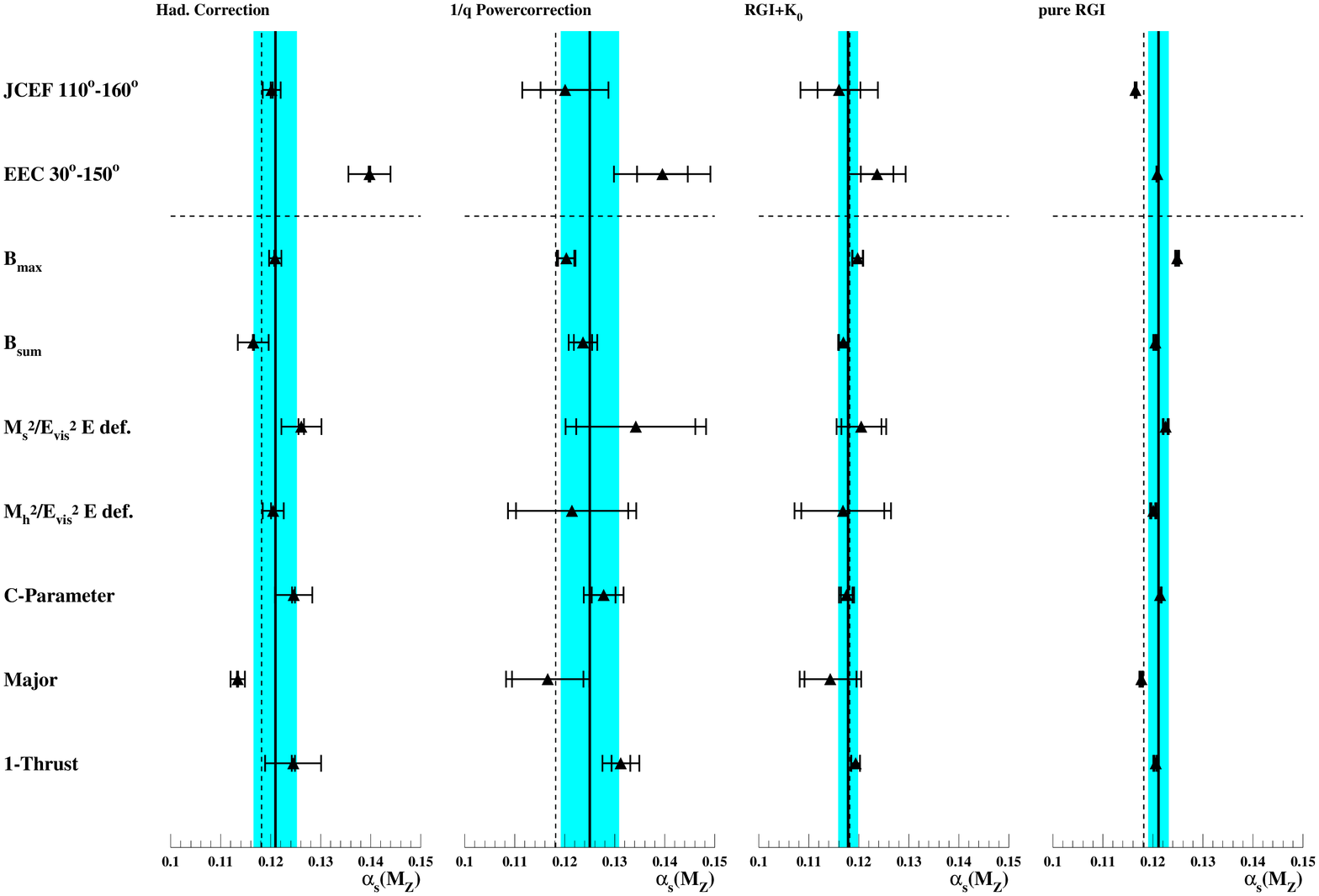,width=20cm}}
    \end{minipage}
    \end{rotate}
}
\end{picture}
}
\caption{Results for \as\ for different observables and
methods. The straight line shows the unweighted mean, the shaded
bar the variance for the fully inclusive observables.
As a reference the world average value of 
$\alpha_s(M_{\mathrm Z})=0.1181\pm0.002$ \cite{PDG2000} is shown as
dashed line. \label{barmeangraphs}}
\end{figure}

\begin{table}[tb]
\begin{center}
\begin{tabular}{|c|r@{.}l@{$\pm$}r@{.}l@{$\pm$}r@{.}l|r@{.}l@{/}l|}\hline
Observable &\multicolumn{6}{c|}{$\alpha_s(M_Z)$} &\multicolumn{3}{c|}{$\chi^2/ndf$} \\ \hline
$\langle 1-T\rangle$  & 0&1199 & 0&0002 &0&0002& 31&9 &  40 \\
$\langle \mbox{Major}\rangle$  & 0&1166 & 0&0001 &0&0003& 10&1 &  14 \\
\hline
$\langle \mbox{C-Parameter}\rangle$  & 0&1205 & 0&0002 &0&0002& 24&6 &  22 \\
\hline
$\langle M_{\mathrm{h}}^2/E_{\mathrm{vis}}^2\rangle$  & 0&1274 & 0&0005 &0&0017& 46&9 &  26 \\
$\langle M_{\mathrm{s}}^2/E_{\mathrm{vis}}^2\rangle$  & 0&1282 & 0&0006 &0&0005& 7&6 &  14 \\
$\langle M_{\mathrm{hE}}^2/E_{\mathrm{vis}}^2\rangle$  & 0&1189 & 0&0004 &0&0005& 9&4 &  14 \\
$\langle M_{\mathrm{sE}}^2/E_{\mathrm{vis}}^2\rangle$  & 0&1216 & 0&0004 &0&0004& 8&1 &  14 \\
\hline
$\langle B_{\mathrm{max}}\rangle$  & 0&1236 & 0&0002 &0&0004& 32&6 &  22 \\
$\langle B_{\mathrm{sum}}\rangle$  & 0&1194 & 0&0001 &0&0004& 21&5 &  22 \\
\hline
$\langle \mbox{EEC}_{70^\circ-110^\circ}\rangle$  & 0&1169 & 0&0002 &0&0001& 18&8 &  14 \\
$\langle \mbox{EEC}_{30^\circ-150^\circ}\rangle$  & 0&1199 & 0&0001 &0&0001& 44&0 &  14 \\
\hline
$\langle \mbox{JCEF}_{110^\circ-160^\circ}\rangle$  & 0&1155 & 0&0002 &0&0001& 31&5 &  14 \\
\hline
weighted mean & 0&1195 & 0&0002 &0&0010& \multicolumn{2}{c}{} &  \\
unweighted mean & 0&1201 & 0&0020 &0&0003& \multicolumn{2}{c}{} &  \\
\hline
\end{tabular}
\caption{\label{simpleRGIresults}
Results for a pure RGI fit to a large set of measurements of different
experiments~\cite{collection_eventshapes}.
For $\ecm\ge M_{\mathrm{Z}}$
only \delphi\ measurements are included in the fit.
The first error is the statistical
uncertainty from the fit, the second one is the systematic uncertainty.
For the mean values only the E--definition Jet Masses have been used
and both EEC and JCEF have been omitted. For the 
definition of the mean values see section \ref{average}.
}
\end{center}
\end{table}
The RGI \as\ results for the different observables are also shown in
Figure~\ref{barmeangraphs} compared to results of  RGI with
power corrections and \as\ values obtained in $\overline{MS}$
with simple power corrections as well as with Monte Carlo hadronisation
corrections.
The last have been calculated as usual by comparing the hadron and parton
level result in a Monte Carlo fragmentation model (\pythia).
It is evident that RGI leads to a far better
consistency  between different observables than the simple power corrections
or the Monte Carlo hadronisation corrections.
In comparison to the RGI plus power correction ansatz
the spread of the \as\ values is not significantly increased.
The unweighted mean and R.M.S. spread of the \as\ values from the fully
inclusive
means shown in Figure~\ref{barmeangraphs} is \as$=0.1201\pm 0.0020$,
using the weighted mean yields \as$=0.1195\pm 0.0002$.
The average from pure RGI is still consistent with that using the RGI plus 
power
correction ansatz. The overall \as\ shift of about 2\% at Z energies indicates
a possible influence of power terms.
The spread of the unweighted mean \as\ values is only of the order of $0.002$.
It indicates possible differences in power terms and uncalculated higher order
contributions which are naively expected to be of the order 
$\alpha_s^3\sim 0.002$.
The strongly differing second order contributions (compare the $B/A$ values 
of the different observables in $\overline{MS}$ shown 
in Figure \ref{abparameter2}) 
as well as the differing
NLLA expansions suggest differing higher order terms for the 
individual observables.

As a further illustration of the validity of RGI one may try to
calculate the size of the power terms observed in the $\overline{MS}$ analysis.
The power term can be approximated as the difference of the RGI and the
$\overline{MS}$ terms:
\begin{equation}
\frac{C_1}{E_{cm}} = R_{\mathrm{RGI}}(E_{cm}) - 
R_{\mathrm{\overline{MS}}}(E_{cm})
\end{equation}
for a fixed $\Lambda_{\overline{MS}}=250\mev$.
This value corresponds to the average \as\ as obtained from the RGI analysis.
As this ansatz is slightly energy dependent, $M_Z$ has been chosen as
the reference energy.
Figure~\ref{abparameter2} shows the comparison between the measured simple
power correction terms and the calculated power term using pure RGI.
The strong correlation between $C_1/A$ and $B/A$ for fully inclusive means
(left side) is well represented by the above ansatz.
It also holds reasonably well for the JCEF and the EEC. Here the
ratio of the first and second order coefficient $B/A$ can be adjusted
by an appropriate choice of the $\chi$ integral.

The agreement of the measured values $C_1/A$ and the RGI curve
depends on the choice of $\Lambda_{\overline{MS}}$. A higher value
of $280\mev$ as indicated by the upper edge of the grey band shown in
Figure~\ref{abparameter2} requires the data points to move down about
by the same amount, leading to a difference between the data and the curve of
about the width of the grey band.

\begin{figure}[tb]
\begin{center}
\mbox{\epsfig{file=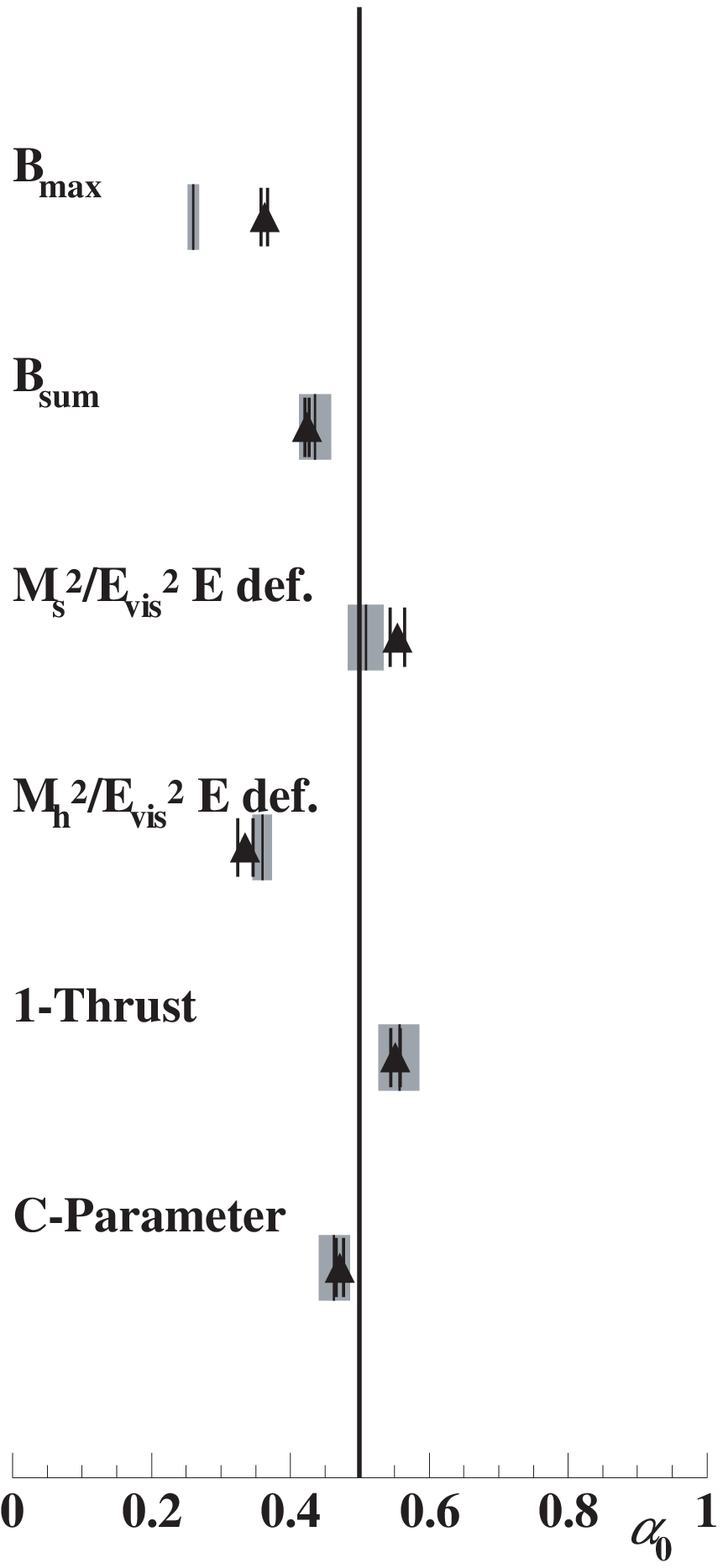,width=7.cm}}
\caption{
Comparison of the results for ${\alpha}_0$ as obtained from fits
with fixed $\Lambda_{\overline{MS}}=0.250$~\gev\ and 
the prediction from pure
RGI. The shaded bands show the variations due to a change of 
$\Lambda_{\overline{MS}}$ between $0.220$~\gev\ and $0.280$~\gev.}
\label{DG_alpha0_DW}
\end{center}
\end{figure}

Similarly to the simple power term $C_1$, the $\alpha_0$ values
of the Dokshitzer-Webber model can be calculated and obviously agree
reasonably with the fitted values, as shown in Figure \ref{DG_alpha0_DW}.
The measured ${\alpha}_0$ values, show a scatter around a
universal value anticipated by the Dokshitzer Webber model.
Even this scatter is described reasonably by the RGI prediction.

Finally a remark on the connection of RGI to optimised scales is in order
here.
Though reference \cite{DG} stresses the difference in philosophy of
RGI perturbation theory and
scale optimisation there is an obvious connection.

RGI perturbation theory describes the data on fully inclusive shape means
well without any renormalisation scale dependence whatsoever.
This implies that a fit of the renormalisation scale using the standard
${\cal O}(\alpha_s^2)$ expansion leads to a scale value where the scale
dependence of the ${\cal O}(\alpha_s^2)$ prediction vanishes. This is
the PMS scale which is close to the ECH scale.
The above contiguity has been observed numerically \cite{dr:reinhardt} and
found to be valid within errors.

\subsection{Measuring the QCD $\mathbf{\beta}$-function based on the RGI
analysis}

Although RGI perturbation theory is based on the RGE the $\beta$-function 
itself is not fixed by the RGI approach.
As the $\beta$-function for an observable $R$ and the QCD coupling is
identical
in second order this implies that the $\beta$-function, or rather the
coefficients $\beta_0$ and $\rho_1$,  can be directly
inferred from the energy evolution of the event shape observable means (see
Equation~\ref{imprunning}).
As $\rho_1$ induces only a small correction (${\cal O}(4\%)$)
and cannot be simultaneously
determined with $\beta_0$ given the current precision of the data, it has been
fixed to the QCD expectation.
Assuming QCD, the number of active flavours, $n_{\mathrm{f}}$, can be
determined as an alternative to determining $\beta_0$.
In these cases $\beta_0$ and $\rho_1$ were calculated from the QCD relations
using the fitted $n_{\mathrm{f}}$ and assuming 
$N_{\mathrm{C}}=C_{\mathrm{A}}=3$. 

The approach allows $\beta_0$ to be measured 
under the assumption that QCD is valid. In order to compare the result with 
predictions from e.g. super symmetric extensions of the standard model one 
of course cannot rely on the QCD calculation for $\beta_1$. Instead  one 
needs to fit the full energy dependence. This analysis is described in section
\ref{lightgluinos}.

The implicit Equation~\ref{imprunning} can be fitted to data on
shape observable means leaving $\Lambda_R$ and $\beta_0$ 
(or $n_{\mathrm{f}}$) as free parameters.
Note that this fit is independent of an $n_{\mathrm{f}}$ dependence
entering implicitly through the second order perturbative coefficient $B$.
This holds as the offset term $\beta_0\ln\Lambda_R$ is unrestricted
because $\Lambda_R$ is a fit parameter and implies that
 a real measurement of $\beta_0$ or $n_{\mathrm{f}}$ can be
performed.

The results of the fits to \delphi\ data are given in Table~\ref{runningres}.
The systematic uncertainties of the data are fully propagated.
The small uncertainty on the approximately inverse power behaved b--mass 
correction 
(see Figure \ref{e_effect}) is given separately.
Further systematic uncertainties due to the possible presence of small power
corrections are not considered here.
Contrary to the case of an \as\ measurement, this
seems to be justified since the energy dependence is even less affected by 
the power correction. 
For all observables the results for $\beta_0$ or $n_{\mathrm{f}}$, 
are fully consistent
with the expected values from QCD.
This observation further strengthens the confidence in RGI perturbation
theory.

\begin{table}[tb]
\begin{center}
\begin{tabular}{|c|r@{$\pm$}l@{$\pm$}l@{$\pm$}l
                  |r@{$\pm$}l@{$\pm$}l@{$\pm$}l
                  |r@{$/$}l
                  |}\hline
Observable & \multicolumn{4}{|c|}{$\beta_0$} & 
             \multicolumn{4}{|c|}{$n_{\mathrm{f}}$}&
             \multicolumn{2}{|c|}{$\chi^2/ndf$} \\ \hline
$\langle \mathrm{Thrust}\rangle$           & 7.7 & 1.1 & 0.2 & 0.1  & 4.6 & 1.3 & 0.3 & 0.1 & 9.2 & 13\\
$\langle \mathrm{C-parameter}\rangle$      & 7.8 & 1.0 & 0.3 & 0.1  & 4.7 & 1.2 & 0.4 & 0.1 & 7.2 & 13\\
$\langle M_{\mathrm{h}}^2/E_{\mathrm{vis}}^2\;\rangle$ E def.
                 & 7.5 & 1.5 & 0.2 & 0.0  & 4.8 & 1.9 & 0.3 & 0.0 & 8.8 & 13\\
$\langle M_{\mathrm{s}}^2/E_{\mathrm{vis}}^2\;\rangle$ E def. 
                 & 7.5 & 1.1 & 0.2 & 0.0  & 5.0 & 1.3 & 0.3 & 0.1 & 7.1 & 13\\
$\langle B_{\mathrm{max}}\rangle$ & 7.7 & 1.4 & 0.1 & 0.1  & 4.7 & 1.9 & 0.1 & 0.2 & 6.3 & 13 \\
$\langle B_{\mathrm{sum}}\rangle$ & 7.7 & 0.9 & 0.1 & 0.1  & 4.8 & 1.2 & 0.1 & 0.2 & 5.9 & 13 \\
$\langle \mathrm{Major} \rangle$            & 8.0 & 1.1 & 0.1 & 0.1  & 4.3 & 1.5 & 0.2 & 0.2 & 9.2 & 13\\
\hline
Weighted Mean   & 7.7 & 0.9 & 0.1 &0.1& 4.7 & 1.2  & 0.1 & 0.1 &\multicolumn{2}{|c|}{}\\
\hline
Theory& \multicolumn{4}{|c|}{7.67} & \multicolumn{4}{|c|}{5}
& \multicolumn{2}{|c|}{}\\ \hline
\end{tabular}
\end{center}
\caption{\label{runningres}Results for the fits of $\beta_0$ or
$n_f$ for different
observables with \delphi\ data. 
The first uncertainty is statistical, the second  systematic, 
and the third indicates the uncertainty due to the b--mass correction.
Here the minimal uncertainties of the individual observables are taken as 
uncertainties of the mean value.
}
\end{table}

\begin{figure}[p]
\mbox{\epsfig{file=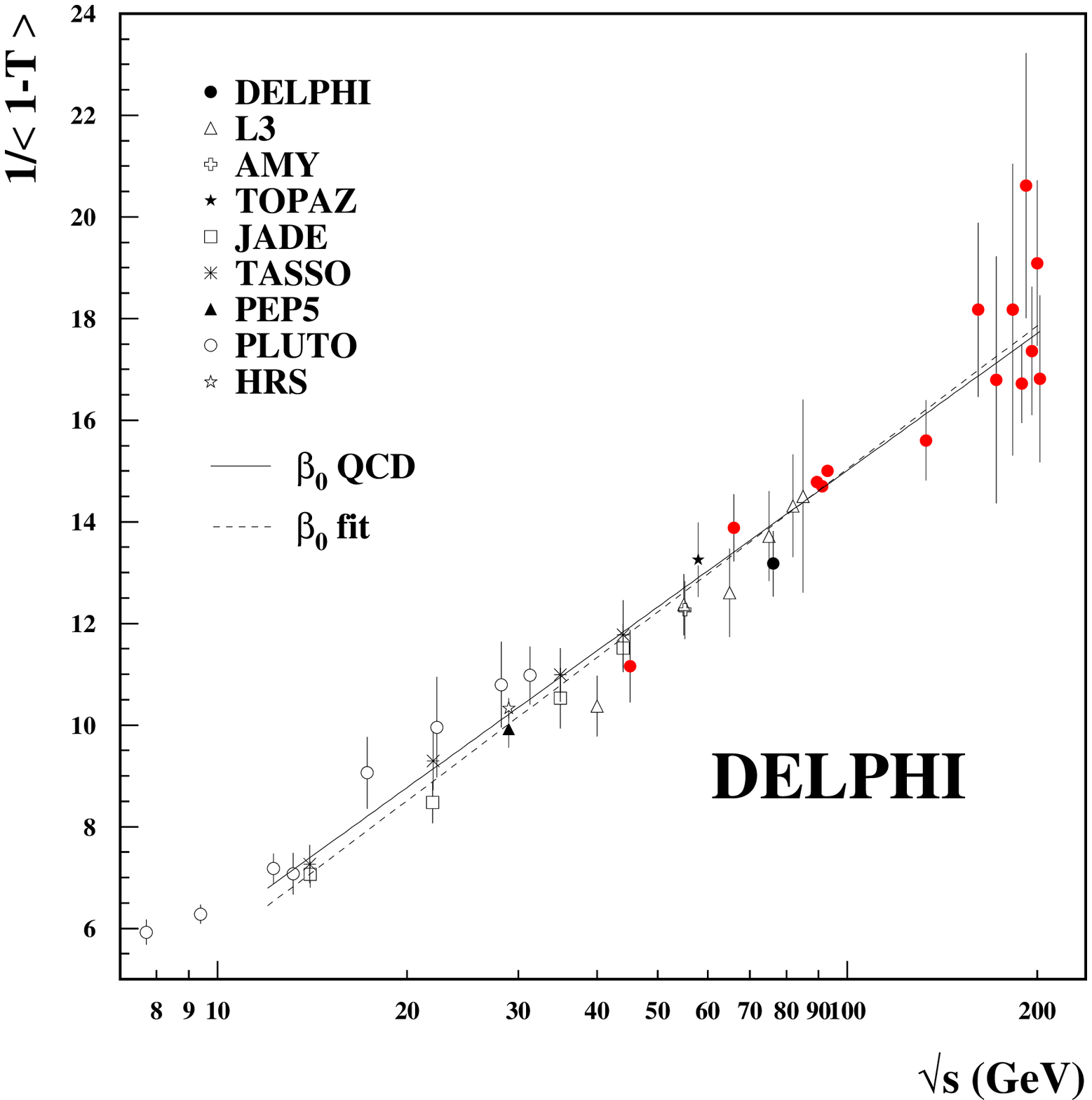,width=15cm}}
\caption{The running of  $1/\langle 1-T \rangle $ against 
$\ln(E_{cm})$. The results of the theoretical
prediction and a combined fit of the data of several experiments is shown. }
\label{thrustecrunning}
\end{figure}

In order to reduce the uncertainty on $\beta_0$ or $n_{\mathrm{f}}$ 
more data, especially from
low energy experiments, needs to be included to increase the fit range.
As $\langle M^2_{\mathrm{h}}/E_{\mathrm{vis}}^2\rangle$ 
receives large power corrections due to mass terms, in
practice only $\langle 1-T\rangle$ is left over as a possible observable.
The data given in \cite{danielpaper,collection_eventshapes,acciarri:2000mk} 
were used in a common fit.
As systematic uncertainties of the low energy data cannot be controlled 
in detail
for this fit, systematic and statistical uncertainties were combined leading to
small $\chi^2/ndf$ values.
Thus the uncertainties on the fit values already include systematic components.
The fit describes the data very well in the energy range
$\sqrt{s}=14$ to $200$\gev.
The slope of the evolution is expected to change at the b production threshold
at $\sqrt{s}\simeq 14$\gev.
The central result of the fit is shown in Figure \ref{thrustecrunning}
and Table~\ref{runningcombined}.
This table also shows the results when the $\langle 1-T \rangle$ data 
of the experiments are fitted separately.
All results are consistent with
the QCD expectation within their fit error.
In order to estimate a systematic uncertainty of $\beta_0$ or $n_{\mathrm{f}}$,
the measurements of one low energy experiment at a time have been left out
from the combined fit.
This leads to a maximum deviation from the central fit of
$\Delta\beta_0=+0.11$, $\Delta n_{\mathrm{f}}=-0.27$ (excluding PLUTO) and
$\Delta\beta_0=-0.08$, $\Delta n_{\mathrm{f}}=+0.17$ (excluding JADE).
Adding the fit error and the systematic uncertainty of the \delphi\ data in
quadrature to the above deviations from the central result
leads to the final result:
\begin{equation*}
\beta_0          = 7.86 \pm 0.32~~~, \nonumber
\end{equation*}
\begin{equation*}
n_{\mathrm{f}}     = 4.75 \pm 0.44~~~. \nonumber
\end{equation*}
These have to be compared with the QCD expectation $\beta_0 = 7.67$ 
or  $n_{\mathrm{f}}=5$.

\begin{table}[b]
\begin{center}
\begin{tabular}{|c|r@{$\pm$}l@{$\pm$}l|r@{$/$}l|r@{$\pm$}l@{$\pm$}l|}\hline
Experiment& \multicolumn{3}{|c|}{$\beta_0$} & \multicolumn{2}{|c|}{$\chi^2/ndf$}& \multicolumn{3}{|c|}{$n_{\mathrm{f}}$}\\ \hline
\delphi & 7.7 & 1.1 &0.1& 9.22 & 13& 4.6 & 1.3 & 0.1\\
L3    & 10.3 & 2.3 &0.0& 0.34 & 4 & 2.3 & 2.9 &0.0\\
JADE  & 7.8 & 0.6 &0.2& 0.92 & 2 & 4.8 & 0.7 &0.2\\
TASSO & 7.7 & 1.2 &0.2& 0.02 & 2 & 5.0 & 1.4 &0.2\\
PLUTO & 8.3 & 1.3 &0.2& 1.00 & 4 & 4.2 & 1.5 &0.2\\ \hline
combined & 7.86 & 0.16 &0.1&  25.7 & 37 & 4.75 & 0.18 &0.1\\ \hline 
Theory& \multicolumn{3}{|c|}{7.67} & \multicolumn{2}{|c|}{}& \multicolumn{3}{|c|}{5}\\ \hline
\end{tabular}
\end{center}
\caption{\label{runningcombined}
Results for the fit of $\beta_0$ for Thrust for different
experiments, and as a combined fit.
The combined fit includes additionally datapoints from
AMY, TOPAZ, MK II and HRS. The first error is statistical, the second is the
uncertainty of the b--mass correction}
\end{table}

As a final cross-check of the RGI method one may try to determine the NNLO
coefficient $\rho_2$ (see Equation \ref{RGE}) from a fit to data.
Small values of $\rho_2$ indicate a good approximation of the 
$\beta$-function by the two loop expansion.

As $\rho_2$ is mainly sensitive to low energy data again only 
$\langle 1-T\rangle$ is used. 
Fixing $b$ and $\rho_1$ to the
values expected  from QCD a two parameter fit yields:
\begin{equation}
\alpha_{\mathrm{s}} = 0.119\pm0.001\pm0.001~~~, \nonumber
\end{equation}
\begin{equation}
\rho_2            = 4 \pm 3 \pm 3.5~~~. \nonumber
\end{equation}
The result for $\rho_2$ is compatible with zero within its uncertainty,
supporting the NLO approximation.

From $\rho_2$ the expected NNLO coefficient of the $\overline{MS}$ expansion
can be deduced:
\begin{equation} 
C=915\pm 40 \nonumber~~~.
\end{equation}
This indicates a sizable correction to the NLO prediction in the 
$\overline{MS}$ scheme. Even for $\rho_2=0$ the NNLO coefficient is 
large: $C=882$.
Using the Pad\'e estimate for the parameter $C$ leads to a value of
$\rho_2=-10.8$ and a corresponding change of the strong coupling
$\Delta \alpha_s=+0.003$ with respect to the case when the NNLO term 
is omitted.

\subsection{Implications for light gluinos \label{lightgluinos}}
Knowledge of the $\beta$ function implies model independent limits on the 
presence of hypothetical particles such as gluinos.
Supersymmetric extensions of the standard 
model predict a weaker energy dependence due 
to the effect of gluino loops. 
In testing this hypothesis one should evidently not presume the QCD prediction 
for $\beta_1$ as  has been done for the $\beta_0$ measurement in the 
previous section. Instead the full logarithmic energy slope
\begin{eqnarray}
\label{logs}
\frac{\mathrm{d}R^{-1}}{\mathrm{d}\ln Q^2}=-\frac{\beta(R)}{R^2}
\end{eqnarray}
has to be measured.
Evaluating this equation in second order implies a small dependence on 
$R(Q)$ or the energy respectively. For our measurements the  choice of the
Z--mass as central scale is reasonable.
For $R={\langle 1-T \rangle}/{A}$ one gets:
$$
\frac{\mathrm{d}R ^{-1}}{\mathrm{d}\ln Q^2}=8.32 
$$
as the QCD prediction. Supersymmetry alters this prediction.
The actual value of the logarithmic slope, however, depends on the mass of the additional particles. Gluinos in the mass range from 5 to 190\gev\ are excluded from direct searches\cite{PDG2000}. Gluinos in the open mass range below 
5\gev\ can be considered massless compared to the energy range of this 
experiment. Therefore one expects $\beta_0^{SUSY}=\frac{27-2N_f}{3}$ and 
$\beta_1^{SUSY}=\frac{81-19N_f}{3}$. This yields the supersymmetry prediction 
for the logarithmic energy slope of:
$$
\frac{\mathrm{d}R ^{-1}}{\mathrm{d}\ln Q^2}=5.66 \quad .
$$ 
Measuring the slope by fitting the function $1/(a+b\ln Q)$ to the 
$\frac{\langle 1-T \rangle}{A}$ data yields:
$$
\frac{\mathrm{d}R^{-1}}{\mathrm{d}\ln Q^2 }=8.70 \pm 0.35 \quad .
$$
The error contains the statistical and systematic uncertainties in a similar 
way as discussed for the $\beta_0$ measurement in the previous section.
The existence of light gluinos can be firmly excluded by our 
result.

\section{Summary and conclusions\label{summary}}
Event shape distributions and their means or integrals
for the observables
$1-T$, $\mbox{C-Parameter}$,
$B_{\mathrm{max}}$, $B_{\mathrm{sum}}$, $\mbox{EEC}$, $\mbox{JCEF}$,
the classical and the $E$- or $p$-scheme definitions of
$M_{\mathrm{h}}^2/E_{\mathrm{vis}}^2$, $M_{\mathrm{s}}^2/E_{\mathrm{vis}}^2$,
have been presented in the wide range of
hadronic centre-of-mass energies from $45\gev$ to $202\gev$.
Measurements at energies below
$\sqrt{s}=M_Z$ were obtained from events containing a hard radiated
photon.

The data can be successfully described over the whole energy range
using Monte Carlo fragmentation models or by analytical power correction
models. The successful description of the $\mbox{EEC}$ by a power correction
ansatz is reported for the first time. Fitting the Dokshitzer--Webber power 
corrections to event shape distributions yields 
$\alpha_s = 0.1110\pm0.0055_{\mathrm{RMS}}$ and 
$\alpha_0=0.559\pm0.073_{\mathrm{RMS}}$. The application to the mean values 
of event shapes yields $\alpha_s=0.1217\pm0.0046_{\mathrm{RMS}}$ and 
$\alpha_0=0.431\pm0.048_{\mathrm{RMS}}$ respectively. 
A measurement of the shift of the Sudakov shoulder of the
$\mbox{C-Parameter}$ indicates an approximately constant power shift
over the whole three jet range.


In addition to the comparison of power correction models, the data for
inclusive shape means of seven observables
depending only on a single energy scale
and integrals of the $\mbox{EEC}$ and $\mbox{JCEF}$ have also been compared to
results of the so--called Renormalisation Group Invariant, RGI,
perturbation theory with and without additional power terms.
This method allows for a measurement of $\beta_0$ which is 
independent of any renormalisation scheme or scale. With respect to $\alpha_s$
the RGI prediction is equivalent to the ECH scheme. 

It has been observed that RGI perturbation theory is able to
describe the energy evolution of these data consistently
with a single value of the strong coupling parameter:
$\alpha_s(M_Z) = 0.1201 \pm 0.0020_{RMS}$. Within RGI perturbation theory 
there is  no need  for power corrections.
The small R.M.S. of the \as\ values obtained from the different
observables indicates an improved decription of the data by  
RGI perturbation theory compared to the standard $\overline{MS}$ treatment.
Furthermore it serves as an important consistency check of the method.

The most important single result of the analysis is the measurement of
the $\beta$-function of strong interactions. 
For the leading coefficient 
, assuming QCD, the average of seven observables as measured by DELPHI is:
\begin{equation*}
\beta_0 =7.7 \pm 0.9 \pm 0.1 ~~~ ,
\end{equation*}
where the first uncertainty is statistical and the second systematic.
This corresponds to the number of active flavours:
\begin{equation*}
n_f= 4.7 \pm 1.2 \pm 0.1  ~~~ .
\end{equation*}
The systematic uncertainty accounts for experimental uncertainties as well as
for the uncertainty induced due to the correction for b hadron decays.
Inclusion of additional low energy data for the observable
$\langle 1-T \rangle$ yields the result:
\begin{equation}
\beta_0 = 7.86 \pm 0.32~~~,\nonumber
\end{equation}
\begin{equation}   
n_f = 4.75 \pm 0.44~~~.\nonumber
\end{equation}
Here the uncertainty includes the  sources mentioned above as
well as an estimate of the systematic error induced by the inclusion of the
low energy data.

Within RGI this quantity can be 
derived without any renormalisation scheme dependence. Power 
corrections have been found to be negligible.  
The precision of this result is greatly increased compared to 
previous determinations of $\beta_0$ from event shape 
observables~\cite{acciarri:2000mk} as well as
to a determination based on the most precise measurements of 
\as\cite{bethke:2000}: $\beta_0=7.76\pm0.44$. 
Further reduction of the uncertainty is to be expected from a proper
combination of
the results of the LEP experiments.
The analysis should then also be repeated for observables other than
$\langle 1-T \rangle$.

Fitting directly the logarithmic energy slope yields
\begin{eqnarray*}
\frac{\mathrm{d}R^{-1}}{\mathrm{d}\ln{Q^2}}=\beta(R) = 8.70 \pm 0.35 ~~~.
\end{eqnarray*} 
This measurement excludes light gluinos in the open mass range below 5\gev\ 
\cite{PDG2000} in a model independent way.

Provided the possible presence of power terms in the event shape means is
clarified by future studies, the good stability of the results for
\as\ for a larger number of event shape observable means observed 
when using RGI perturbation theory  
may indicate the possibility of improved measurements of the strong coupling
\as.


\subsection*{Acknowledgements}
\vskip 3 mm
 We are greatly indebted to our technical 
collaborators, to the members of the CERN-SL Division for the excellent 
performance of the LEP collider, and to the funding agencies for their
support in building and operating the DELPHI detector.\\
We acknowledge in particular the support of \\
Austrian Federal Ministry of Education, Science and Culture,
GZ 616.364/2-III/2a/98, \\
FNRS--FWO, Flanders Institute to encourage scientific and technological 
research in the industry (IWT), Belgium,  \\
FINEP, CNPq, CAPES, FUJB and FAPERJ, Brazil, \\
Czech Ministry of Industry and Trade, GA CR 202/99/1362,\\
Commission of the European Communities (DG XII), \\
Direction des Sciences de la Mati$\grave{\mbox{\rm e}}$re, CEA, France, \\
Bundesministerium f$\ddot{\mbox{\rm u}}$r Bildung, Wissenschaft, Forschung 
und Technologie, Germany,\\
General Secretariat for Research and Technology, Greece, \\
National Science Foundation (NWO) and Foundation for Research on Matter (FOM),
The Netherlands, \\
Norwegian Research Council,  \\
State Committee for Scientific Research, Poland, SPUB-M/CERN/PO3/DZ296/2000,
SPUB-M/CERN/PO3/DZ297/2000, 2P03B 104 19 and 2P03B 69 23(2002-2004)\\
JNICT--Junta Nacional de Investiga\c{c}\~{a}o Cient\'{\i}fica 
e Tecnol$\acute{\mbox{\rm o}}$gica, Portugal, \\
Vedecka grantova agentura MS SR, Slovakia, Nr. 95/5195/134, \\
Ministry of Science and Technology of the Republic of Slovenia, \\
CICYT, Spain, AEN99-0950 and AEN99-0761,  \\
The Swedish Natural Science Research Council,      \\
Particle Physics and Astronomy Research Council, UK, \\
Department of Energy, USA, DE-FG02-01ER41155. \\
We thank A.A. Pivovarov for numerous enlightening discussions
and helpful explanations.


\clearpage
\newcommand{\esdcollection}{
 AMY Coll., Y.K. Li et al. {\em Phys. Rev.} {\bf D41} (1990) 2675. \\
 HRS Coll., D. Bender et al. {\em Phys. Rev.} {\bf D31} (1985) 1.\\
 P.A. Movilla Fernandez, et. al. and the JADE Coll.
 {\em Eur. Phys. J.} {\bf C1} (1998) 461.\\
 PEP5 Coll., A. Peterson et al. {\em Phys. Rev.} {\bf D37} (1988) 1. \\
 PLUTO Coll., C. Berger et al. {\em Z. Phys.} {\bf C12} (1982) 297. \\
 TASSO Coll., W. Braunschweig et al. {\em Z. Phys.} {\bf C47} (1990) 187.\\
 TOPAZ Coll., Y. Ohnishi et al. {\em Phys. Lett.} {\bf B313} (1993) 475.
}

\end{document}